\documentclass[twocolumn,amstext,amssymb,amsmath,superscriptaddress,showpacs,nofootinbib,aps,prd,preprintnumbers,floatfix]{revtex4-1}
\usepackage{lmodern}
\usepackage[T1]{fontenc}
\usepackage[utf8]{inputenc}
\usepackage{amstext,amssymb,amsmath}
\usepackage{hyperref}
\usepackage[table]{xcolor}
\usepackage{multirow}
\usepackage{graphicx}
\usepackage{slashed}
\usepackage{enumitem}
\usepackage[normalem]{ulem}
\usepackage{comment}
\usepackage{placeins}

\definecolor{LightGray}{gray}{0.91}
\definecolor{LightBlue}{rgb}{0.87, 0.94, 1}

\newcommand{\eq}[1]{\eqref{eq:#1}}
\newcommand{\fig}[1]{Fig.~\ref{fig:#1}}
\newcommand{\tab}[1]{Tab.~\ref{tab:#1}}
\newcommand{\tr}{\mathrm{tr}}

\newcommand{\TeV}{\,\text{TeV}}

\newcommand{\Sec}[1]{Sec.~\ref{sec:#1}}
\newcommand{\Subsec}[1]{Sec.~\ref{subsec:#1}}
\newcommand{\App}[1]{App.~\ref{sec:#1}}
\newcommand{\MPl}{M_\text{Pl}}

\newcommand{\mhPrsq}{m_{h^\prime}^2}
\newcommand{\msPrsq}{m_{s^\prime}^2}
\newcommand{\N}{\mathcal{N}}

\makeatletter
    \def\CT@@do@color{%
      \global\let\CT@do@color\relax
            \@tempdima\wd\z@
            \advance\@tempdima\@tempdimb
            \advance\@tempdima\@tempdimc
    \advance\@tempdimb\tabcolsep
    \advance\@tempdimc\tabcolsep
    \advance\@tempdima2\tabcolsep
            \kern-\@tempdimb
            \leaders\vrule
                    \hskip\@tempdima\@plus  1fill
            \kern-\@tempdimc
            \hskip-\wd\z@ \@plus -1fill }
    \makeatother

\begin{document}

\title{Vacuum Stability in the Standard Model and Beyond}
\author{Gudrun~Hiller}
\affiliation{TU Dortmund University, Department of Physics, Otto-Hahn-Str.4, D-44221 Dortmund, Germany}
\affiliation{Department of Physics and Astronomy, University of Sussex, Brighton, BN1 9QH, U.K.}
\author{Tim~H\"ohne}
\affiliation{TU Dortmund University, Department of Physics, Otto-Hahn-Str.4, D-44221 Dortmund, Germany}
\author{Daniel~F.~Litim}
\affiliation{Department of Physics and Astronomy, University of Sussex, Brighton, BN1 9QH, U.K.}
\author{Tom~Steudtner}
\affiliation{TU Dortmund University, Department of Physics, Otto-Hahn-Str.4, D-44221 Dortmund, Germany}
\begin{abstract}
We revisit the  stability of the Standard Model  vacuum, and investigate its quantum effective potential using  the highest available  orders in perturbation theory and  the most accurate determination of input parameters to date. We observe that  the stability of the electroweak vacuum centrally depends on the values of the top mass and the strong coupling constant.  
We estimate that reducing  their uncertainties by a factor of  two to three is sufficient to establish or refute  SM vacuum stability  at the $5\sigma$ level. We further investigate vacuum stability for a variety of singlet scalar field extensions  with and without flavor using the Higgs portal mechanism.  We identify the  BSM parameter spaces for stability and find sizable room for new physics. We further study the phenomenology of Planck-safe models at colliders, and determine the impact on the Higgs trilinear,  the Higgs-to-electroweak-boson, and the Higgs quartic couplings, some of which can be significant. The former two  can be probed at the HL-LHC,  the latter requires a future collider with sufficient energy and precision such as the FCC-hh.
\end{abstract}
\maketitle
\tableofcontents
\flushbottom

\section{Introduction}
It is widely appreciated that the 
Standard Model (SM) of particle physics is incomplete.  
Yet, clear-cut signatures for new physics  at the electroweak scale are missing, despite of extensive experimental searches and a variety of anomalies in the data. 
Also, theory guidance for top-down model building from first principles   beyond the paradigm of effective theories is scarce.
Therefore, it has  been proposed to revisit the  metastability of the SM vacuum \cite{Buttazzo:2013uya} and to turn it into a bottom-up model building task  \cite{Hiller:2022rla,Hiller:2023bdb}. 
While the onset of metastability in the SM around $10^{10}\,\text{GeV}$
is a high energy effect -- though still far below the quantum gravity scale -- 
its remedy may arise from new physics at any scale below,  possibly as low as a  TeV. At the same time, new physics modifications of the Higgs potential may also affect interaction vertices of the Higgs particle,  thus offering additional opportunities for indirect tests of stability at  present and future colliders such as the (high-luminosity) Large Hadron Collider LHC (HL-LHC) \cite{ATLAS:2019mfr}, the Future Circular Collider (FCC) \cite{FCC:2018byv},
the Chinese Electron Positron Collider (CEPC) \cite{CEPCStudyGroup:2018ghi},
the International Linear Collider Project (ILC) \cite{ILC:2019gyn}
or a muon collider \cite{Casarsa:2023vqx}.
In this light, the aim of this paper is threefold:

Firstly, we revisit  the  stability of the SM vacuum, and investigate its quantum effective potential using  the highest available  orders in perturbation theory and  the most accurate determination of input parameters to date.
We find that  the stability  centrally depends on the values of the top mass and the strong coupling constant, their uncertainties~\cite{Workman:2022ynf}, and correlation~\cite{CMS:2019esx}. We  estimate that reducing  their uncertainties by a factor of  two to three is sufficient to establish or refute   stability  at the $5\sigma$ level. 
 In principle, this goal can be achieved with threshold scanning at $e^+ e^-$-colliders, for which a projected total uncertainty  of around  $75 \, \text{MeV}$ for the top mass is deemed achievable \cite{Simon:2019axh}. 

Secondly, we  systematically investigate the stability of  scalar field extensions  with and without flavor, using the Higgs portal mechanism. 
This includes the addition of real, complex, $O(N)$, or $SU(N)\times SU(N)$ symmetric singlet scalar  fields $S$, their self-interactions, and their renormalizable portal couplings $\sim (H^\dagger H)( S^\dagger S)$ with the Higgs $H$. 
For either of these, we identify the  BSM parameter spaces of masses and couplings for SM extensions to be ``Planck-safe'' -- meaning stable up to or at the Planck scale, and free of subplanckian Landau poles -- and  uncover sizable room for new physics. 
Results are achieved by extensive studies of the two-loop RG running of 
couplings between the  scale of new physics  and the Planck scale using the precision tool ARGES \cite{Litim:2020jvl}.
Our methodology has been developed in a series of 
 earlier works  \cite{Hiller:2019mou,Hiller:2020fbu,Bissmann:2020lge,Bause:2021prv,Hiller:2022rla,PlanckSafeQuark}, inspired by models of particle physics with controlled interacting UV fixed points \cite{Litim:2014uca,Litim:2015iea,Bond:2016dvk,Buyukbese:2017ehm,Bond:2017lnq,Bond:2017suy,Bond:2017tbw,Bond:2017wut,Kowalska:2017fzw,Bond:2018oco,Bond:2019npq,Fabbrichesi:2020svm,Bond:2021tgu}.
Previous studies of the Higgs portal 
include BSM models with a  real~\cite{Falkowski:2015iwa,Khan:2014kba,Han:2015hda,Garg:2017iva}, complex~\cite{Gabrielli:2013hma,Elias-Miro:2012eoi,Gonderinger:2012rd,Costa:2014qga,Khoze:2014xha,Anchordoqui:2012fq}, or charged scalar(s)  \cite{Bandyopadhyay:2016oif,Bandyopadhyay:2021kue,Chakrabarty:2020jro,Hamada:2015bra},  models with additional BSM Yukawa couplings
~\cite{Latosinski:2015pba,Hiller:2019mou,Hiller:2020fbu,Bause:2021prv,Xiao:2014kba,Dhuria:2015ufo,Salvio:2015jgu,Son:2015vfl,DuttaBanik:2018emv,Borah:2020nsz} and 
two-Higgs-doublet models~\cite{Chowdhury:2015yja,Khan:2015ipa,Ferreira:2015rha,Ferreira:2015pfi,Bhattacharya:2019fgs,Swiezewska:2015paa,Chakrabarty:2016smc,Schuh:2018hig,Bagnaschi:2015pwa}  (for an overview see \cite{Hiller:2022rla} and references therein).

Thirdly, we investigate  the modifications of the Higgs potential dictated by Planck-safe SM extensions and their phenomenology at colliders. In particular, we  determine the impact of new physics on the Higgs trilinear,  the Higgs-to-electroweak-boson, and the Higgs quartic couplings, which can be sizable. We show that the former two  can already be probed at the HL-LHC \cite{ATLAS:2019mfr},  whereas the latter will require a future collider with sufficient energy and precision such as the FCC-hh \cite{FCC:2018byv}.

The paper is organized as follows.
We begin with an update of  the SM quantum effective potential and  its stability in terms of the most critical input parameters 
 (\Sec{SM-stability}). This is followed by a study of the Higgs portal 
 for a variety of singlet scalar field extensions with and without flavor,
 and their BSM parameter spaces for  safe and stable extensions up to the Planck scale  (\Sec{hp}).
   We further work out the phenomenology of stable SM extensions,  in particular their impact on the Higgs trilinear, quartic, and couplings to the $Z$ boson, and their signatures at present and future colliders (\Sec{Pheno}).
    We  close with a brief discussion of results and some conclusions (\Sec{Conclusion}). 
    Four appendices contain further details of the SM stability analysis (\App{SM-stab-details}), conventions and terminology (\App{PS}), scalar mixing  (\App{Mix}), and constraints from unitarity (\App{uni}).

\section{Revisiting SM Vacuum Stability}\label{sec:SM-stability}

Since the discovery of the Higgs boson~\cite{ATLAS:2012yve,CMS:2012qbp} the meta-stability of its potential has been evidenced~\cite{Degrassi:2012ry,Buttazzo:2013uya}. 
The metastability appears to be non-pathological in that the tunnel rate is  many orders of magnitude larger than the age of the universe~\cite{Buttazzo:2013uya}. Also, absolute stability has neither been excluded conclusively due to underlying uncertainties of SM observables.
Ever since these early works, important strides have been made in both theory and experiment improving this prediction. 
Thus, a high-precision determination of the region of vacuum stability in the SM is warranted and will be conducted here, enhancing previous studies, e.g.~\cite{Degrassi:2012ry,Bezrukov:2012sa,Alekhin:2012py,Buttazzo:2013uya,Andreassen:2014gha,Bednyakov:2015sca,Chigusa:2018uuj}  .\footnote{Determining the tunnel rate into the false vacuum, and verifying that the lifetime is indeed  much larger than the age of the universe is beyond the scope of this work.}

\subsection{Input}
 Progress on the experimental side consists of improved precision for all input observables that determine vacuum stability in the SM.
These include the Higgs, top and $Z$ pole masses $M_{h,t,Z}$, the 5-flavor strong coupling $\alpha_s^{(5)}(\mu=M_Z)$, the fine-structure constant $\alpha_e$ and the hadronic contribution to its running $\Delta\alpha_e^{(5)}(\mu=M_Z)$, Fermi's constant $G_F$, $\overline{\text{MS}}$ quark masses $m_b(\mu=m_b)$, $m_c(\mu=m_c)$, $m_{u,d,s}(\mu=2\,\text{GeV})$ and lepton pole masses $M_{e,\mu,\tau}$. 
Their central values and uncertainties are taken from the  2024 update of the PDG~\cite{PDG2024}.

Progress on the theory side consists of several streams.
Specifically, using \cite{Alam:2022cdv}, all running SM parameters and their uncertainties are determined from the input observables at a reference scale
\begin{equation}\label{eq:mu_ref}
    \mu_\text{ref} = 200~\text{GeV}\,.
\end{equation}
The precision of this procedure is an improvement upon \cite{Buttazzo:2013uya}, an overview of all loop contributions is found in \cite{Martin:2019lqd}.
In particular, five loop logarithmic resummations are conducted~\cite{Baikov:2012zm,Baikov:2016tgj,Herzog:2017ohr,Luthe:2017ttg,Chetyrkin:2017bjc} while
threshold and matching corrections to the light quark masses and gauge couplings are considered up to four loops in QCD~\cite{Melnikov:2000qh,Martin:2018yow}. The pole masses of the top quark, $Z$ and Higgs boson are matched at full two loop precision with leading three-loop corrections~\cite{Martin:2014cxa,Martin:2015rea,Martin:2016xsp,Martin:2022qiv}.

 SM vacuum stability is established provided the quantum effective potential is bounded from below and devoid of deeper secondary minima. For the SM effective potential, we employ the following RG-improved ansatz 
\begin{equation}\label{eq:Veff}
 \frac{1}{(4\pi)^2 } V_\text{eff} = \frac{1}{4} \alpha_{\lambda,\text{eff}}(h) e^{4  \bar{\Gamma}(h)} h^4\,.
\end{equation}
Here $h$ denotes the Higgs field, the factor $\bar{\Gamma}(h)$  arises from resumming the Higgs anomalous dimension, and $\alpha_{\lambda,\text{eff}}(h)$ relates to the effective Higgs coupling extracted from the effective potential. The latter  signals vacuum stability provided that
\begin{equation}
  \alpha_{\lambda,\text{eff}} > 0
\end{equation} 
for all field values. In our computation, the coupling $ \alpha_{\lambda,\text{eff}} $  is obtained at a constant field value $h=h_0$ by matching the effective potential \eq{Veff} to fixed-order perturbative calculations of the effective potential~\cite{Ford:1992pn,Martin:2013gka,Martin:2014bca,Martin:2015eia,Martin:2017lqn,Martin:2018emo}. 
  We work in Landau gauge. While the effective potential, and hence  $ \alpha_{\lambda,\text{eff}} $,  is gauge-dependent, the depth of its minimum is not, see App.~\ref{sec:SM-stab-details} for details.
  
At one-loop, we have
\begin{equation}\label{eq:lambda_0-1L}
  \begin{aligned}
    & \alpha_{\lambda,\text{eff}}(h_0)   = \alpha_\lambda  \\
    &{}\quad + \frac{3}{16} (\alpha_1 + \alpha_2)^2 \left(\ln\frac{\alpha_1 + \alpha_2}{2 \alpha_t} - \frac56\right)   \\
     &{}\quad +  4\,\alpha_\lambda^2 \left( \ln\frac{4\alpha_\lambda}{\alpha_t} - \frac32\right) + \frac{3}{8} \alpha_2^2 \left(\ln\frac{\alpha_2}{2 \alpha_t} - \frac56\right)  \\
    &{}\quad - \sum_{f} N_f  \alpha_f^2 \left(\ln\frac{\alpha_f}{\alpha_t} - \frac{3}{2}\right) 
    + \mathcal{O}(\text{2 loop})\,,
  \end{aligned}
  \end{equation}
 where the running couplings are given by
  \begin{equation}
    \alpha_i=\frac{z_i^2}{(4 \pi)^2} \text{ with } z_i = g_{1,2,3}, \text{ or } y_{t,b,c,s,u,d,\tau,\mu,e}\,.  \label{eq:gy}
  \end{equation}
  Further, $\alpha_\lambda = \lambda/(4\pi)^2$ denotes the Higgs self-interaction of the tree-level potential.
  The sum in the last line of \eq{lambda_0-1L} runs over all fermions flavors $f=t,b,c,s,u,d,\tau,\mu,e$, where $N_f=3$ for quarks and  $N_f=1$ for leptons.
  In addition to the well-known one-loop terms \eq{lambda_0-1L}, we include electroweak, strong, top and scalar contributions at two loops~\cite{Ford:1992pn}, strong and top contributions at three loops~\cite{Martin:2013gka,Martin:2017lqn} and QCD contributions at four loops~\cite{Martin:2015eia} with Goldstone resummation~\cite{Martin:2014bca,Martin:2017lqn}.

   A virtue of the ansatz \eqref{eq:Veff} is that large logarithms   such as $\log h/\mu_\text{ref}$ are resummed (see \App{SM-stab-details} for details).\footnote{In this regard, our ansatz for the effective potential is slightly different from those used in earlier works, e.g.~\cite{Casas:1994qy,Buttazzo:2013uya,Andreassen:2014gha,Bednyakov:2015sca}. We have checked that results are equivalent within errors.}
   The matching \eq{lambda_0-1L} is additionally optimized by choosing the initial field value $h_0$ such that  $\mu_\text{ref}^2 = 8 \pi^2 \alpha_t(\mu_\text{ref}) h_0^2$, which eliminates logarithms stemming from top loops.  
Lastly, the coupling $\alpha_{\lambda,\text{eff}}(h)$ for larger fields $h>h_0$ is obtained via a resummation procedure detailed in \App{SM-stab-details}. At leading order, this is equivalent to an RG evolution of \eq{lambda_0-1L}. 
   To implement the resummation, we have used four-loop beta functions for gauge couplings~\cite{Davies:2019onf,Davies:2021mnc,Bednyakov:2021qxa}, three-loop for Yukawa~\cite{Chetyrkin:2012rz,Bednyakov:2012en,Davies:2021mnc,Bednyakov:2021qxa} as well as quartic~\cite{Chetyrkin:2012rz,Chetyrkin:2013wya,Bednyakov:2013cpa} interactions. On top of that, 5-loop QCD corrections for the strong coupling beta function~\cite{Baikov:2016tgj,Herzog:2017ohr,Luthe:2017ttg,Chetyrkin:2017bjc} as well and four loops for Yukawas and quartic~\cite{Chetyrkin:2016ruf} are utilized. This is accompanied by three-loop anomalous dimensions of the Higgs field~\cite{Chetyrkin:2012rz}.

\subsection{Results}

\begin{figure}
  \centering
  \begin{tabular}{c}
    \includegraphics[scale=.55]{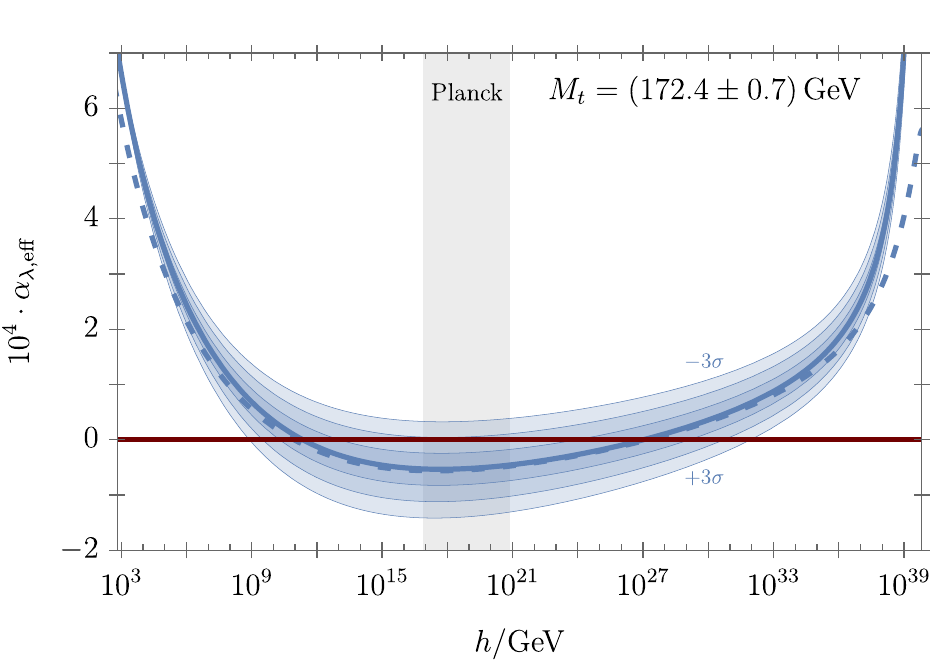} 
  \end{tabular}
    
  \caption{
  Shown is the coupling $\alpha_{\lambda,\text{eff}}$ characterizing the quantum effective potential at the field value $h$ where the top mass is given by the PDG 2024 central value $M_t^{\sigma}$ (thick blue line) as well as three standard deviations in either direction (thin lines).
  For comparison, we also  show the resummed tree-level potential with the central value $M_t^{\sigma}$ as the top mass (thick dashed line). The gray band indicates the Planck scale. Effective potentials for which $\alpha_{\lambda,\text{eff}} $ stays positive for all field values $h$ are stable.
 }
  \label{fig:SM-eff-stab}
\end{figure}

We find that at  the reference scale $\mu_\text{ref}$  the effective quartic $\alpha_{\lambda,\text{eff}}$ is enhanced by $17\%$ over its tree level value  $\alpha_{\lambda}(\mu_\text{ref})$.
 Inspecting the effective potential, instability is already manifest in resummed tree-level potential where $\alpha_{\lambda,\text{eff}}(h_0) = \alpha_\lambda$ and the potential turns negative at $\Lambda_{0} \approx 2.1\cdot 10^{11}$~GeV\footnote{Note that this scale is gauge-dependent, see \App{SM-stab-details} for details.} using PDG central values.
Taking into account quantum corrections to the potential, $\alpha_{\lambda,\text{eff}}(h)$ still turns negative before the Planck scale around $\Lambda_{0, \text{eff}} \approx 5.3\cdot 10^{11}$~GeV.
However, the coupling remains only slightly negative $\alpha_{\lambda,\text{eff}} \gtrsim - 10^{-4}$, hinting at metastability of the potential. If quantum gravity effects are neglected, $\alpha_{\lambda,\text{eff}}$ turns positive again after the Planck scale.
The field dependence of $\alpha_{\lambda,\text{eff}}$ (solid, including $3 \sigma$ uncertainty bands form $M_t$) and the resummed tree-level coupling $\alpha_{\lambda}$ (dashed, using central values) are displayed in \fig{SM-eff-stab}. Overall, loop corrections to the effective potential at the reference scale are subleading to RG running effects for position and depth of the minimum.  

\begin{table}[h!]
  \centering
  \begin{tabular}{|c|l|rr|rr|}
       \hline
       \rowcolor{LightBlue} Obs. & Value & \multicolumn{2}{|c|}{$\alpha_\lambda > 0$} &  \multicolumn{2}{|c|}{$\alpha_{\lambda,\text{eff}} > 0$}\\ 
       \hline 
      PDG 2024 \cite{PDG2024}: &  &  & &  & \\
      $M_h/\text{GeV}$&$ 125.20(11)$ & $127.97$ & $+ 25.2\sigma$& $127.85$ &$+24.0\sigma$ \\
      $M_t^\sigma/\text{GeV}$ & $172.4(7)$ & $171.04$ & $- \phantom{0}1.9\sigma$ & $171.10$ & $- \phantom{0}1.9 \sigma$ \\
       $M_t^\text{MC}/\text{GeV}$ &$ 172.57(29)$ & & $- \phantom{0}5.3 \sigma$ &  & $- \phantom{0}5.1 \sigma$ \\
       $m_t/\text{GeV}$ &$162.5(^{+2.1}_{-1.5}) $ &$161.3\phantom{0}$&$- \phantom{0}0.8 \sigma$ &$161.4\phantom{0}$ &$- \phantom{0}0.7 \sigma$ \\
       $\alpha_s^{(5)}(M_Z) $ & $ 0.1180(9)$ & $0.1215$ & $+\phantom{0}3.9\sigma$ & $0.1213$ & $+\phantom{0}3.7 \sigma$ \\
       \hline
       PDG 2023 \cite{Workman:2022ynf}: &  &  & &  & \\
       $M_h/\text{GeV}$&$125.25(17)$ & $128.17$ & $+ 17.2\sigma$& $128.05$ &$+16.5\sigma$  \\
       $M_t^\sigma/\text{GeV}$ & $172.5(7)$ & $171.06$ & $- \phantom{0}2.1\sigma$ & $171.13$ & $- \phantom{0}2.0 \sigma$ \\
       $M_t^\text{MC}/\text{GeV}$ & $172.69(30)$ & & $- \phantom{0}5.4 \sigma$ &  & $- \phantom{0}5.2 \sigma$ \\
        $m_t/\text{GeV}$ &$162.5(^{+2.1}_{-1.5}) $ &$161.4\phantom{0}$&$- \phantom{0}0.8 \sigma$ &$161.4\phantom{0}$ &$- \phantom{0}0.7 \sigma$  \\
        $\alpha_s^{(5)}(M_Z) $ & $ 0.1180(9)$ & $0.1217$ & $+\phantom{0}4.1\sigma$ & $0.1215$ & $+\phantom{0}3.9 \sigma$ \\
        \hline
        CMS \cite{CMS:2019esx}:  &  &  & &  & \\
        $M_t/\text{GeV}$ & $170.5(8)$ &$169.25$&$- \phantom{0}1.6 \sigma$ &$169.31$ & $- \phantom{0}1.5 \sigma$ \\
        $\alpha_s^{(5)}(M_Z) $ & $ 0.1135(^{+21}_{-17})$ & $0.1167$ &$+\phantom{0}1.5\sigma$ & $0.1165$  & $+\phantom{0}1.4 \sigma$ \\
       \hline
  \end{tabular}
  \caption{
  Strong coupling, Higgs or top mass and their uncertainty from PDG 2024~\cite{PDG2024} and 2023~\cite{Workman:2022ynf} updates  as well as CMS analysis~\cite{CMS:2019esx}. 
  For each observable, the shift around the central values from required to stabilize the tree level $(\alpha_\lambda > 0)$ or quantum effective potential $(\alpha_{\lambda,\text{eff}} > 0)$ in the SM before the Planck scale is given. In the PDG, top masses from cross-section measurements $(M_t^\sigma)$, Monte-Carlo generators $(M_t^\text{MC})$ as well as $\overline{\text{MS}}$ mass of the top at the scale $(m_t)$ are distinguished. For the study~\cite{CMS:2019esx}, other input observables are taken from~\cite{PDG2024}.
  }
  \label{tab:uncertainties}
\end{table}

An overview of the most sensitive observables to determine the stability of the SM potential is collected in \tab{uncertainties}.
Recalling the direct link  between the mass of the Higgs  and its quartic, $\lambda \simeq G_F/\sqrt{2} M_h^2$, the Higgs pole  mass is key to extract $\alpha_\lambda(\mu_\text{ref})$ and thus its uncertainty has the highest impact at tree-level for $\alpha_{\lambda,\text{eff}}(\mu_\text{ref})$. However, the uncertainty of $M_h$  is of minor importance for the RG running of   $\alpha_\lambda$. The reasons are  {\it i)} the $M_h$ uncertainty is small, 
 about a permille,   {\it ii)}   the value of the quartic $\alpha_\lambda(\mu_\text{ref})\approx 7.8 \cdot  10^{-4}$ is small, and smaller than other couplings involved and discussed below,
 $\alpha_s$ and $\alpha_t$,
 and that  {\it iii)} the beta function of  $\alpha_\lambda$  does not vanish when $\alpha_\lambda$ vanishes.
 This  makes 
 the impact  of  the Higgs quartic, and hence the uncertainty of $M_h$,  on the RG-induced loss of vacuum stability very small.
As  shown in the first row of  \tab{uncertainties}, a Higgs mass  about  $24\,\sigma$ higher than the present value, above $127.85$~GeV  would be required to achieve stability (for central values of other couplings).
On the other hand,  $\alpha_t$ and $\alpha_3$, corresponding to the top mass and $\alpha_s^{(5)}(M_Z)$,  are the largest couplings at $\mu_\text{ref}$ and dominate the running. 
To stabilize the Higgs, 
$\alpha_s^{(5)}(M_Z)$ requires an $3.7\sigma$ upward shift from the 2024 PDG world average.
Note however that many individual studies summarized in \cite{Workman:2022ynf} quote much larger uncertainties. 
The critical role of the strong coupling may appear surprising given that its influence  on the effective potential as well as the running is loop-suppressed. 
This suppression, however, is effectively compensated 
by $\alpha_3$ (and $\alpha_t$) being numerically large compared to the other SM gauge and Yukawa couplings.\footnote{This  effect has previously been noticed in the context of the strong gauge portal for stability~\cite{Hiller:2022rla}.}

Alternatively, a smaller value of the top mass may also entail vacuum stability in the SM.
The PDG provides three world averages for the top mass. The  pole mass  extracted from cross section measurements, $M_t^\sigma = (172.4 \pm 0.7)~\text{GeV}$, 
implies that a $1.9 \sigma$ downward shift from its central  value stabilizes the Higgs potential.
A secondary top mass estimate $M_t^\text{MC} = (172.57 \pm 0.29)$~GeV stems from template fits of kinematic distributions sensitive to
the top-quark pole mass \cite{Hoang:2020iah}. These template fits are based on a modeling of top-quark production and decay dynamics in Monte Carlo event generators.
The small uncertainty 
requires a $5.1\sigma$ shift to achieve stability. The recent update~\cite{Dehnadi:2023msm} suggests that the difference in uncertainties of both pole mass predictions is similar to the uncertainty differences among the various Monte-Carlo generators. 
 Finally, the PDG also quotes an $\overline{\text{MS}}$ top mass value $m_t(\mu=m_t)$, which has a large uncertainty of about $1\%$ that would allow stabilization of the Higgs potential with a $0.7\sigma$ shift.

 \begin{figure}[t]
  \centering
  \includegraphics[scale=.55]{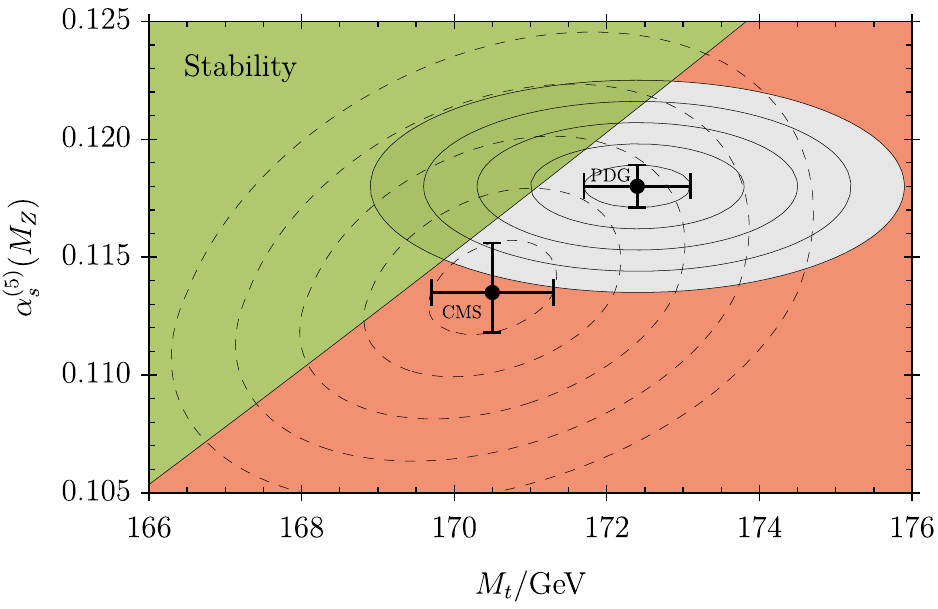}
  \includegraphics[scale=.55]{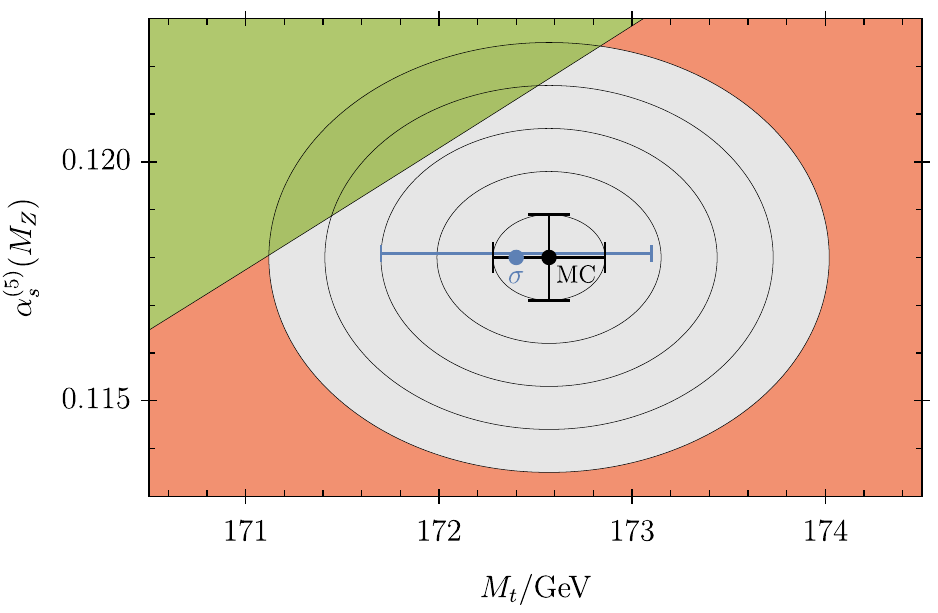}
  \caption{  
Regions of stability for the SM Higgs potential as a function of the top mass $M_t$ and the strong coupling constant $\alpha_s^{(5)}(M_Z)$. 
Color-coding indicates stability ($\alpha_{\lambda,\text{eff}}(\mu)\ge 0$, green) or otherwise (red/gray).
Uncertainties are derived by combining the $1\sigma-5\sigma$ intervals of both observables (dashed or solid rings). 
Upper panel: 2024 PDG values with top mass $M_t^\sigma$ (gray background) and uncorrelated uncertainties as well as CMS analysis~\cite{CMS:2019esx} with correlated uncertainties.
Lower panel: PDG 2024 central value for the top mass $M_t^\text{MC}$ (black dot) and uncertainties added in quadrature. For comparison, we also show $M_t^\sigma$ (blue dot) and  its $1\sigma$ uncertainty range (blue crosshair).
}
\label{fig:SM-stab-region}
\end{figure}

If PDG uncertainties in both top mass  $M_t^\sigma$ and strong coupling are  combined in quadrature less then a $2\sigma$ deviation from the central values is required to achieve stability,
see \fig{SM-stab-region}. 
However, this neglects correlations between the observables, which may play a critical role in determining the stability of the SM Higgs vacuum. 

The correlation was taken into account by the CMS analysis~\cite{CMS:2019esx}. 
Here the central value of $\alpha_s^{(5)}(M_Z)$ and $M_t$ are smaller than the PDG world averages, and the uncertainty for $\alpha_s^{(5)}(M_Z)$
is significantly larger.  
Moreover, $\alpha_s^{(5)}(M_Z)$ and $M_t$ in \cite{CMS:2019esx} are individually further away from the region of absolute stability than in combination.

\fig{SM-stab-region} shows the stability region (green) in the $(M_t,\alpha_s^{(5)})$ plane~\cite{Andreassen:2017rzq}, and compares the PDG 2024 central values using the pole mass $M_t^\sigma$ and uncorrelated errors (top panel, full lines), the CMS determinations of  $\alpha_s$ versus $M_t$ and correlated errors (top panel, dashed lines), and  the PDG 2024 central values using the Monte-Carlo mass $M_t^{\rm MC}$ and uncorrelated errors (bottom panel). The bottom panel also indicates the $M_t^\sigma$ central value and $1\sigma$ error bar  (blue horizontal crosshair).
We observe from  \fig{SM-stab-region}  that all central values reside outside the stability region. 
Given the sizable uncertainty in the pole mass $M_t^\sigma$, the PDG central values are less than $2\sigma$  away from stability when using  uncorrelated errors.
In turn, the  smaller uncertainty in  the Monte-Carlo mass indicates a  $3\sigma$  tension with vacuum stability when using uncorrelated errors. 
We emphasize that correlations matter, for which  the CMS analysis provides evidence, albeit with large uncertainties.

Given the present state of affairs, one may also turn the question around and ask how much accuracy is required individually in the determination of the top mass and the strong coupling constant   to establish the absence of SM Higgs stability at the  $5\sigma$ level or above. Assuming that central values do not change, 
we find that  a $5\sigma$ signature necessitates the uncertainty in the top mass $M_t$ to come down to the $200-300$~MeV range. 
 Future $e^+  e^-$-colliders can clearly reach this  goal with  projected uncertainties below $100$ MeV  \cite{Simon:2019axh}.
Similarly, 
the uncertainty in  $\alpha_s^{(5)}$ would have to come down to the $(6-7)\cdot10^{-4}$ range. 
 In the nearer term future progress from the HL-LHC  \cite{Azzi:2019yne} towards these targets can be expected.

We briefly comment on the influence of other SM input. 
 The electroweak gauge couplings $\alpha_{1,2}(\mu_\text{ref})$  are larger than $\alpha_\lambda(\mu_\text{ref})$, and  therefore have sizable impact on its RG evolution. However, the values of $\alpha_{1,2}$ are extracted from electroweak precision observables, $\alpha_e$, $G_F$, $M_Z$ and $\Delta \alpha_\text{had}^{(5)}(M_Z)$ to the running fine structure constant, which have too small of an uncertainty to
 matter at the level of $M_{t,h}$ and $\alpha_s^{(5)}(M_Z)$ for the evolution of $\alpha_\lambda$ and its sign.
In particular, the current tension~\cite{Parker:2018vye,Morel:2020dww} in the determination of the fine structure constant is insignificant in the context of vacuum stability. 
The $W$ pole mass has a  larger uncertainty than the one of the $Z$, and is therefore not included in the analysis.
The remaining input parameters, such as quark and lepton masses, are too small  to play a role for the effective potential and stability analysis.

In summary,  the fate of the standard model vacuum remains an important open question,
which  requires higher precision in the determinations of the top mass, the strong coupling, and their uncertainties and  correlation.
We also emphasize that stability  is largely dominated by RG effects, and  that the role of finite-order corrections to the effective potential is minor. This supports the view that the stability of  SM extensions 
may very well be extracted from RG studies of the tree-level potential.
This will be our main search strategy from here on.

\section{Stability via  Higgs Portals}
\label{sec:hp}

In this section, we change gear and take the metastability of the SM as  a model building task to find stable vacua all the way up to the Planck scale. Our focus is on a variety of singlet scalar field extensions with and without flavor, and the prospects for stability through the Higgs portal mechanism.

\subsection{Higgs Portal Mechanism}

We consider models in which the SM is extended by a scalar singlet sector under the SM $SU(3)_C \times SU(2)_L\times U(1)_Y$ gauge group.
Such scalars, generically denoted by real components $S_i$, allow for a well-known renormalizable portal interaction 
\begin{align} \label{eq:generalPortal}
{\cal L} \supset \sum_i \delta_i\, (H^\dagger H) (S_i\! {}^\intercal\! S_i)
\end{align}
with the SM Higgs doublet $H$ via canonically marginal
portal couplings $\delta_i$.
This Higgs portal most directly affects the Higgs sector and can cure metastability~\cite{Machacek:1984zw}.
Previous works considering such models include e.g.~\cite{Falkowski:2015iwa,Khan:2014kba,Han:2015hda,Garg:2017iva,Gabrielli:2013hma,Elias-Miro:2012eoi,Gonderinger:2012rd,Costa:2014qga,Khoze:2014xha,Ibrahim:2022cqs}.

In particular, the Higgs portal affects the RG evolution of Higgs quartic $\lambda$, which enters in the scalar potential term as $\lambda (H^\dagger H)^2$, at 1-loop order
\begin{equation}  \label{eq:betalambda}
\beta_\lambda=\beta^{\text{SM}}_\lambda+ \sum_i 2\,N_i \,\alpha_{\delta_i}^2  \, , 
\end{equation}
where $\beta_\lambda\equiv \frac{\text{d} \alpha_\lambda}{\text{d} \ln \mu} $ denotes the beta-function of  $\alpha_\lambda$,  and 
$N_i$ denotes the number of real scalar components in each $S_i$, such that \eq{generalPortal} is compatible with an $O(N_i)$ symmetry for each $S_i$.
Here and in the following we denote for any quartic coupling 
\begin{equation}
\alpha_q=\frac{q}{(4 \pi)^2}, \quad \text{ with } q=\lambda, \delta,... 
\end{equation}
and  \eq{gy} for the gauge and Yukawa couplings.
Therefore, at a scale $\Lambda$ at or above the electroweak one,
\begin{equation} 
\alpha_\lambda(\Lambda)-\alpha_\lambda^{\text{SM}}(\Lambda) \propto  \sum_i 2\,N_i \,\alpha_{\delta_i}^2 >0 \, ,
\end{equation}
that is, the portal contribution increases the Higgs quartic relative to its SM value.

The $\beta$-function of the Higgs portal coupling is technically natural, i.e. $\beta_\delta \propto \alpha_\delta$. Thus, if vanishing, $\alpha_\delta$ cannot be switched on by quantum fluctuations. As a consequence there cannot be any RG induced sign changes of $\alpha_\delta$ in the running either. As further discussed in concrete models in the following,  the RG evolution of $\alpha_\delta$ is governed by both the SM as well as the BSM sector.
The details of the  latter depend on symmetries; in general there exist several quartic interactions among the BSM scalars. However, the pure BSM quartics  enter $\beta_\lambda$ directly only starting from three loops, and are always mediated by the portal coupling $\alpha_\delta$~\cite{Machacek:1984zw}. Therefore, the pure BSM scalar potential decouples from  the Higgs portal mechanics  for sufficiently small couplings.
On the other hand, we find that, due to the interplay of BSM quartics with the portal, even small values of $\delta$ induce stability once the pure BSM quartics are sufficiently large.

For the  RG-analysis and numerics to identify stability regions we follow closely \cite{Hiller:2022rla} to which we refer for further details. In particular, we make use of the SM couplings from \cite{Alam:2022cdv} and apply full 2-loop running of all couplings in our BSM models.
The central goal of this work is to identify parameter configurations in scalar SM extensions that allow for \textit{Planck safety}, that is a RG flow from the new physics scale $\mu_0 = M_s \gtrsim 1\, \text{TeV}$ - where $M_s$ denotes the physical mass of the BSM scalar(s) - up to the Planck scale $M_\text{Pl} \simeq 10^{19} \, \text{GeV}$ without poles and vacuum instabilities. These Planck-safe parameter configurations span the \textit{BSM critical surface} at the matching scale.
If in the RG evolution all (tree-level) vacuum stability conditions are fulfilled all the way to the Planck scale we speak of \textit{strict Planck safety}. On the other hand, if a RG flow features intermediate metastabilities in the Higgs potential but a stable potential at the Planck scale we speak of \textit{soft Planck safety}, see \App{PS} for details. 

The remainder of this section deals with 
vacuum stability in BSM extensions  with $O(N)$ global symmetry (Sec.~\ref{sec:orthogonal}),
models with  a single scalar singlet field (Sec.~\ref{sec:singlet}),  the incompatibility  of a negative portal  with Planck-safety (Sec.~\ref{subsec:negativeportal}),  flavorful BSM extensions with more delicate symmetries (Sec.~\ref{sec:flavor}) and  the availability of negative  BSM quartics for these (\Subsec{negquart}).

\subsection{\texorpdfstring{${O(N_S)}$}{O(Ns)} Symmetric Scalars  \label{sec:orthogonal}}

\begin{figure*}
  \centering
  \renewcommand*{\arraystretch}{0}
  \begin{tabular}{cc}
    \includegraphics[width=.75\columnwidth]{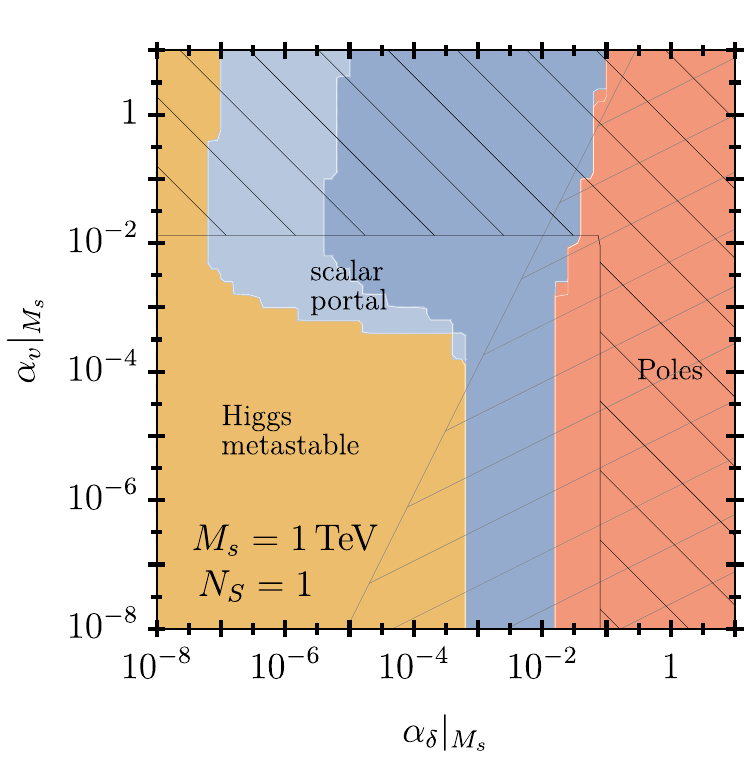} &
    \includegraphics[width=.75\columnwidth]{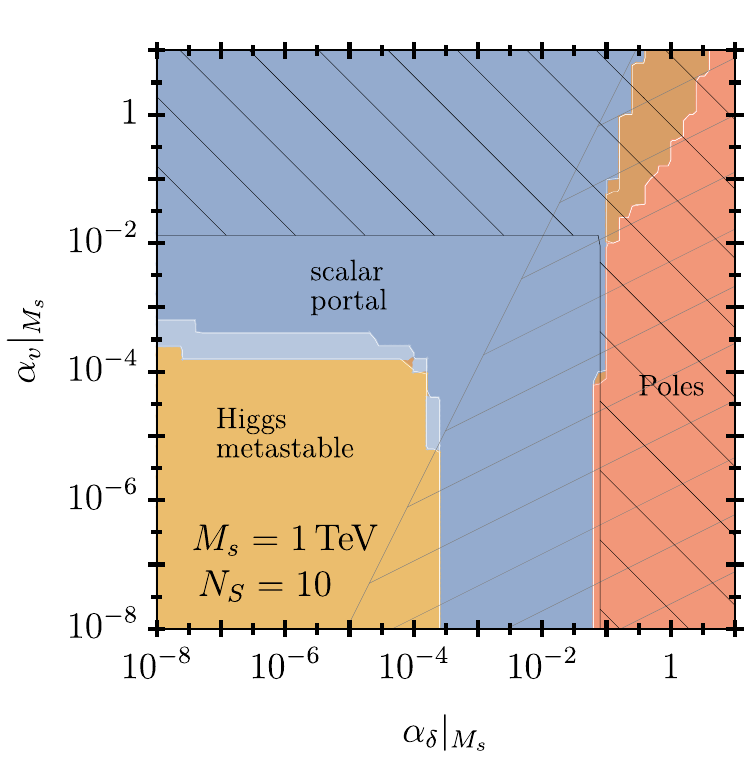} \\
    $\quad$ & $\quad$\\
    \includegraphics[width=.75\columnwidth]{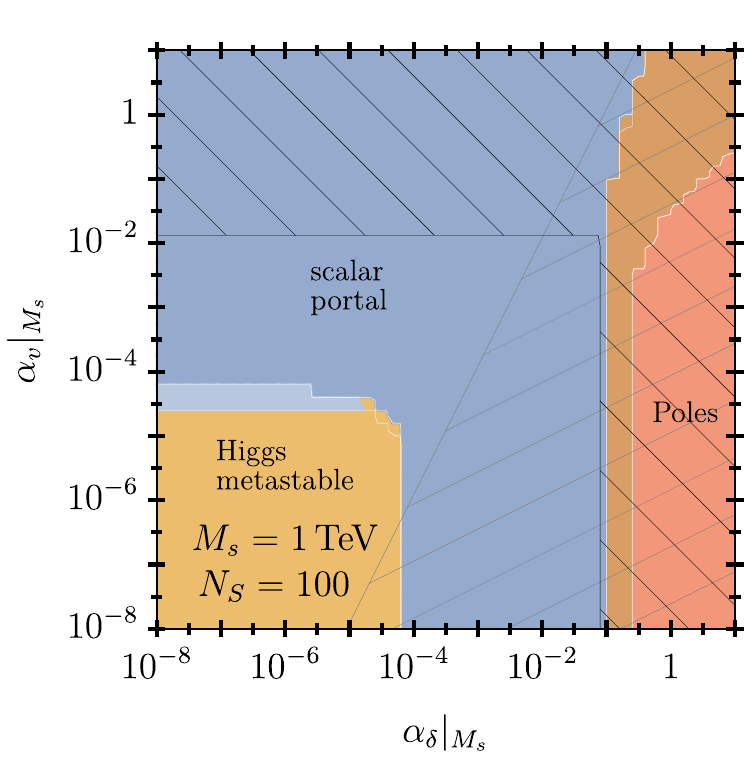} &
    \includegraphics[width=.75\columnwidth]{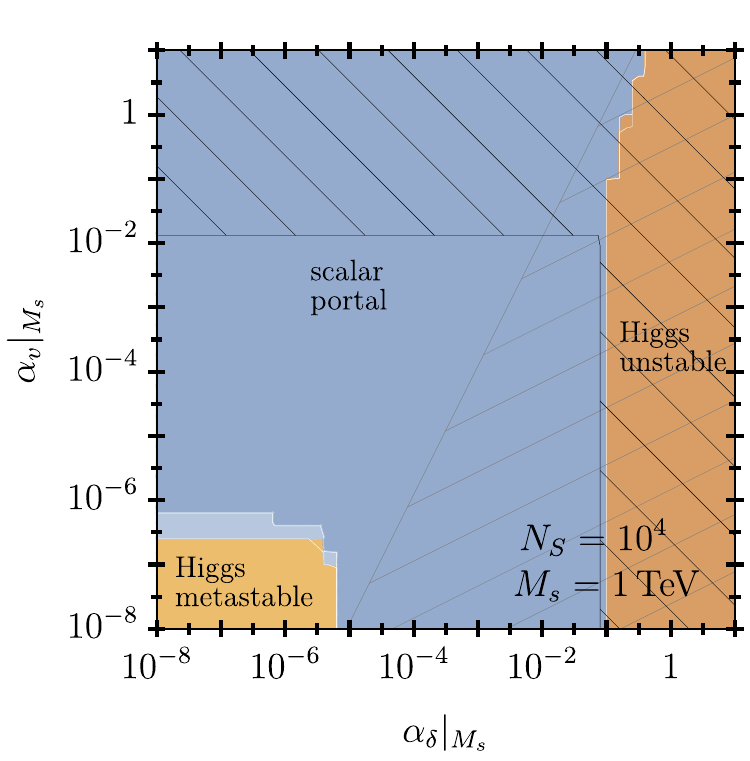} \\
  \end{tabular}
\caption{BSM critical surface for the $O(N_S)$ model with $N_S=1,10,100$ and $10^4$ generations of real scalar singlets spanned by the pure BSM quartic $\alpha_v(M_s)$ and the Higgs portal $\alpha_\delta(M_s)$. Red areas correspond to a RG evolution featuring Landau poles below the Planck scale. Brown indicates instabilities in the Higgs potential ($\min \alpha_\lambda(\mu) \leq -10^{-4}$). Gray regions feature an unstable BSM scalar potential i.e.~violation of  \eq{Vstab1} at some scale $\mu$ before $\MPl$. Yellow regions correspond to a SM-like RG evolution with $-10^{-4} \leq \min \alpha_\lambda(\mu),\,\alpha_\lambda(M_\text{Pl}) \leq 0$ hinting a metastable Higgs. Dark (light) blue areas  correspond to a strictly (softly) Planck-safe RG evolution with a stable potential all the way up to (at) $M_\text{Pl}$, see \App{PS}. Hatched black areas violate tree-level perturbative unitarity \eq{unitarity}, while gray, lightly  hatched ones are experimentally excluded by  \eq{betabound} on the scalar mixing angle, see \Sec{Pheno} for details.}
\label{fig:surface-O(N)}
\end{figure*}

The simplest global symmetry group a SM extension with $N_S$ real BSM scalars can exhibit is $O(N_S)$, which implies a mass parameter $\mu_S$, a  portal coupling $\delta$ as well as BSM quartic $v$.
The scalar potential
\begin{equation}\label{eq:VHS1}
\begin{aligned}
V_{O(N_S)} =&  - \mu_H^2 H^\dagger H -\frac{1}{2} \mu_S^2 S^T S +\lambda (H^\dagger H)^2\\
& + v (S^T S)^2   + \delta \, (H^\dagger H)  (S^T S)
\end{aligned}
\end{equation}
is stable if 
\begin{equation} \label{eq:Vstab1}
\begin{aligned}
        & \qquad  \lambda > 0, \qquad    v > 0, \qquad   \delta >- 2\sqrt{ \lambda v}.
      \end{aligned}
\end{equation}
For $N_S = 1$ the global symmetry becomes a $\mathbb{Z}_2$. 
The setup \eq{VHS1} describes also $\tfrac12 N_S$ complex scalars with  a $U(N_S/2)$ global symmetry. 
For more delicate global symmetries, there might be additional  quartic interactions beyond the BSM scalar portal $\delta$ and  self-coupling $v$. However, the potential \eq{VHS1} can be recovered by switching off all couplings that violate the $O(N_S)$ symmetry.

The RG-analysis and numerics how to identify stability regions is described in  \App{PS}  and \cite{Hiller:2022rla}, to which we refer for further details. The Higgs portal mechanics arises already at 1-loop via \eq{betalambda}, with $i=1$ and $N_1 = N_S$. Including also the two-loop contribution $\beta_\lambda$ reads
\begin{equation}  \label{eq:betalambdaONS}
\begin{aligned}
\beta_\lambda=&\beta^{\text{SM}}_\lambda+ \beta^{\text{BSM},(1)}_\lambda + \beta^{\text{BSM},(2)}_\lambda,\\
\beta^{\text{BSM},(1)}_\lambda =& + 2 N_S \,\alpha_\delta^2, \\
\beta^{\text{BSM},(2)}_\lambda =& - 4\,N_S (4 \,\alpha_\delta + 5 \alpha_\lambda) \alpha_\delta^2. 
\end{aligned}
\end{equation}
The pure BSM quartic $\alpha_v$  does not contribute to $\beta_\lambda$ at 1- and 2-loop \cite{Machacek:1984zw}. However it contributes positively at 1-loop to $\beta_\delta$ via
\begin{equation}  \label{eq:betadeltaONS}
\begin{aligned}
\beta_\delta=& \beta^{(1)}_\delta + \beta^{(2)}_\delta,\\
\beta^{(1)}_\delta \subset & [+8 (2+ N_S)\alpha_v] \alpha_\delta, \\
\beta^{(2)}_\delta \subset & [-160 (2+ N_S) \alpha_v^2  - 96 N_S\, \alpha_v \alpha_\delta] \alpha_\delta. 
\end{aligned} 
\end{equation}
Thus, a sizable $\alpha_v$ can indirectly also cause an uplift of $\alpha_\lambda$ by inducing an uplift in $\alpha_\delta$. Recall that the full $\beta_\delta \propto \alpha_\delta$ is technically natural, we just omitted the terms involving SM couplings as they are not relevant for demonstrating the discussed stabilization mechanisms.
For completeness we also give the beta function at one and two loops for the pure BSM quartic
\begin{equation}  \label{eq:betavONS}
\begin{aligned}
\beta_v & = \beta^{(1)}_v + \beta^{(2)}_v\,,\\
\beta^{(1)}_v &=  (8 N_S+64) \alpha_v^2 + 2 \alpha_\delta^2\,, \\
\beta^{(2)}_v &= -(2688 +576 N_S) \alpha_v^3 - 16 \alpha_\delta^3 - 80 \alpha_v \alpha_\delta^2 \\
&\phantom{= \ } + 4 (\alpha_1 + 3 \alpha_2 - 3 \alpha_b- 3 \alpha_t) \alpha_\delta^2 \,. 
\end{aligned} 
\end{equation}

 \begin{figure}
    \centering
    \begin{tabular}{l}
    \hspace*{-1cm}
    \includegraphics[width=1.2\columnwidth]{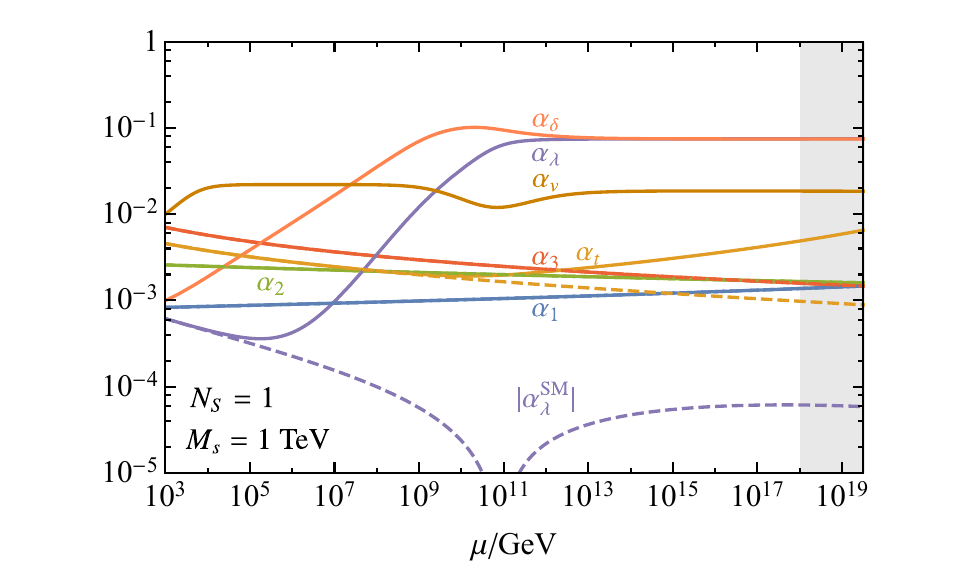}\\
    \hspace*{-1cm}
    \includegraphics[width=1.2\columnwidth]{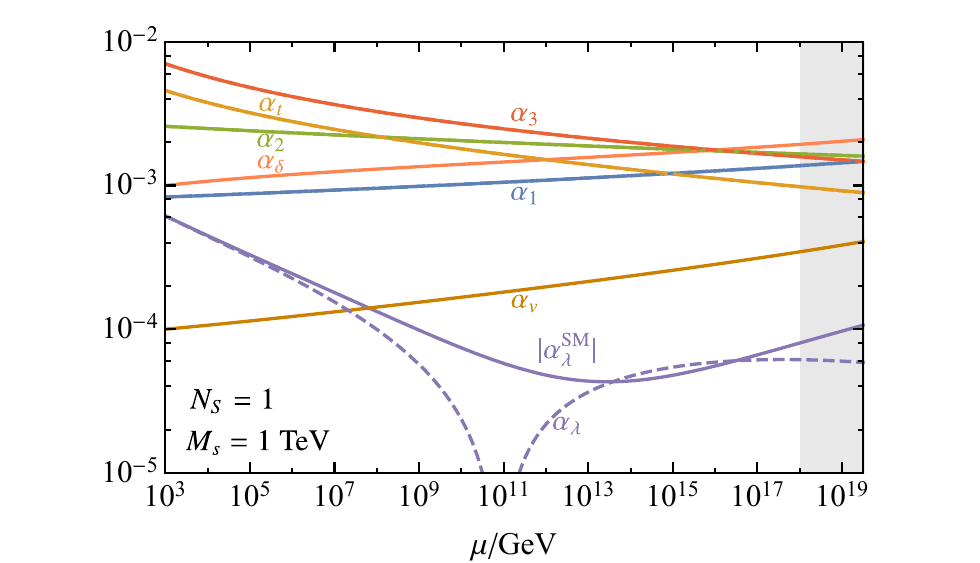}
    \end{tabular}
    \caption{Two-loop renormalization group flow in the SM (dashed lines) and the scalar $O(N_S)$ model (solid lines) for $N_S=1$, $M_s=1\text{ TeV}$, $(\alpha_\delta,\,\alpha_v)|_{M_s}= (10^{-3}, \,10^{-2})$ (top) and $(\alpha_\delta,\,\alpha_v)|_{M_s}= (10^{-3}, \,10^{-4})$ (bottom). In the upper plot quartic couplings are trapped in a walking regime before the Planck scale, whereas in the lower plot the running to $\MPl$ occurs within a weakly coupled regime without walking.}
    \label{fig:Running_ON}
\end{figure}

\begin{figure}
  \centering
  \renewcommand*{\arraystretch}{0}
  \begin{tabular}{c}\textbf{}
    \includegraphics[width=.75\columnwidth]{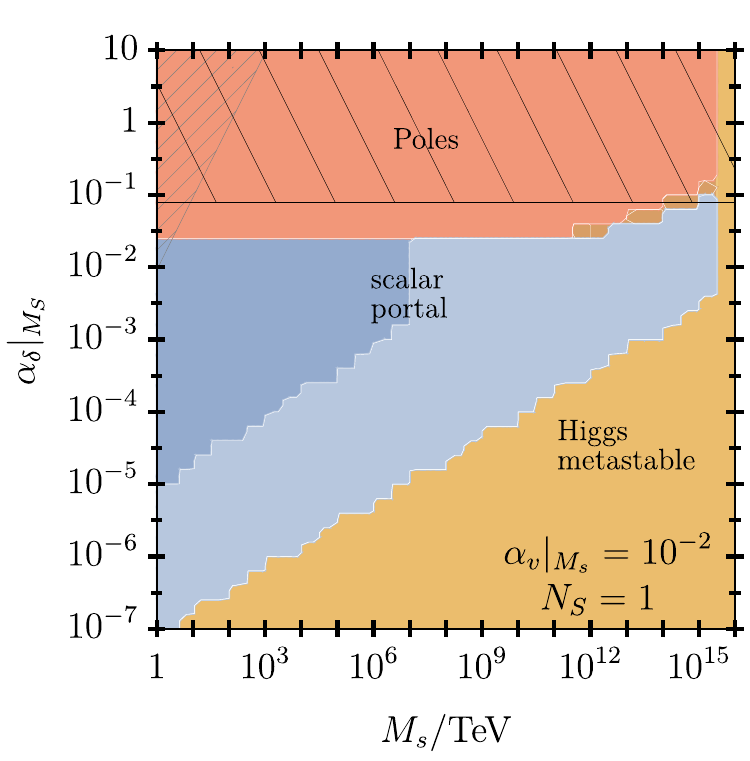}\\
     \includegraphics[width=.75\columnwidth]{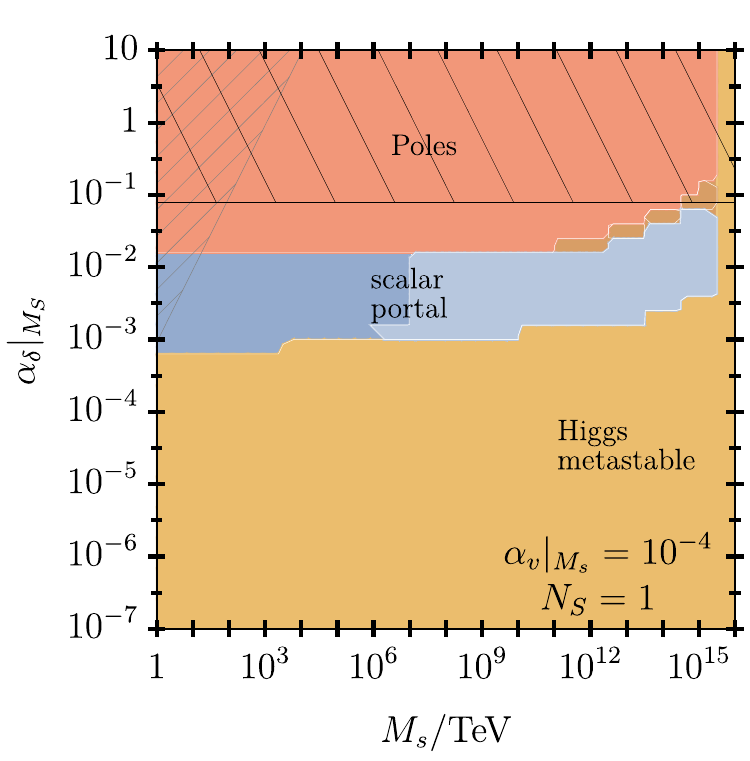}
  \end{tabular}
  \caption{BSM critical surface in the $M_s-\alpha_\delta(M_s)$-plane in a minimal BSM model featuring $N_S=1$ real BSM scalars and $O(N_S)$ symmetry for fixed $\alpha_v(M_s)=10^{-2}$ (top) and $\alpha_v(M_s)=10^{-4}$ (bottom). Same color coding as \fig{surface-O(N)}. }
  \label{fig:surfaceNS1}
\end{figure}

In \fig{surface-O(N)}, the BSM critical surface in $\alpha_\delta-\alpha_v$ space is displayed for different field multiplicities $N_S$ and mass $M_s=1\,\text{TeV}$.
Generically, the metastability of the Higgs potential is cured if $\alpha_v$ or $\alpha_\delta$ are sufficiently large. However, too large values of $\alpha_\delta$ lead to the loss of Higgs stability and to subplanckian Landau poles. 
Due to  the term $N_S \,\alpha_\delta^2$ in \eq{betalambdaONS} an increase of  $N_S$ generically lowers the minimal values of $\alpha_\delta$ and $\alpha_v$ required to stabilize the Higgs. 
We emphasize that even a feeble portal coupling suffices to achieve stability due to the indirect stabilization from a large $\alpha_v(M_s)$.

Also shown (dark hatched) in  \fig{surface-O(N)} and subsequent figures are constraints from 
tree-level perturbative unitarity
\cite{Dawson:2021jcl}
\begin{equation} \label{eq:unitarity}
\alpha_\lambda \lesssim \frac{1}{6\pi}, \quad \alpha_v\lesssim \frac{1}{24 \pi}, \quad  \alpha_\delta  \lesssim \frac{1}{4\pi}, \quad
\end{equation}
as well as constraints from mixing between the Higgs and the BSM scalar (lighter, gray hatched).
The latter requires the BSM scalar to acquire a vacuum-expectation value (VEV), see \App{Mix} and \App{uni} for details.
While the unitarity bounds limit couplings to be rather perturbative, outside of which theoretical control ceases anyway, the implications of the Higgs being an admixture 
can lead to additional constraints.

Depending on the  parameters $\alpha_\delta,\,\alpha_v$ and $N_S$ the stabilization of the Higgs potential may occur either within a weakly coupled or a walking regime. The latter refers to an RG flow where a subset of couplings is interlocked at sizable, fixed values for several orders of magnitude, see \fig{Running_ON} (top figure). Walking regimes are a well-known phenomenon that has already been observed in earlier works \cite{Hiller:2020fbu,Hiller:2022rla,Heikinheimo:2017nth}. They occur due to the presence of a pseudo fixed point, i.e.~a nearby fixed point of the RG flow in the complex plane of couplings. The walking greatly enhances the parameter space that exhibits Planck safety, and its occurrence with respect to non-walking is more pronounced at larger $\alpha_\delta,\,\alpha_v$ and $N_S$.

Let us briefly  make some general remarks on the  robustness of results,
that are  achieved within  perturbation theory.
First of all, we use state-of-the-art higher loop contributions for the RG-analysis. Once couplings reach unitarity limits \eq{unitarity}, of course,
results  need to be taken with a grain of salt, however, there is large parameter space under perturbative control, see for instance \fig{surface-O(N)}.
In walking regimes,  see the upper plot in \fig{Running_ON} where some couplings become borderline perturbative,  one is  in the deep-stable region safely
obeying \eq{Vstab1}. So while the regions between boundaries of stability and instability will receive larger uncertainties towards larger couplings,
the generic structure of  vacuum stabilization via the Higgs portal remains intact for wider ranges. It would be interesting to revisit this analysis in the future
once three-loop beta-functions for the BSM quartics are available.

As for the impact of  higher loops on the scalar potential in the concrete BSM parameter points, we are encouraged by the small impact within the SM, i.e., the fact that
stability or instability is  decided by running, and therefore safely assume for the scans
that corrections are too small to have an impact also in BSM. Once the BSM parameter space is more pinned down, a computation of the effective potential
would be very much motivated to facilitate precision study.

We find that for TeV-ish BSM scalars a stable potential \eq{window} at large $N_S$ is roughly obtained for 
\begin{equation}
\begin{aligned} \label{eq:ON-cond}
\quad & 10^{-3}/\sqrt{N_S} \lesssim \alpha_\delta \lesssim 10^{-1} \\
 \text{or} \quad & N_S\, \alpha_v \gtrsim 10^{-2} ~ \text{and} ~ \log \alpha_\delta \gtrsim -0.16\, N_S -6.6\\
& \qquad \qquad \qquad \qquad   \; \, (\log \alpha_\delta \gtrsim -0.93 \,N_S-6.1)
\end{aligned}
\end{equation}
for strict (soft) Planck safety, see also \fig{surface-O(N)}. The second condition in \eq{ON-cond} corresponds to an indirect stabilization of the Higgs potential. The relatively large value of $\alpha_v(M_s)$ induces a significant RG growth in $\alpha_\delta$ which is then sufficient to render $\alpha_\lambda$ positive up to $\MPl$ even for tiny $\alpha_\delta(M_s)$. 
Note that the allowed ranges for $\alpha_v(M_s)$ and $\alpha_\delta(M_s)$ in \eq{ON-cond} generically shrink for larger $M_s$, as there is less RG time from the matching scale $\mu_0 \simeq M_s$ to the Planck scale left to stabilize the Higgs.

\subsection{Single Scalar Singlet  \label{sec:singlet}}

Next, we discuss stability for  the common scenario with a single real BSM scalar and a $\mathbb{Z}_2$-symmetry.
The BSM critical surface is shown in the $\alpha_{\delta}$--$\alpha_v$ plane in the upper left panel of \fig{surface-O(N)}, as well as the $M_s$--$\alpha_\delta(M_s)$ plane in \fig{surfaceNS1}. 
For  feeble $\alpha_v$ and lower values of $\alpha_\delta(M_s)$ the full scalar potential remains metastable at the Planck scale. 
Moderate values of $\alpha_\delta(M_s)$ allow for Planck safety, largely independently of the size of $\alpha_v(M_s)$. Concretely, we find for $M_s=1\,\text{TeV}$
\begin{equation} \label{eq:window}
\begin{aligned}
9.2 \cdot 10^{-4} \lesssim & \alpha_\delta(M_s) \lesssim 1.9 \cdot 10^{-2},\\
(8.2 \cdot 10^{-4} \lesssim & \alpha_\delta(M_s) \lesssim 2.0\cdot 10^{-2}),
\end{aligned}
\end{equation}
for strict (soft) Planck safety. While $\alpha_\delta$ is the main actor behind the stabilization of the potential, this may occur within a regime of running (smaller $\alpha_\delta$, see \fig{Running_ON}, lower panel) or walking couplings (larger $\alpha_\delta$, see \fig{Running_ON}, upper panel). In general, the RG flow enters walking regimes at lower scales the larger either or both $|\alpha_{\delta,v}(M_s)|$ are. 
More concretely, the onset of walking may range from close above the matching scale $M_s$, to far beyond the Planck regime or even the hypercharge Landau pole around $\mu\approx 10^{41}$~GeV for very small values of $\alpha_{\delta,v}(M_s)$.
On the other hand, too large values of $\alpha_\delta(M_s)$ give rise to subplanckian Landau poles in $\alpha_\delta$. This phenomenon has little sensitivity to the numerical values of $\alpha_v(M_s)$ (cf.~\fig{surface-O(N)}) or $M_s$ (cf.~\fig{surfaceNS1}), but rather to $\alpha_\delta(M_s)$ itself.

For sufficiently large values of the pure BSM quartic $\alpha_v(M_s) \gtrsim 3 \cdot 10^{-3}$, $\alpha_v$ may also drive the stabilization of the Higgs potential through the portal, even if the latter is feeble. 
Therefore, the range of $\alpha_\delta(M_s)$ that allows for Planck safety grows, especially towards lower coupling values. 
For $M_s \simeq 1$~TeV, the range of viable $\alpha_\delta(M_s)$ increases  to roughly $(10^{-7})\,10^{-5} \lesssim \alpha_\delta(M_s) \lesssim 10^{-1}$ for strict (soft) Planck safety.
However, as $\alpha_\delta$ is technically natural even the stabilization mechanism via sizable $\alpha_v(M_s)$ must cease to occur once  $\alpha_\delta(M_s)$ is too small. 
With fixed $\alpha_v(M_s)$ but increasing $M_s$ larger minimal values of $\alpha_\delta(M_s)$ are required to bound the potential from below, as less RG time is left to do so, see \fig{surfaceNS1}. 
Note that for strict Planck safety generically  an upper bound on the scale of new physics exists, $M_s  \sim \mu_0 \lesssim  10^{7}\,\text{TeV}$ corresponding to the scale 
where $\alpha_\lambda$ gets negative in the SM.

\subsection{Negative Portals are Not Safe}
\label{subsec:negativeportal}

 In principle, the vacuum stability conditions \eq{Vstab1} allow  for a  negative  portal coupling $\alpha_\delta(\mu_0)<0$ at the matching scale. In practice, however, this turns out to be in conflict with Planck safety. Specifically, 
for $-\alpha_\delta(\mu_0)>0$ too large,  the stability condition \eq{Vstab1} is already violated at the matching scale $\mu_0$.
Then, for smaller $-\alpha_\delta(\mu_0)>0$, the RG evolution drives the portal quickly towards more negative values and subsequently  into a pole, again in violation of \eq{Vstab1}.
This pattern is understood  by recalling  that the portal coupling is technically natural,
\begin{align}
\beta_\delta =  X \cdot  \alpha_\delta \,, 
\end{align}
where the proportionality factor $X=X(\alpha_i)$ is polynomial in the couplings. It can be read off, for instance,  from \eq{betadeltaONS} and turns out to be positive, $X>0$, to leading order in perturbation theory.
Consequently, for slowly varying $X$, we observe a power-law growth of the portal coupling towards the UV,  $\alpha_\delta(\mu)  \sim \alpha_\delta(\mu_0) (\mu/\mu_0)^X$, irrespective of its sign.
On the other hand, for tiny $|\alpha_{\delta,v}(\mu_0)|$, the portal contribution to \eq{betalambdaONS} is too small to prevent $\alpha_\lambda$ from turning negative. 
This pattern is not improved by increasing the BSM quartic $\alpha_{v}(\mu_0)$, which enhances $X$,
invariably leading to a violation of the third condition in \eq{Vstab1} due to the faster growth of  $|\alpha_\delta(\mu)|$. 

We have also searched for sweet spots by taking $\alpha_\delta(\mu_0)<0$   the least negative for given $\alpha_{v}(\mu_0)\ge 0$, and demanding that $\alpha_\lambda\ge 0$ for all scales below $\MPl$. However, we  find that there are none, meaning that it is not possible to simultaneously satisfy the three  stability conditions \eq{Vstab1} for all scales $\mu_0\le \mu\le \MPl$. 
Ultimately, the  reason for this is that the instability scale of the SM is too far away from the Planck scale.
We  conclude that BSM models with a negative portal coupling  are not Planck-safe.

\subsection{Flavorful Matrix Scalars  \label{sec:flavor}}

In this section we discuss models featuring a flavorful, complex scalar matrix field $S_{ij}$, with $i,j=1 \ldots N_F$ and $N_F > 1$ being the number of flavors, thus the number of real degrees of freedom is $2N_F^2$. In addition to the portal coupling $\delta$, the potential 
\begin{equation}\label{eq:VHSi}
\begin{aligned}
V_{SU(N_F)^2} =& -  \mu_H^2 H^\dagger H -\mu_S^2 \,\tr\left[\,S^\dagger S\right] \\
& + u\,\tr\left[S^\dagger S S^\dagger S\right]  + v \left[ \tr \,S^\dagger S\right]^2   \\ 
& +\lambda (H^\dagger H)^2 + \delta \, (H^\dagger H) \,\tr\left[S^\dagger S\right] ,\\
\end{aligned}
\end{equation}
contains two pure BSM  quartics $u,v$. 
Here, traces are in flavor space; note the potential is invariant under the  $SU(N_F)_L \times SU(N_F)_R$ flavor symmetry under which $S \to V^\dagger S U $ and $V,U \subset SU(N_F)_L,  SU(N_F)_R$, respectively.
The flavor symmetry greatly reduces the number of couplings and beta functions.
Note that for  $u=0$ the global symmetry is enhanced to $O\left(2\,N_F^2\right)$. Therefore $\alpha_u$ is 
technically natural  just like  the portal coupling $\alpha_\delta$. 

The reasons for  considering  scalar matrix fields with $N_F^2$ complex scalars instead of simply studying $N_F$ copies of a single scalar are manyfold:
Firstly, quadratic field multiplicities $\propto N_F^2$  rather than linear ones enhance the impact of the scalar sector on the model, an effect that has been crucial in constructing  exact asymptotically safe models \cite{Litim:2014uca}, and concrete safe SM extensions \cite{Bond:2017wut}.
Secondly, the flavor symmetry in the scalars can be linked to SM flavor,  a possibility  explored in \cite{Hiller:2019mou,Hiller:2020fbu,Bissmann:2020lge,Bause:2021prv,PlanckSafeQuark} choosing $N_F=3$.
Thirdly, the flavor structure of (\ref{eq:VHSi}) allows for different ground states. Depending on the sign of $u$, two non-trivial ground states $V^\pm$  exist with stability conditions
\begin{equation}\label{eq:vstab}
\begin{aligned}
\lambda > 0, & \qquad \Delta>0, \qquad \delta >  -2 \sqrt{\lambda \Delta},\\
 V^+ : \quad & u > 0, \quad \Delta = u/N_F + v \,, \\
V^- :  \quad & u < 0, \quad \Delta = u + v \,.
\end{aligned}
\end{equation}
 Notably, $V^-$ breaks flavor universality spontaneously \cite{Hiller:2019mou,Hiller:2020fbu},
a unique feature of these models.
Thus, they provide novel BSM explanations for ongoing flavor anomalies that suggest a violation of lepton flavor universality e.g. to the muon anomalous magnetic moment~\cite{Hiller:2019mou}
or an excess of $B \to K \nu\bar \nu$ branching ratio \cite{Belle-II:2023esi} suggesting taus to couple differently than the other leptons \cite{Bause:2023mfe}.
Lastly, from the connection to SM flavor genuine new, testable  signatures for colliders  arise that are flavorful yet not in conflict with severe FCNC constraints \cite{Bissmann:2020lge}.

The Higgs portal is very potent in these  models as it is quadratically  enhanced  by $2 N_F^2$ as in \eq{betalambda}.\footnote{Note that there are $2\,N_F^2$ real BSM scalar components but the portal term in \eq{VHSi} is differently normalized than in \eq{generalPortal}.}
Specifically, the beta function of the Higgs quartic reads
\begin{equation}  \label{eq:betalambdaSij}
\begin{aligned}
\beta_\lambda=&\;\beta^{\text{SM}}_\lambda+ \beta^{\text{BSM},(1)}_\lambda + \beta^{\text{BSM},(2)}_\lambda,\\
\beta^{\text{BSM},(1)}_\lambda =& + N_F^2 \,\alpha_\delta^2, \\
\beta^{\text{BSM},(2)}_\lambda =& - 2 N_F^2 \, (2\, \alpha_\delta + 5\, \alpha_\lambda) \alpha_\delta^2.
\end{aligned}
\end{equation}

\begin{figure}
  \centering
  \renewcommand*{\arraystretch}{0}
  \begin{tabular}{c}\textbf{}
    \includegraphics[width=.75\columnwidth]{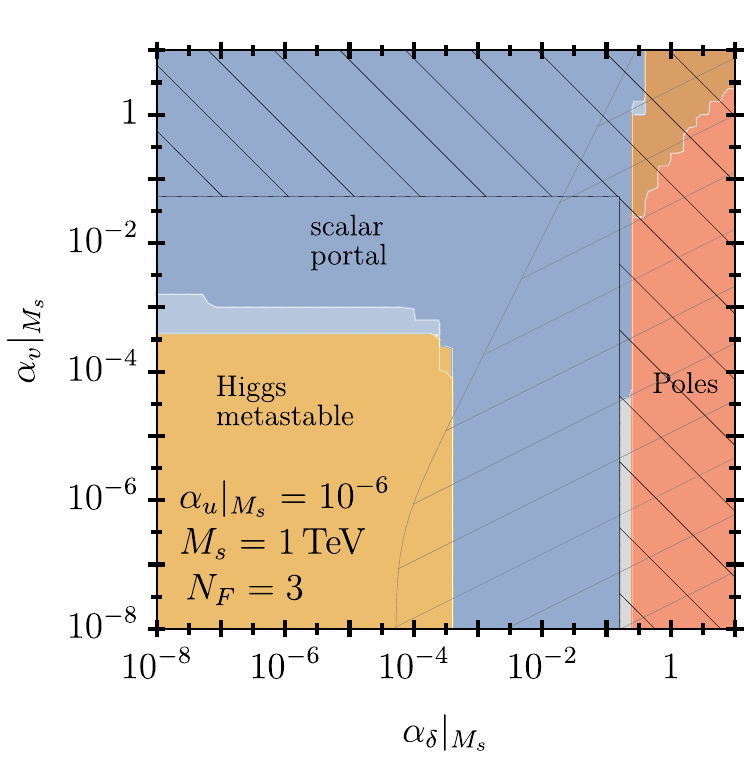}\\
     \includegraphics[width=.75\columnwidth]{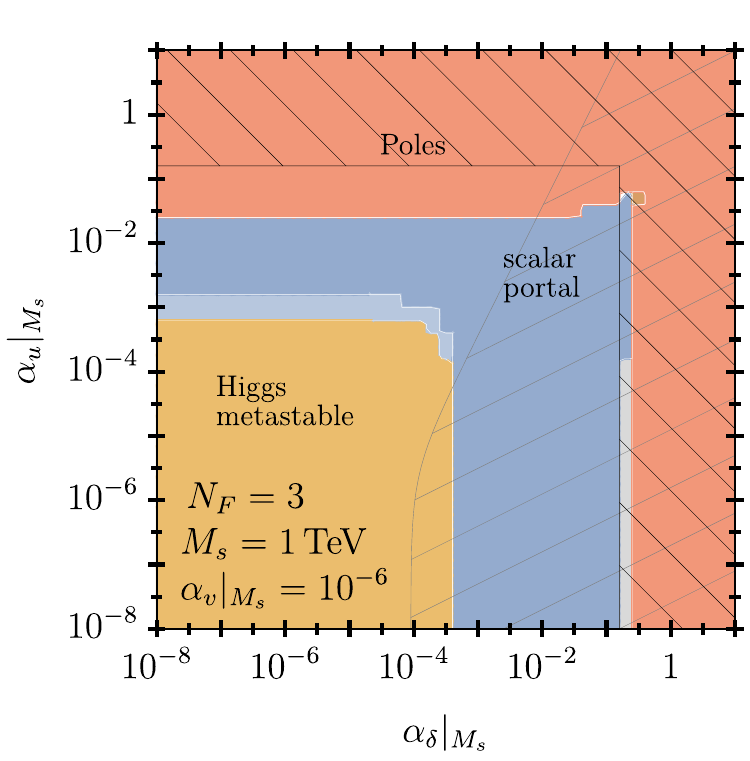} \\
      \includegraphics[width=.75\columnwidth]{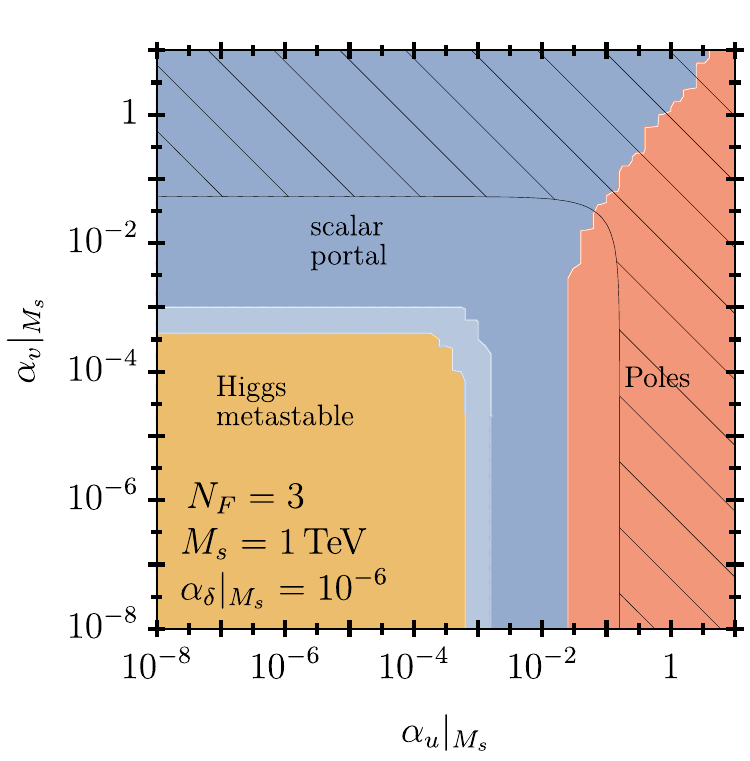}
  \end{tabular}
  \caption{BSM critical surface in the $SU(N_F) \times SU(N_F)$ model  in the $\alpha_\delta(M_s)$--$\alpha_v(M_s)$- (top) , $\alpha_\delta(M_s)$--$\alpha_u(M_s)$- (middle)  and $\alpha_u(M_s)$--$\alpha_v(M_s)$-plane (bottom) with the remaining quartic feeble, at $10^{-6}$, and  $N_F=3$ and $M_s=1\,\text{TeV}$.
  Same color coding as \fig{surface-O(N)} with  blue corresponding to the vacuum  $V^+$. }
  \label{fig:Sij-delta-v}
\end{figure}

The pure BSM quartics $u,v$ contribute to $\beta_\lambda$ only beyond 2-loop order \cite{Machacek:1984zw,Hiller:2020fbu}.
Thus, as in the $O(N_S)$ model their influence is channeled through their contribution to $\beta_\delta$, and of minor direct importance for $\alpha_\lambda$.
However, the BSM quartics $\alpha_{u,v}$ matter  for the RG evolution of the portal coupling $\alpha_\delta$. They contribute positive at one-loop to the running of the portal
\begin{equation}  \label{eq:betadeltaSij}
  \begin{aligned}
  \beta_\delta=&\; \beta^{(1)}_\delta + \beta^{(2)}_\delta,\\
  \beta^{(1)}_\delta \subset & +[8 N_F \alpha_u + 4(N_F^2+1)\alpha_v] \alpha_\delta, \\
  \beta^{(2)}_\delta \subset & -[20 (N_F^2+1) \alpha_u^2 + 48 N_F \alpha_u \alpha_\delta + 80 N_F \alpha_u \alpha_v \\
  & \quad+ 20(N_F^2+1)\alpha_v^2 + 24(N_F^2+1)\alpha_v \alpha_\delta]\alpha_\delta. 
  \end{aligned} 
  \end{equation}
For fixed values of $N_F$, $M_s$, the BSM critical surface depending on $\alpha_\delta(M_S)$ and $\alpha_{u,v}(M_s)$ is displayed in \fig{Sij-delta-v}.
For feeble $\alpha_u(M_s)$ (upper plot) there is a striking resemblance to the $O(N_S)$ symmetry case shown in \fig{surface-O(N)}. For both $\alpha_{\delta,v}(M_s)$ feebly small, the running is SM-like and the Higgs potential remains metastable. 
Increasing the values of $\alpha_{\delta}(M_s)$ promotes the scalar portal itself to the primary actor stabilizing $\alpha_\lambda$. This is achieved at first within a weakly coupled, and later with increasing portal coupling a walking regime. Even larger $\alpha_{\delta}(M_s)$ drives itself into a Landau pole.

  \begin{figure}
    \centering
    \renewcommand*{\arraystretch}{0}
    \includegraphics[width=.75\columnwidth]{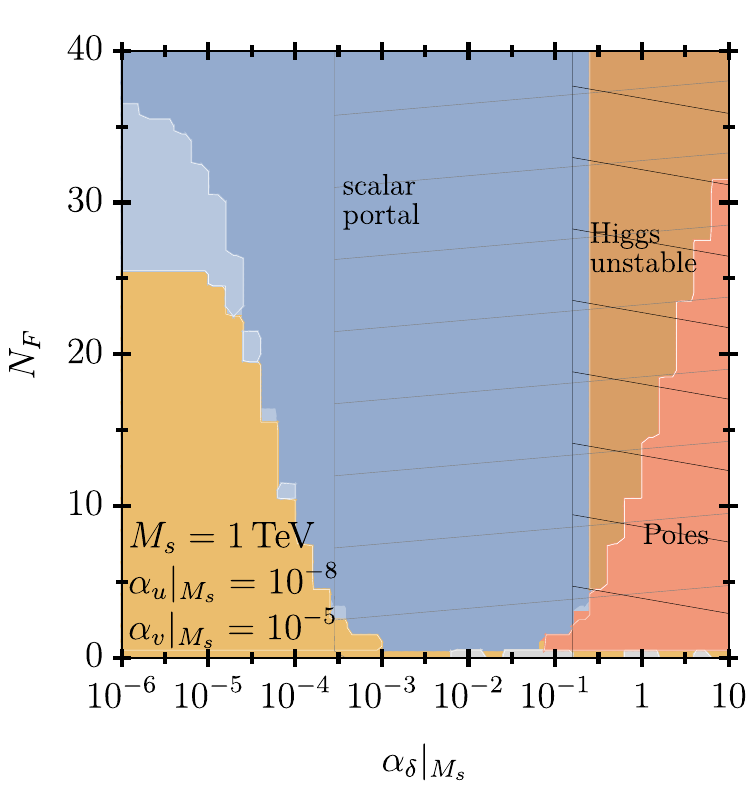}
    \caption{BSM critical surface in the $SU(N_F) \times SU(N_F)$ model in the $\alpha_\delta(M_s)-N_F$ plane for $M_s=1\,\text{TeV}$, $\alpha_{u}(M_s) = 10^{-8}$ and $\alpha_{v}(M_s) = 10^{-5}$. Same color coding as \fig{Sij-delta-v}. }
    \label{fig:Sij-delta-NF}
  \end{figure}

\fig{Sij-delta-NF} displays the interplay between  $N_F$ of the stabilization via $\alpha_\delta$, with tiny  $\alpha_{u,v}(M_s)$. For  larger $N_F$, the stability  window is widened. Approximately, we find
\begin{equation} \label{eq:adboundSij}
    \frac{1.4 \cdot 10^{-3} }{N_F} \lesssim  \, \alpha_\delta(M_s) \lesssim 
    \begin{cases}
(0.05 \dots 0.08) \, N_F, \, \, N_F \lesssim 10 \\
0.25, \qquad \qquad \qquad N_F\gtrsim 10 
\end{cases}
\end{equation}
which is largely independent of $\alpha_{u,v}(M_s)$ and $M_s$ unless both quartics are sizable $\alpha_{u,v}(M_s) = \mathcal{O}(1)$ or $M_s$ is close to $\MPl$.
Notice that the required minimal value of $\alpha_\delta(M_s)$ can always be reduced by increasing $N_F$ and vice versa. 
The maximum possible value on the other hand for $N_F \lesssim 10$ is dictated by avoiding subplanckian Landau poles in $\alpha_\delta$. 
This changes for $N_F \gtrsim 10$ where we obtain roughly $\alpha_\delta \lesssim 0.25$ independent of $N_F$, as for larger values in $\alpha_\delta$ the Higgs potential is destabilized. This can be understood from \eq{betalambdaSij}. The numerically dominant terms for large $\alpha_\delta$ and $N_F$ are $N_F^2 \alpha_\delta^2 - 4 N_F^2 \alpha_\delta^3$, where the first is from one- and the second from two-loop. This contribution is negative for $\alpha_\delta >1/4$ independent of $N_F$, resulting in an unstable Higgs.

Constraints from tree-level perturbative unitarity \eq{unitarity-uni}, displayed by the dark hatched regions,  disfavor Planck safety from large $\alpha_\Delta \gtrsim 5 \cdot 10^{-2}$. The $\alpha_\delta$ bound on the other hand is not really relevant as too large values of $\alpha_{\delta}$ anyway induce poles.

\begin{figure}
  \centering
  \hspace*{-1cm}
  \includegraphics[width=1.2\columnwidth]{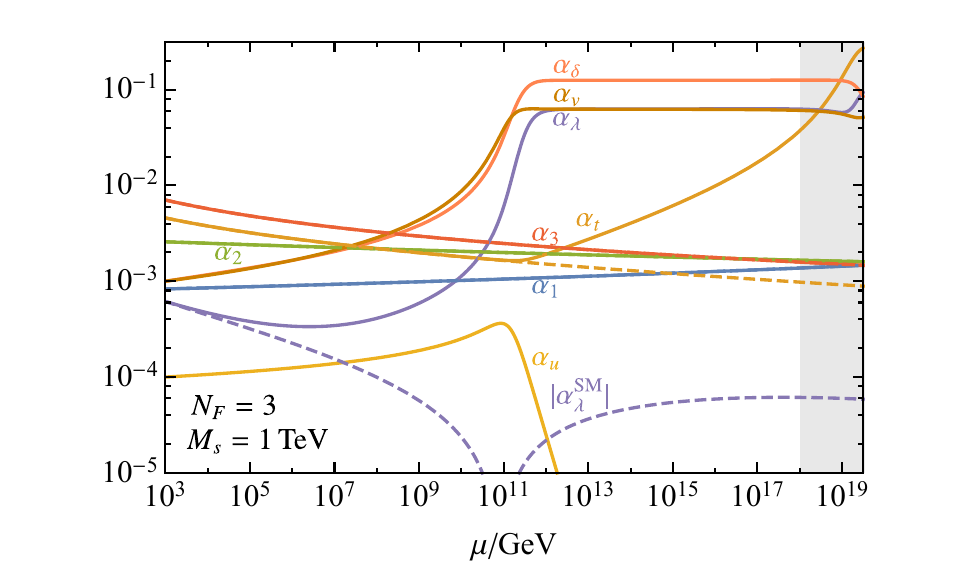}\\
  \caption{Two-loop renormalization group flow in the SM (dashed lines) and in the scalar $SU(N_F) \times SU(N_F)$ model (solid lines) for $N_F=3$ and $M_s=1\,\text{TeV}$ as well as $\alpha_{\delta,u,v}(M_s)= (10^{-3}, \,10^{-4}, \,10^{-3})$. All quartic couplings are trapped in a walking regime before the Planck scale, except for $\alpha_u$ which asymptotically approaches zero.}
  \label{fig:Sij-walking}
\end{figure}

Returning to the upper plot of \fig{Sij-delta-v}, larger values of $\alpha_{v}(M_s)$ improve the stability of the scalar potential due to the occurrence of a walking regime. 
In particular, the walking corresponds to the same pseudo fixed point of the $O(2N_F^2)$ symmetry case. Therefore, it is expected that $\alpha_v$ reaches sizable values while $|\alpha_u| \to 0$ as the walking regime is entered, irrespective of the sign of $\alpha_u$, but without transitions between the vacua $V^\pm$ \eq{vstab}. This is displayed in \fig{Sij-walking}. Stabilization of the Higgs within a walking regime is also possible within a range of sizable $\alpha_u(M_s)$. 
Thus, sufficiently sizable $\alpha_{u,v}(M_s)$ may stabilize the Higgs potential, even if $\alpha_\delta(M_s)$ itself is too small to allow this, see lower plot in \fig{Sij-delta-v}.
In this case, the pure BSM quartics are driving force behind the running of $\alpha_\delta$ and through it the stabilization of $\alpha_\lambda$. In this case we roughly obtain
\begin{align}
(6\dots8) \cdot 10^{-3} \lesssim & \; N_F^2 \,\alpha_v(M_s),\label{eq:avboundSij}\\
3 \cdot 10^{-3} \lesssim & \; N_F \, \alpha_u(M_s) \lesssim (6\dots8) \cdot 10^{-2} \label{eq:auboundSij} 
\end{align}
for feeble $10^{-10} \lesssim \alpha_\delta(M_s) \lesssim 10^{-5}$, vanishing $\alpha_{u/v}(M_S)$ and TeV-ish $M_s$. The upper bound on $\alpha_u(M_s)$ is due to the occurrence of subplanckian Landau poles in $\alpha_u$ which constitutes an important qualitative difference to $\alpha_{v}$-induced Planck safety. When increasing $M_s$ by a few orders of magnitude also the lower bounds on $\alpha_{u,v}(M_s)$ increase, as there is less RG time to rescue the potential and the effect is channeled though $\alpha_\delta$. This is similar to the $O(N_S)$ model.

With increasing $N_F$, the window of stabilization for $\alpha_v(M_s)$ is generically larger than the one for $\alpha_u(M_s)$. 
This follows from the leading-$N_F$ contributions to \eq{betadeltaSij}, which are of order $\propto N_F^2 \alpha_v$ and $\propto N_F \alpha_u$, respectively.

\begin{figure*}
  \centering
  \renewcommand*{\arraystretch}{0}
  \begin{tabular}{cc}
   \includegraphics[width=.75\columnwidth]{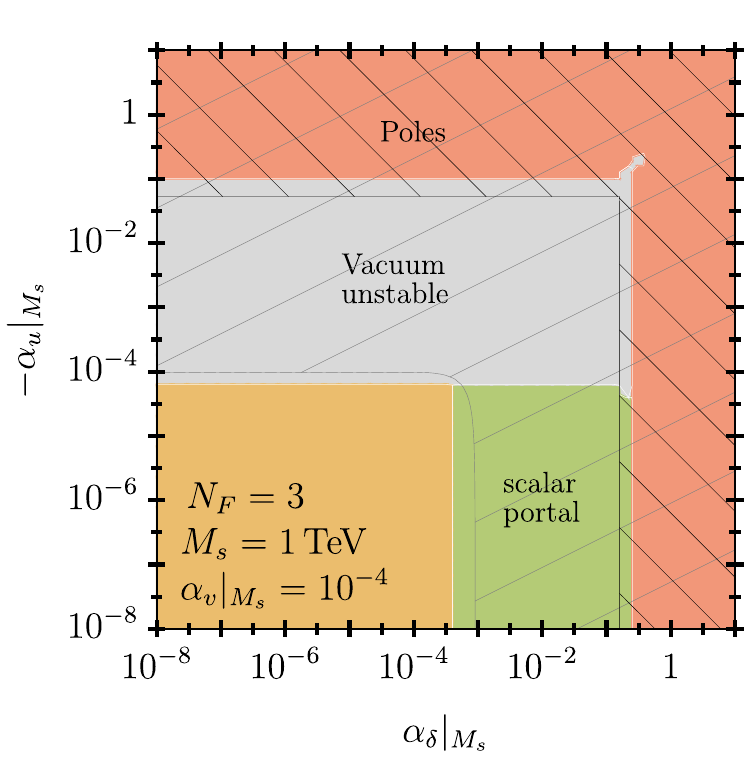} &
     \includegraphics[width=.75\columnwidth]{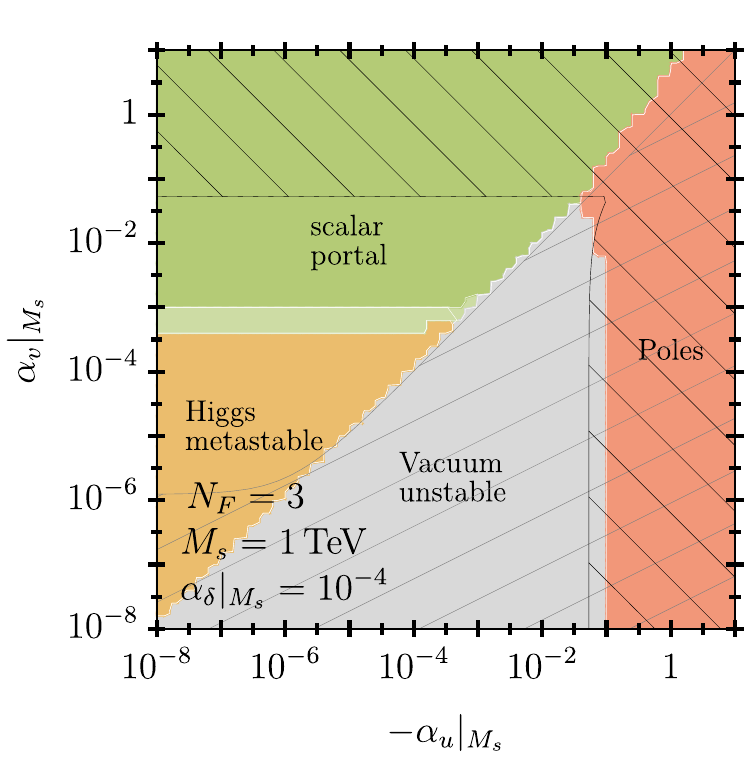} \\
    $\quad$ & $\quad$\\
    \includegraphics[width=.75\columnwidth]{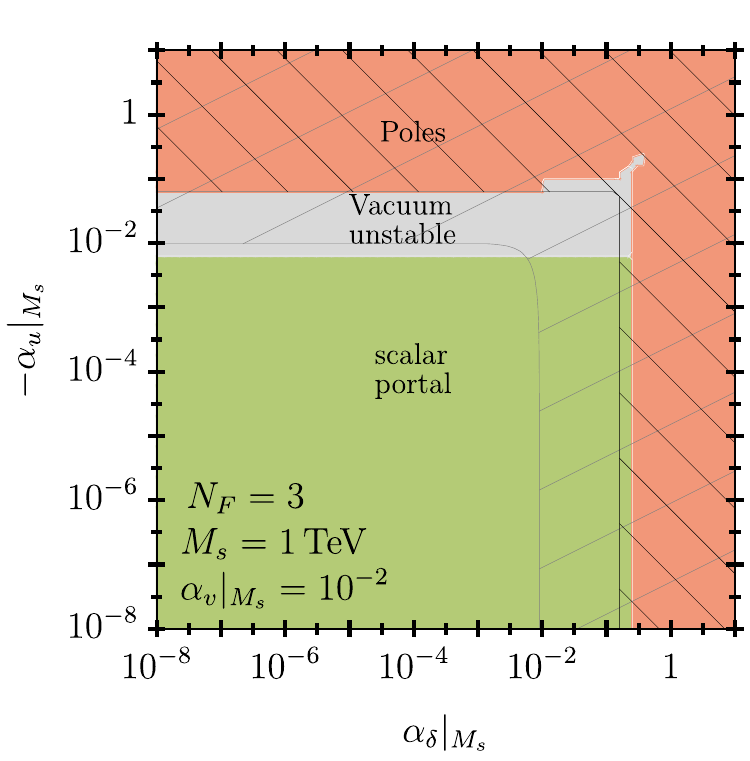} &
   \includegraphics[width=.75\columnwidth]{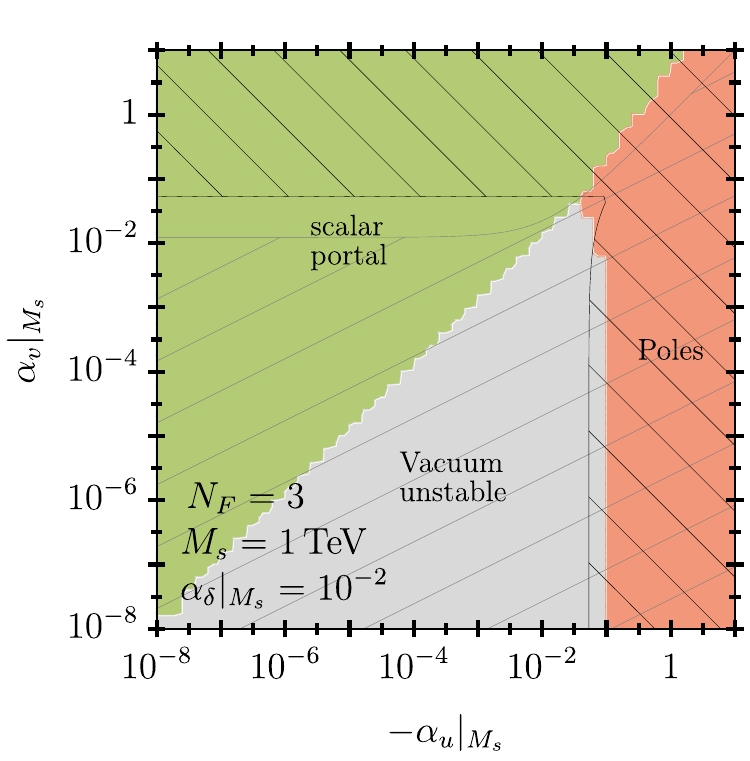} \\
  \end{tabular}
  \caption{BSM critical surface of the $SU(N_F) \times SU(N_F)$ model for negative $\alpha_u(M_s)$
   with  $M_s=1\,\text{TeV}$, $N_F=3$   in the $\alpha_\delta(M_s)-\alpha_u(M_s)$ plane and $\alpha_{v}(M_s)=10^{-4}$ ($\alpha_{v}(M_s)=10^{-2}$) in the upper (lower) plot on the left,
   and  in the $\alpha_u(M_s)-\alpha_v(M_s)$ plane for $\alpha_{\delta}(M_s)=10^{-4}$ ($\alpha_{\delta}(M_s)=10^{-2}$) in the upper (lower) plot on the right.
  Dark (light) green corresponds to strict (soft) Planck safety in vacuum  $V^-$, with other colors as  in \fig{surface-O(N)}. 
}
  \label{fig:surfaceSij3NeguFixedv}
\end{figure*}

\subsection{Negative BSM Quartics}
\label{subsec:negquart}

Here, we  investigate whether Planck safety can be realized in the $SU(N_F) \times SU(N_F)$ model for negative BSM quartics at the matching scale. 
The tree-level stability conditions \eq{vstab} in principle allow all BSM quartics to be negative as long as  $\alpha_u$ and $\alpha_v$ are not  negative at the same time. 
Depending on the sign of $\alpha_u(\mu_0)$ stability can be realized in the two different vacuum configurations $V^\pm$. Recall also that $\alpha_u$ is technically natural. Thus, no RG induced sign changes in $\alpha_u$ occur, and  for negative (positive) $\alpha_u(\mu_0)$ Planck safety can only be realized in  $V^-$ ($V^+$). This is an important difference to models with additional Yukawa couplings \cite{Hiller:2019mou,Hiller:2020fbu,PlanckSafeQuark} which spoil the technical naturalness of $\alpha_u$ and can induce transitions between the vacua.

\begin{figure*}
  \centering
  \renewcommand*{\arraystretch}{0}
  \begin{tabular}{cc}
   \includegraphics[width=.75\columnwidth]{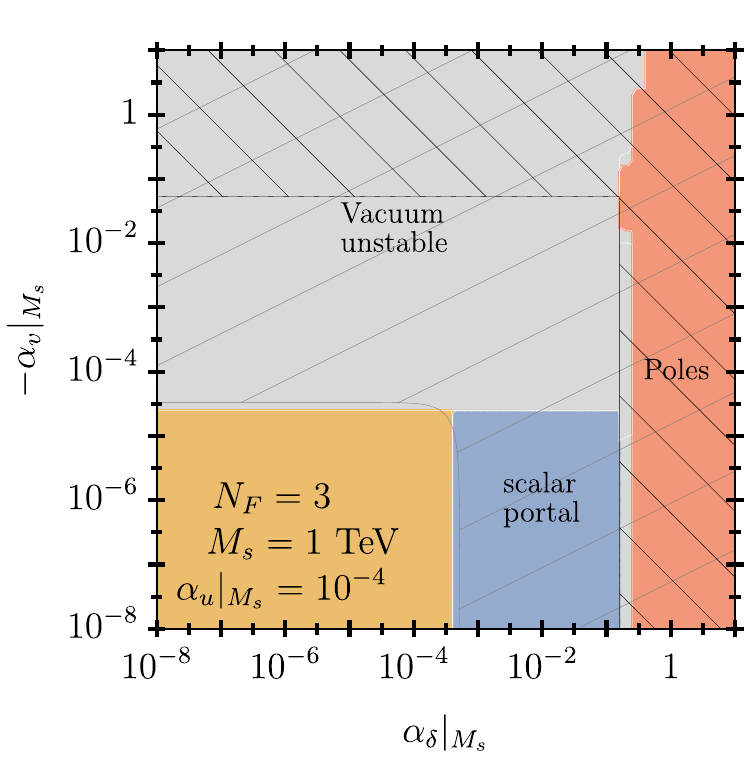} &
    \includegraphics[width=.75\columnwidth]{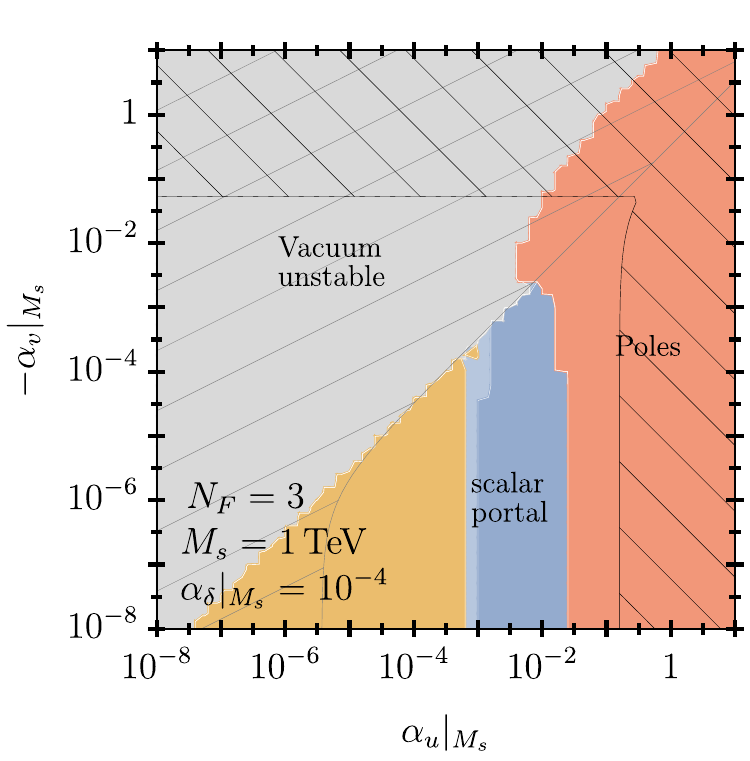} \\
      $\quad$ & $\quad$\\
   \includegraphics[width=.75\columnwidth]{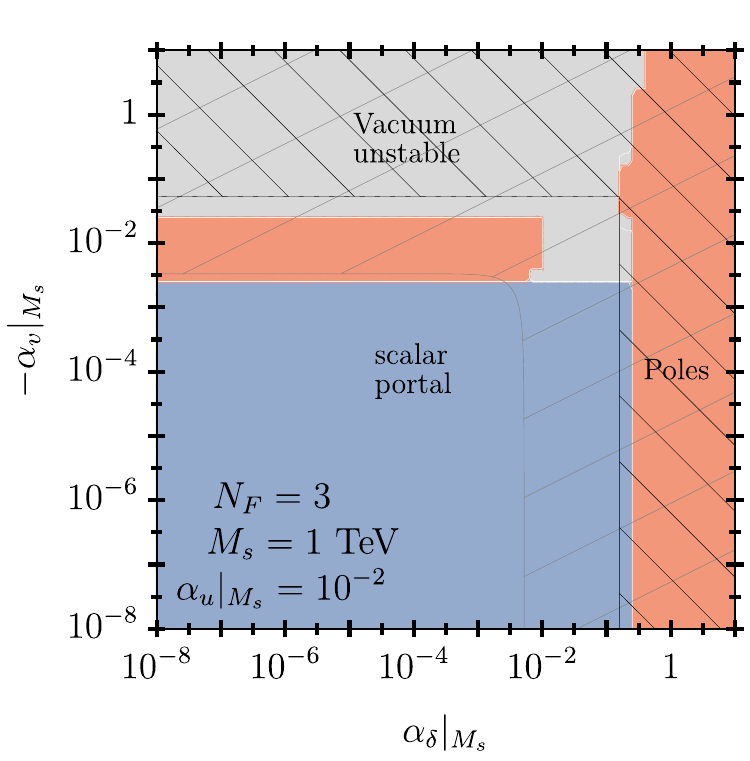} &
      \includegraphics[width=.75\columnwidth]{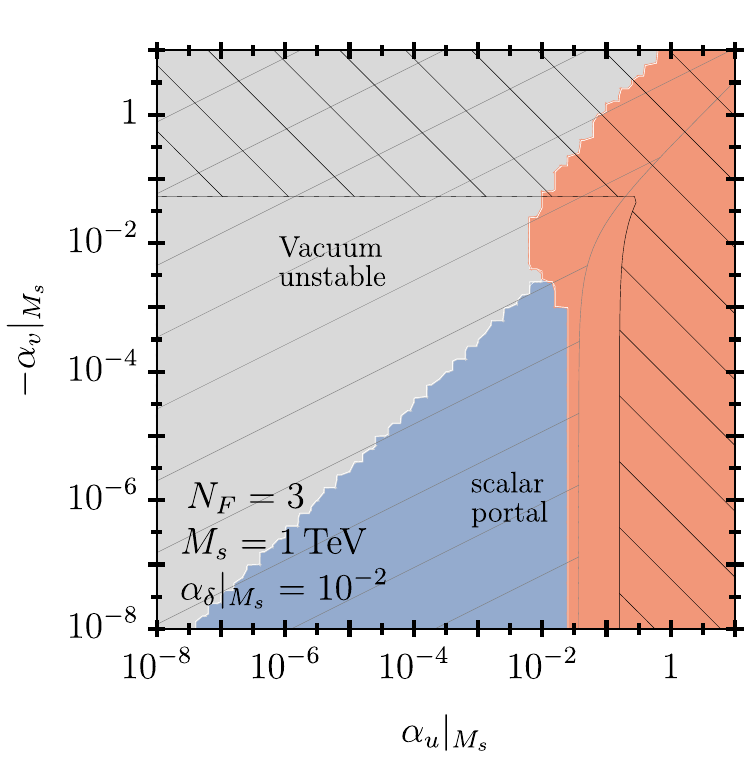} \\
  \end{tabular}
  \caption{BSM critical surface of the $SU(N_F) \times SU(N_F)$ model for negative $\alpha_v(M_s)$ with  $M_s=1\,\text{TeV}$, $N_F=3$ in the $\alpha_\delta(M_s)-\alpha_v(M_s)$ plane for  $\alpha_{u}(M_s)=10^{-4}$ ($\alpha_{u}(M_s)=10^{-2}$) in the upper (lower) plot on the left, and in the $\alpha_u(M_s)-\alpha_v(M_s)$ plane for  $\alpha_{\delta}(M_s)=10^{-4}$ ($\alpha_{\delta}(M_s)=10^{-2}$) in the upper (lower) plot on the right.
 Same color coding as \fig{Sij-delta-v}. }
  \label{fig:surfaceSij3NegvFixedu}
\end{figure*}

We start by analyzing  a negative Higgs portal coupling.
Although in principle allowed by the stability conditions \eq{vstab} we find that  $\alpha_\delta(\mu_0)<0$  is  excluded by  a Planck-safe RG evolution. This result is independent of the signs of $\alpha_{u,v}(\mu_0)$ 
and analogous  to our findings in the $O(N_S)$ model, discussed in Sec.~\ref{subsec:negativeportal}:
also in the  $SU(N_F) \times SU(N_F)$ model the coefficient in the one-loop beta function for the portal, $X$, is positive, see  \eq{betadeltaSij}, for any $N_F \geq 1$, or otherwise inconsistent with  stability \eq{vstab}.
Therefore, with arguments analogous to the $O(N_S)$ model, for portal couplings $\alpha_\delta(\mu_0)<0$ one always encounters vacuum instabilities or poles below $\MPl$.

For the pure BSM quartics $\alpha_{u,v}$, on the other hand,  we  find significant Planck-safe regions in parameter space  for either negative $\alpha_u(\mu_0)$ or  $\alpha_v(\mu_0)$.
In  general, $\alpha_u(\mu_0) <0$ requires  $\alpha_v(\mu_0) > |\alpha_u(\mu_0)|$, see  \eq{vstab}, to realize stability in the vacuum $V^-$, also illustrated in  \fig{surfaceSij3NeguFixedv}.
Likewise, $\alpha_v(\mu_0)<0$ implies positive $\alpha_u(\mu_0) >  N_F |\alpha_v(\mu_0)|$ to satisfy \eq{vstab} in $V^+$, consistent with \fig{surfaceSij3NegvFixedu}.
Other features are similar to the previous discussion of non-negative quartics.

\section{Probing  the Higgs Potential}
\label{sec:Pheno}

The Higgs portal coupling in combination with spontaneous symmetry breaking in the SM and BSM sector induces mixing between the SM Higgs and the BSM scalar. This directly affects the phenomenology of the 125 GeV Higgs boson.
Here we discuss the constraints on the models' parameters from the Higgs width and its couplings. In particular, we work out the impact of our models on the coupling of the Higgs boson to a pair of $Z$ bosons and on its self-couplings
 at leading order.

\subsection{Higgs-BSM Mixing}

We  discuss scalar mixing, and refer  to \App{Mix} for further details.
In addition to electroweak symmetry breaking,  with a VEV $v_h$ for the Higgs as $H=\frac{1}{\sqrt{2}} (h+v_h)$, BSM symmetry breaking may occur and 
the BSM scalar $S$ acquires a VEV, $v_s$, as
\begin{equation} \label{eq:brokenfields}
\begin{aligned}
O(N_S): &\; S_i=(s+v_s) \delta_{i1} + \dots, \\
V^+: &\; S_{ij}=\frac{\delta_{ij}}{\sqrt{2 N_F}}(s+v_s +  i\, \tilde{s})  + \dots \\
V^-: &\; S_{ij}=\frac{\delta_{i1} \delta_{j1}}{\sqrt{2}}(s + v_s + i\, \tilde{s}) + \dots \,.\\
\end{aligned}
\end{equation}
with the real SM and BSM Higgs modes $h$ and $s$, respectively, as propagating degrees of freedom.
Here $V^\pm$ refer to the different vacuum configurations in the $SU(N_F) \times SU(N_F)$ model. 
The ellipsis refer to additional components of the broken scalar fields that are irrelevant for the mixing between $s$ and $h$, while $\tilde{s}$ is an additional pseudoreal scalar singlet which appears for complex fields. 
We assume that the dangerous Goldstone bosons  receive additional, symmetry breaking mass terms to meet phenomenological constraints \cite{Hiller:2020fbu}. A detailed analysis considering also opportunities  from light scalars for cosmology  is beyond the scope of this work.

Plugging \eq{brokenfields} into the potentials \eq{VHS1} and \eq{VHSi} one obtains the scalar potential, which depends only on $h$, $s$ and the VEVs,
\begin{equation} \label{eq:Vbroken}
\begin{aligned}
V(h,s) = &-\frac{\mu_H^2}{2} (h+v_h)^2 - \frac{1}{2} \mu_S^2 (s+v_s)^2+ \frac{\lambda}{4} (h+v_h)^4\\
&+ \frac{\Delta}{\N^2} (s+v_s)^4 + \frac{\delta}{2\N}(h+v_h)^2(s+v_s)^2 \, , 
\end{aligned}
\end{equation}
where the model-specific expressions are given for 
\begin{equation}
\begin{aligned} \label{eq:ModelMatch}
O(N_S): &\quad \N=1,\quad \Delta=v,\\
V^+: &\quad \N=2,\quad \Delta=\frac{u}{N_F} + v,\\
V^-: &\quad \N=2,\quad \Delta=u + v.\\
\end{aligned}
\end{equation}

The fields $h$ and $s$ in the gauge basis mix into the mass eigenstates $h^\prime$ and $s^\prime$,  \eq{rot1}, \eq{rot2}.We identify the masses of these physical fields with the Higgs mass $M_h = m_{h^\prime} = 125$ GeV and the BSM scalar $M_s = m_{s^\prime} >M_h$.
The scalar mixing angle $\beta$ can be written as
\begin{equation} \label{eq:beta}
\tan 2 \beta =\frac{\alpha_\delta}{\sqrt{\alpha_\lambda \alpha_\Delta}} \frac{m_h m_s}{m_s^2-m_h^2} \, ,
\end{equation}
and  is induced by the portal coupling; $m_h \propto v_h$ and  $m_s \propto v_s$ denote mass terms  in the gauge basis, cf. \App{Mix}.

The five (six) \textit{a priori} independent model parameters $\mu_H^2,\,\mu_S^2,\,\lambda,\,\delta,\,v$ (and $u$) in the potential of the $O(N_S)$ ($SU(N_F) \times SU(N_F)$) model are correlated by experimental determinations of  $M_h$ 
and $v_h \simeq 246\,\text{GeV}$ (e.g.~via Fermi's constant $G_F=(\sqrt{2}v_h^2)^{-1}$). Hence, the models are controlled by  three (four) BSM degrees of freedom, for which we choose  $M_s,\,\alpha_\delta(M_s),\,\alpha_v(M_s)$ (and $\alpha_u(M_s)$)  in addition to the number of flavors  $N_S$ ($N_F$).  
This implies also that  the Higgs quartic coupling $\lambda(M_s)$  becomes a function of  these parameters, see \eq{lambda} for the tree-level matching, and in general deviates from its SM value $\lambda_\text{SM}=\tfrac{M_h^2}{2v_h^2}$.
While this effect can be sizable, with observable consequences worked out in  Sec.~\ref{subsec:signatures}, it has only a minor effect on the running, cf. \Sec{SM-stability}

In general, mixing  
reduces  the decay width of the physical Higgs $h^\prime$ to SM final states $\{f\}$ compared  to the SM by a global factor
\begin{equation}
\Gamma(h' \rightarrow \{f\})= \cos^2 \beta \; \Gamma^\text{SM}(h \rightarrow \{f\})\, ,
\label{eq:Higgswidth}
\end{equation}
subject to  a  model-independent 95\% c.l. limit \cite{Dawson:2021jcl}
\begin{equation} \label{eq:betabound}
    |\sin \beta| \leq 0.2 
\end{equation}
from combined Higgs signal strength measurements from ATLAS \cite{ATLAS:2019nkf} and CMS \cite{CMS:2020gsy}.

\subsection{Trilinear, Quartic and \texorpdfstring{$ZZh$}{Zhh} Higgs Couplings}

The scalar mixing affects all couplings of the physical Higgs boson. In particular, $g_{h^\prime ZZ}$,
the coupling to  $Z$'s  is reduced in comparison to the SM one, $g_{hZZ}^\text{SM}$, with
relative shift 
\begin{equation} \label{eq:dghZZ}
\delta g_{hZZ} = \frac{g_{h^\prime ZZ}}{g_{hZZ}^\text{SM}}-1 =\cos \beta -1\, ,
\end{equation}
with the second expression holding at tree-level.
The Higgs coupling to $W$'s is analogously affected by mixing, however,  experimentally weaker constrained  and has smaller projected sensitivity than the $Z$ \cite{FCC:2018byv}, so we focus on the latter.
Currently, $\delta g_{hZZ}$  is experimentally constrained at the level of 6\% by ATLAS \cite{ATLAS:2022vkf} and 7\% by CMS \cite{CMS:2022dwd}. There is also a constraint from ATLAS assuming equal coupling modifiers to pairs of $Z$ and $W$ bosons with an accuracy of 3.1\% \cite{ATLAS:2022vkf}, which would still result in a slightly weaker bound on $\beta$ than \eq{betabound} from the Higgs signal strength. 
However, the projected sensitivities to $\delta g_{hZZ}$ of 1.5\% at HL-LHC \cite{Cepeda:2019klc} and 0.16\% at FCC-ee \cite{FCC:2018byv}  improve the bound on $|\sin \beta|$ to 0.17 and 0.06, respectively.
The ILC at 500 GeV (1 TeV) center of mass energy has a sensitivity of  $0.3 \% \, (0.17 \%)$ in the $hZZ$ coupling \cite{ILC:2019gyn}.

Our models also impact the triple self coupling $\kappa_3$ of the Higgs boson.
The SM tree-level expression
$V^{(3)}_\text{SM} =  \lambda_\text{SM} v_h h^3 \equiv \kappa_3^\text{SM} h^3$
is modified by mixing.
The cubic terms in the scalar potential in the gauge basis read
\begin{equation}
\begin{aligned}
V^{(3)}(h,s) = \lambda v_h h^3 + \frac{\delta}{\N} v_s h^2 s + \frac{\delta}{\N} v_h h s^2 + \frac{4 \Delta}{\N^2} v_s s^3
\end{aligned}
\end{equation}
which after rotating to the mass basis via \eq{rot2} yields
\begin{equation}
\begin{aligned}
V^{(3)}(h^\prime,s^\prime) \supset & \; \kappa_3 h^{\prime 3} \\
=& \left(\lambda v_h \cos^3 \beta - \frac{\delta v_s}{\N} \cos^2 \beta \sin \beta \right.\\ 
& \left. + \frac{\delta v_h}{\N} \cos\beta \sin^2 \beta - \frac{4 \Delta  v_s}{\N^2} \sin^3\beta \right)h^{\prime 3} \\
\end{aligned}
\end{equation}
and thus
\begin{equation} \label{eq:kappa3}
\begin{aligned}
\frac{\kappa_3}{\kappa_3^\text{SM}}=& \frac{\lambda}{\lambda_\text{SM}} \cos^3 \beta - \frac{\delta v_s}{\N \lambda_\text{SM} v_h} \cos^2 \beta \sin \beta \\
&+ \frac{\delta}{\N \lambda_\text{SM}} \cos\beta \sin^2 \beta - \frac{4 \Delta v_s}{\N^2 \lambda_\text{SM} v_h} \sin^3\beta  \, . 
\end{aligned}
\end{equation}
Experimentally, $\kappa_3$ is currently only  poorly constrained by ATLAS and CMS with $-1.5 < \kappa_3/\kappa_3^\text{SM} < 6.7$ \cite{ATLAS:2021jki} and $-1.24 < \kappa_3/\kappa_3^\text{SM} < 6.49$ \cite{CMS:2022dwd}, respectively. Bounds are expected to significantly  improve in the future with projected sensitivities of 50\% at HL-LHC, 44\% at FCC-ee and 5\% at FCC-hh \cite{FCC:2018byv}.
The ILC at 500 GeV (1 TeV) center of mass energy has a sensitivity of  $27 \% \, (10 \%)$ in the trilinear ratio \eq{kappa3} \cite{ILC:2019gyn}.

We also consider BSM effects in the Higgs quartic self-coupling, encoded in $V^{(4)}(h^\prime,s^\prime) \supset \kappa_4 h^{\prime 4}$, where in the SM $\kappa_4^\text{SM}=\tfrac{\lambda_\text{SM}}{4}$, and
\begin{equation} \label{eq:kappa4}
\begin{aligned}
\frac{\kappa_4}{\kappa_4^\text{SM}}\!= \!\frac{\lambda}{\lambda_\text{SM}}\cos^4 \beta\! +\!\frac{2 \delta}{\N \lambda_\text{SM}} \cos^2\beta \sin^2\beta \!+\!\frac{4 \Delta}{\N^2 \lambda_\text{SM}} \sin^4 \beta \, .
\end{aligned}
\end{equation}
No LHC constraints on $\kappa_4$ have been reported to date \cite{Workman:2022ynf}. The FCC-hh has a projected sensitivity of $-4\lesssim \kappa_4/\kappa_4^\text{SM} \lesssim 10$ for $0\lesssim\kappa_3/\kappa_3^\text{SM} \lesssim 1$ \cite{FCC:2018byv}.
Note that unitarity  provides an upper limit $\kappa_4/\kappa_4^\text{SM} \lesssim 65$, see App. \ref{sec:uni}.

BSM effects in \eq{dghZZ}, \eq{kappa3}, and \eq{kappa4} have been implemented at the leading order. An analysis of the Higgs observables in the SM and BSM  beyond  tree-level as well as
a global higher order analysis of electroweak precision observables in the scalar singlet models such as in \cite{Dawson:2021jcl} is desirable but beyond the scope of this work.

\begin{figure}
\centering
\renewcommand*{\arraystretch}{0}
\begin{tabular}{c}
\includegraphics[width=\columnwidth]{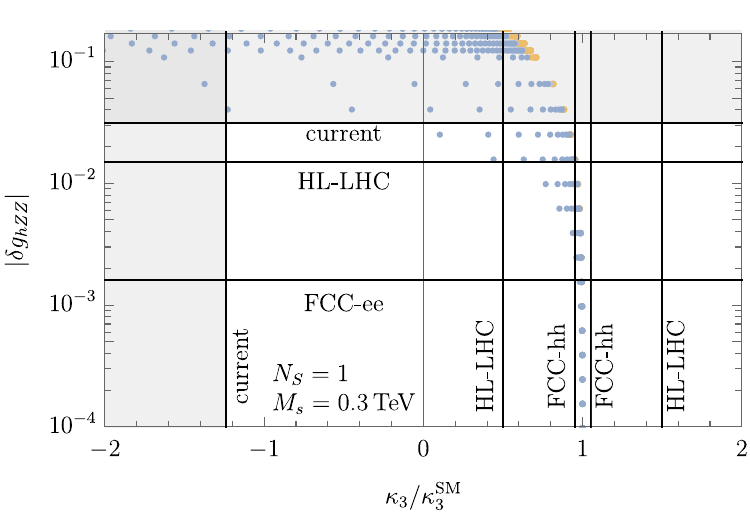}\\
\includegraphics[width=\columnwidth]{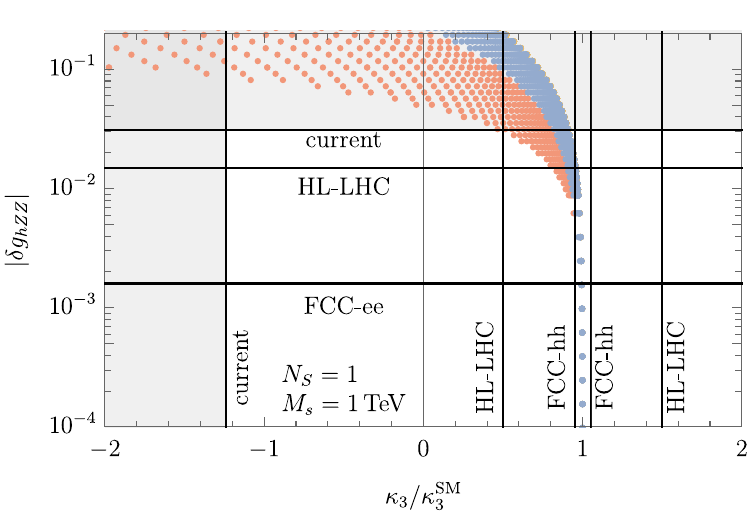}
\end{tabular}
\caption{Higgs couplings $\delta g_{hZZ}$ \eq{dghZZ}  versus $\kappa_3$  \eq{kappa3}  in the $N_S=1$ scalar singlet model for $M_s=0.3 \TeV$ (top), $M_s=1 \TeV$ (bottom). The scattered dots correspond to different BSM coupling values $\alpha_\delta|_{M_s}$ and $\alpha_v|_{M_s}$ as in \fig{surface-O(N)}. The color indicates the Planck fate, same color coding as \fig{surface-O(N)}, blue means safety. Black horizontal and vertical lines correspond to experimental sensitivities at current and future colliders. Gray regions are  excluded by current data.}
\label{fig:ONSPheno}
\end{figure}
\subsection{Signatures of Safe Scalar Singlet Extensions}
\label{subsec:signatures}

We begin with a few analytical approximations, which explain the main features of the phenomenology of Higgs couplings  in Planck-safe scalar singlet extensions.
For small mixing angles $\beta \ll 1$, one obtains
\begin{align}\label{eq:beta-approx}
\beta \simeq \frac{\delta}{2 \sqrt{\lambda \Delta}} \frac{m_{h}}{M_{s}} \, .
\end{align}
Expanding in small $\beta$ and using \eq{lambda} we find that in the trilinear ratio \eq{kappa3} the $\mathcal{O}(\beta)$-term cancels against the shift in $\lambda$. 
Hence, the leading effect arises at $\mathcal{O}(\beta^2)$
\begin{align}
 \frac{\kappa_3}{\kappa_3^\text{SM}} \simeq 1 - \beta^2 \left(\frac{3}{2} \frac{\lambda}{\lambda_\text{SM}} -\frac{\delta}{\N \lambda_\text{SM}} \right)\, .
\end{align}
For the quartic ratio \eq{kappa4} we obtain, using the same approximations
\begin{equation} \label{eq:k4approx}
\frac{\kappa_4}{\kappa_4^\text{SM}} \simeq 1 + \frac{\delta^2}{4 \Delta \lambda_\text{SM}} \, ,
\end{equation}
which is unsuppressed by neither small $\beta$ nor large BSM masses.
Within Planck-safe regions with $\delta \gtrsim 2 \sqrt{\lambda \Delta}$,  this is a positive and sizable  correction of order unity  to the quartic ratio, see \fig{surface-O(N)} in the $O(N_S)$ and \fig{Sij-delta-v} in the matrix scalar model. The BSM shift is enhanced by $M_s$ for fixed $\beta$, and  one therefore expects $\kappa_4$ to be more sensitive to larger masses
than $\kappa_3$.

\begin{figure*}
  \centering
  \renewcommand*{\arraystretch}{0}
  \begin{tabular}{cc}
  \includegraphics[width=\columnwidth]{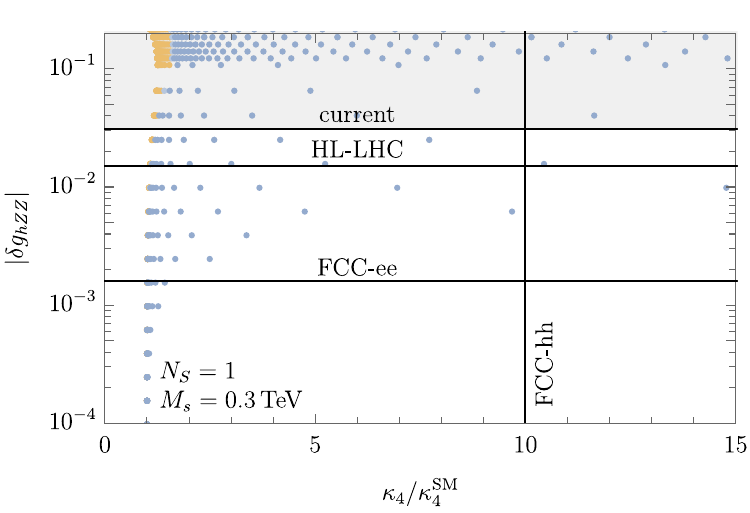}&
  \includegraphics[width=\columnwidth]{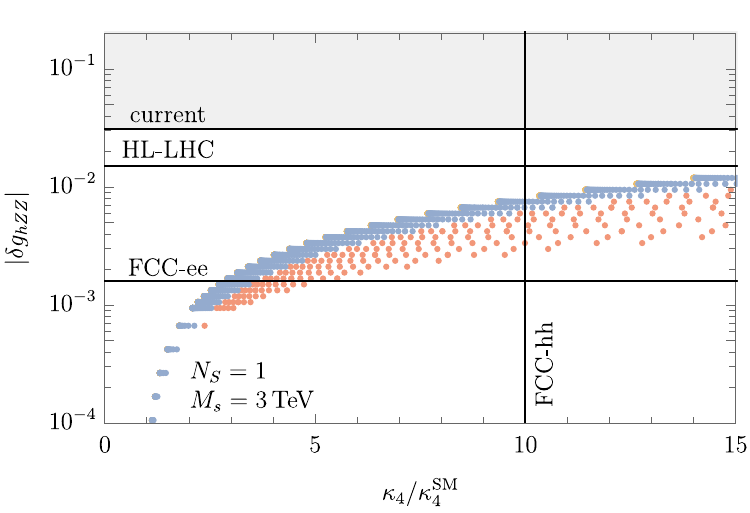}\\
 \includegraphics[width=\columnwidth]{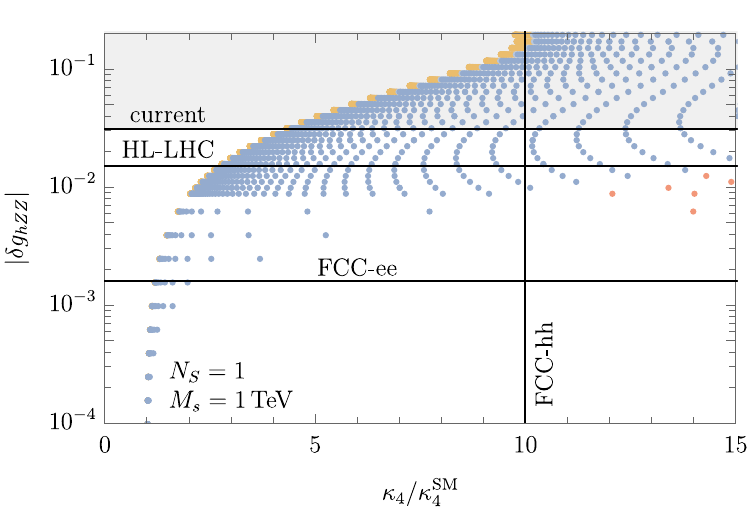}&
   \includegraphics[width=\columnwidth]{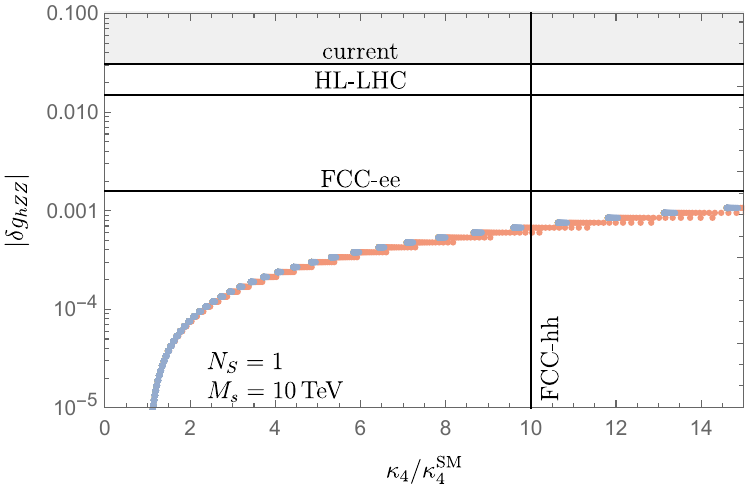}\\
  \end{tabular}
  \caption{Higgs couplings  $\delta g_{hZZ}$ \eq{dghZZ}  versus $\kappa_4$  \eq{kappa4}  in the  scalar singlet model $(N_S=1)$ for different scalar field masses, showing $M_s=0.3 \TeV$ (top left), $M_s=1 \TeV$ (bottom left), $M_s=3 \TeV$ (top right), and $M_s=10 \TeV$ (bottom right); color-coding and exclusion regions  as in \fig{ONSPheno}. The admissible band of Planck-safe models becomes increasingly narrower with increasing mass.}
  \label{fig:k3k4}
\end{figure*}

For the diboson-Higgs  coupling we expect small BSM effects of the order
\begin{align} \label{eq:corr}
\delta g_{hZZ} \simeq -\frac{\beta^2}{2} 
\, . 
\end{align}
Unlike in $\kappa_4$, both BSM shifts in $\kappa_3$ and $\delta g_{hZZ}$ are suppressed by $M_s^2$, see \eq{beta-approx}, i.e., they are smaller for larger BSM scalar masses. 

In the following we  validate these features in the numerical analysis, which is based on the full expressions without expansions in small mixing angles or mass ratios. 

The  couplings $\delta g_{hZZ}$ \eq{dghZZ} and $\kappa_3$ \eq{kappa3}, as well as $\delta g_{hZZ}$ \eq{dghZZ}  against $\kappa_4$ \eq{kappa4}  in the $N_S=1$ scalar singlet extension are shown in \fig{ONSPheno} and \fig{k3k4} for $M_s=0.3 \TeV$ and $M_s=1 \TeV$. Due to its higher sensitivity to larger $M_s$ we
also show BSM effects in $\kappa_4/\kappa_4^\text{SM}$ for $3 \TeV$ and $10 \TeV$. For  larger $M_s$ the band becomes increasingly  narrower and tuned, and decouples from
$\delta g_{hZZ}$  and $\kappa_3$. Blue dots correspond to Planck-safety.

We observe that
safe scalar singlet models are already probed, with some part of the parameter space ruled out by the LHC
(gray area in \fig{ONSPheno} and  \fig{k3k4}).
The bound on $\delta g_{hZZ}$ excludes large deviations in $\kappa_3/\kappa_3^\text{SM}$.
In the nearer future  the HL-LHC \cite{Cepeda:2019klc} will improve the limit on $\delta g_{hZZ}$ by roughly a factor of two, and could observe the trilinear for smaller $M_s$; the projected reach in parameter space is larger with $\delta g_{hZZ}$ than  with $\kappa_3$. For TeV-ish BSM masses and above 
BSM effects  in $\kappa_3$ are within $20 \%$, and within $10 \%$ after the HL-LHC  phase due to the correlation with $hZZ$-limits, and
detectable at  the future hadron collider FCC-hh \cite{FCC:2018byv}. For a future FCC-ee collider \cite{FCC:2018byv}  one expects an order of magnitude improvement in $\delta g_{hZZ}$ from the current limit.
 In contrast, the enhancement of $\kappa_4/\kappa_4^\text{SM}$ can be sizable,  typically a factor of a  few,  and even $O(10)$ factors are  achievable. 
 Consequently, for larger $M_s \approx 10 \TeV$, signatures slip out of reach of  the FCC-ee  due to the smallness of $\delta g_{hZZ}$. We stress that   an enhanced Higgs quartic may well be  the only signature of the model, which is in reach of the FCC-hh (see \fig{k3k4}).

Changing  $N_S$ but keeping $\alpha_\delta,\,\alpha_v$ and $M_s$ would give identical values of $\kappa_3$ and $\delta g_{hZZ}$, so the plots hold generically in $O(N_S)$ models.
While  for larger $N_S$ Planck safety can be obtained already for lower coupling values and hence, points corresponding to a metastable Higgs can convert to Planck-safe ones in \fig{ONSPheno}, this does not increase to region of available, and testable parameter space.

The phenomenology of the $SU(N_F) \times SU(N_F)$ model in Planck-safe parameter configurations is
very similar to the $O(N_S)$ model, and therefore not shown.
Note that Planck-safe parameter configurations in the  vacua $V^\pm$ map onto similar points in the Higgs coupling space and require additional observables such as those related to the flavor structure to disentangle them.

In this section we studied the modifications of the Higgs couplings from the portal
to BSM scalars. Models allow for further phenomenology from the BSM sector.
Possible signatures  include decays of the BSM scalars  to SM fermions or
gauge bosons via the Higgs portal and mixing, giving rise  to  resonance structures in diboson, dijet or dilepton distributions  around the mass of the BSM scalar.  
However, the production rate of the scalar $s$ is suppressed by $\beta^2$, although could be larger in models with vector-like fermions with SM charges and Yukawa coupling to the scalar singlet  and SM fermions \cite{Bissmann:2020lge}.
 A detailed study of the reach  at colliders and other searches
is beyond the scope of this works  
For further phenomenological discussion of the $O(N_S)$ model and the  $SU(N_F) \times SU(N_F)$ model we refer to e.g. \cite{Dawson:2021jcl,Huang:2016cjm} 
and \cite{Hiller:2020fbu}, respectively.

\section{Conclusions}\label{sec:Conclusion}

We revisited the stability analysis of the SM Higgs potential with the state-of-the art precision in theory and experiment.
The mass of the  top quark  followed by the strong coupling constant are the most critical observables to determine vacuum stability, 
with their correlation shown in \fig{SM-stab-region}. 
The region  (green) in which the Higgs quartic remains positive is away from the PDG 2024 averages of
$M_t$ and $\alpha_s$  by $-1.9 \sigma$ and  $+3.7\sigma$, respectively.
Correlations matter and ideally should be provided by combined experimental determinations, such as from CMS  \cite{CMS:2019esx},
which finds lower central values  closer to the stability region and more sizable uncertainties. 
An overview of shifts using different inputs for the top mass  is provided in \tab{uncertainties}. 
 We estimate that reducing   uncertainties by a factor of  two to three is sufficient to establish or refute  SM vacuum stability  at the $5\sigma$ level. 
  Incidentally,  the Monte-Carlo mass  $M_t^{\text{MC}}$, whose value excellently agrees with the pole mass, is measured already today with  such high precision, see \tab{uncertainties}.
 Sharpening its interpretation would  hence also be a useful endeavor.
  Our results encourage further  combined  analyses  of the top quark  mass together with the strong coupling constant
to shed light on the stability of the SM vacuum.

In addition, we  have investigated the Higgs portal mechanism  as a minimally invasive route towards stability beyond the SM.
We systematically studied a variety of  singlet scalar field extensions  covering real, complex, vector and matrix scalar fields, with and without flavor symmetries, thereby also extending earlier works. Particular emphasis has been given to the interplay of the Higgs, portal, and BSM quartic couplings and their higher order RG evolution up to the Planck scale.

We find that a TeV-ish (but possibly much heavier)  BSM scalar singlets with modest portal coupling within $10^{-3} \lesssim  \alpha_\delta(\mu_0) \lesssim 10^{-2}$ suffices to achieve stability. 
For larger mass the required value of $\alpha_\delta(\mu_0)$ slowly increases as there is less RG time left to rescue the potential (\fig{surface-O(N)}).
The value of the Higgs portal coupling can also be quite feeble by increasing the number of BSM scalars, or for sizable pure BSM quartic couplings by indirect stabilization from RG-mixing \eq{betadeltaONS}, \eq{betadeltaSij}.
Negative values of the portal $\alpha_\delta(\mu_0)$ -- although in principle allowed  -- are  inconsistent with Planck safety because they run into trouble, either into subplanckian Landau poles or RG-induced vacuum instabilities. 
The $SU(N_F) \times SU(N_F)$ models, on the other hand,   feature two pure BSM quartics for which we find sizable Planck-safe areas in the parameter space with one of them negative (\fig{surfaceSij3NeguFixedv} and  \fig{surfaceSij3NegvFixedu}). These  models can also break flavor universality spontaneously.

There is sizable room for new physics in the  scalar singlet models that can be tested at the HL-LHC \cite{Cepeda:2019klc} in Higgs couplings to electroweak bosons, its trilinear coupling, and for sub-TeV BSM masses (\fig{ONSPheno}).
A future lepton collider, such as the FCC-ee \cite{FCC:2018byv}  can explore these models further by reaching deeper into regions with smaller Higgs-BSM mixing 
due to an order of magnitude higher sensitivity to  $\delta g_{hZZ}$.
BSM effects in the quartic Higgs self-coupling \eq{kappa4}, on the other hand, can experience  significant enhancements by a factor of up to ten (\fig{k3k4}), which 
require a future collider with sufficient energy and precision such as the
FCC-hh \cite{FCC:2018byv}.

The Higgs-to-electroweak-boson
and the trilinear self-couplings  provide  promising  nearer term  collider observables of  planck-safe flavorful scalar singlet extensions. 
Further studies exploring  the Higgs couplings are encouraged.

\acknowledgments

We thank Jonas Lindert, Martin Schmaltz, Kai Spychala, and Emmanuel Stamou for discussions and comments on the manuscript. This work is supported by the \textit{Studienstiftung des Deutschen Volkes} (TH), and the Science and Technology Research Council (STFC) under the Consolidated Grant ST/X000796/1 (DFL). This work was  performed in part at Aspen Center for Physics (GH, DFL), which is supported by National Science Foundation grant PHY-2210452, and was partially supported by a grant from the Bernice Durand Fund (GH).

\FloatBarrier
\appendix

\section{SM Effective Potential} \label{sec:SM-stab-details}

In this appendix, we recall the main steps to find the SM effective potential following e.g.~\cite{Ford:1992pn,Martin:2013gka,Martin:2014bca,Martin:2015eia,Martin:2017lqn,Martin:2018emo}.
Classically, the stability of the Higgs potential only depends on the sign of its quartic interaction $\propto \lambda (H^\dagger H)^2$. At quantum level, additional corrections arise 
as the classical potential is improved to the effective potential, including  quantum effects and operators of higher powers in the field.

Technically, the quantum effective potential $V_\text{eff}$ is obtained by expanding the Higgs $H = \frac{1}{\sqrt{2}} h + \hat{H}$ around a constant, classical field  $h$ and integrating out the quantum field $\hat{H}$.
The RG equation for the effective potential reads
\begin{equation}
    0 = \left(\frac{\partial}{\partial \ln\mu} + \gamma \frac{\partial}{\partial \ln h} + \sum_i \beta_i \frac{\partial}{\partial \alpha_i}\right) V_\text{eff}\,,
\end{equation}
where $\gamma$ is the anomalous dimension of the classical field $h$ and $\alpha_i$ are all SM couplings with their beta functions $\beta_i$. 
In the following, we neglect terms $\propto h^2$ and $\propto h^3$ in the scalar potential, which are irrelevant for the discussion of a false minimum at large field values.
As the classical potential is $V= \tfrac14 \lambda h^4$, we relate the effective potential to a dimensionless parameter $\lambda_h$ via 
\begin{equation}
    V_\text{eff} = \tfrac{1}{4} \lambda_h (\alpha_i,z_h)\, h^4\quad \text{ where } \quad z_h = h/\mu
\end{equation} without loss of generality.
Thus the total scale and field dependence in $\lambda_h$ is given by  
\begin{equation}\label{eq:lambda_h}
  \frac{\partial \lambda_h}{\partial \ln z_h} = \left(\frac{4\gamma}{1-\gamma} + \sum_i \frac{\beta_i}{1- \gamma} \frac{\partial}{\partial \alpha_i} \right) \lambda_h\,.
\end{equation}
 In the following, we  fix the RG-scale $\mu = \mu_\text{ref}$ and resum the effective potential in $h$, starting from some initial field value $h_0$. Eq.~\eq{lambda_h} implies that this can be achieved via the relation 
\begin{equation}
  \frac{\partial \lambda_h}{\partial \ln h} = \left(\frac{4\gamma}{1-\gamma} + \sum_i \bar{\beta}_i \frac{\partial}{\partial \bar{\alpha}_i} \right) \lambda_h\,.
\end{equation}
Here we have introduced redefined couplings $\bar{\alpha}_i(h)$ that implicitly depend on the field value. The $\bar{\alpha}_i(h)$  are related to the usual couplings at the initial field value $h=h_0$: 
\begin{equation}
  \bar{\alpha}_i(h_0) = \alpha_i(\mu_\text{ref})\,,
\end{equation}
and implement the resummation in $h$ via the modified running
\begin{equation}\label{eq:betabar}
   \bar{\beta}_i \equiv \frac{\mathrm{d}}{ \mathrm{d} \ln h} \bar{\alpha}_i(h) =\frac{\beta_i (\bar{\alpha})}{1 - \gamma(\bar{\alpha})} \,.
\end{equation}
Note that for resumming the effective potential it is not sufficient to just use conventional RG running of all couplings to the scale $\mu = h$, i.e. evolving $\alpha_i(\mu_\text{ref})$ to $\alpha_i(h)$. 
The difference is the anomalous dimension term in the denominator of \eq{betabar}, which corrects the resummation by accounting the explicit $h$ dependence of the potential, leveraging its overall RG invariance. However, this subtlety only enters at two-loop in the potential. 

Next, we refine the ansatz via
\begin{equation}\label{eq:Gamma}
  \lambda_h = e^{4 \bar{\Gamma}}  \lambda_\text{eff}  \quad \text{ with } \quad  \bar{\Gamma}(h) = \int_{h_0}^{h} \frac{\mathrm{d}h'}{h'}\frac{\gamma}{1- \gamma}\,.
\end{equation}
Since the exponential term is always positive, vacuum stability solely depends on the sign of the quantity $\lambda_\text{eff}$, which is resummed by
\begin{equation}
  \frac{\partial \lambda_\text{eff}}{\partial \ln h} = \sum_i \bar{\beta}_i \frac{\partial}{\partial \bar{\alpha}_i} \lambda_\text{eff}\,.
\end{equation}
In order to obtain a starting value $\lambda_\text{eff}(h_0)$ the effective potential 
\begin{equation}\label{eq:Veff-ansatz}
  V_\text{eff} = \frac14 \lambda_\text{eff}(h) e^{4\bar{\Gamma}} h^4
\end{equation}
is compared against fixed-order calculations~\cite{Ford:1992pn,Martin:2013gka,Martin:2014bca,Martin:2015eia,Martin:2017lqn,Martin:2018emo} at constant field value $h=h_0$. By definition, $\bar{\Gamma}(h_0) = 0$ and 
\begin{equation}\label{eq:lambda_0-1L-z0}
  \begin{aligned}
    \lambda_\text{eff}(h_0)  &{}= \,  \lambda  + \frac{1}{(4\pi)^2}  \bigg[   4\lambda^2 \left( \ln\frac{2\lambda h_0^2}{\mu_\text{ref}^2} - \frac32\right) \\
     &{}  + \frac{3}{8} g_2^4 \left(\ln\frac{g_2^2 h_0^2}{4 \mu_\text{ref}^2} - \frac56\right)  \\
    & + \frac{3}{16} (g_1^2 + g_2^2)^2 \left(\ln\frac{(g_1^2 + g_2^2)h_0^2}{4 \mu_\text{ref}^2 } - \frac56\right)\\
    & - \sum_f N_f  y_f^4 \left(\ln\frac{y_f^2 h_0^2}{2 \mu_\text{ref}^2} - \frac{3}{2}\right)
    \bigg] + \mathcal{O}(\text{2 loop})\,,
  \end{aligned}
  \end{equation}
  where the last line is summed over all fermion flavors  with   $N_f=3$ for quarks and $N_f=1$ for leptons.
  
  We chose $h_0$ to minimize the size of loop corrections to $\lambda_\text{eff}(h_0)$ by controlling the size of logarithmic terms. Around the electroweak scale, the strong gauge coupling $g_3$ is most sizable, but only appears at two loops in $\lambda_\text{eff}$ and outside of logarithms. The next largest coupling is the top Yukawa $y_t$. Therefore, logarithms stemming from top corrections are switched off by choosing 
  \begin{equation} \label{eq:h0}
    h_0 = \frac{\sqrt{2} \,\mu_\text{ref}}{ y_t(\mu_\text{ref})} \approx 1.53\, \mu_ \text{ref}\,,
  \end{equation}
  which yields \eq{lambda_0-1L}.
  Notice that other logarithms in \eq{lambda_0-1L-z0} become more sizable by this choice. However, these terms remain nevertheless subleading owing to the smallness of couplings in front of them.
  This choice has been adopted in \Sec{SM-stability} with the reference scale \eq{mu_ref} and $\alpha_{\lambda,\text{eff}}(h) = \lambda_\text{eff}(h)/(4\pi)^2$.

  We have restricted ourselves to Landau gauge $\xi = 0$ for which the specific fixed-order results~\cite{Ford:1992pn,Martin:2013gka,Martin:2014bca,Martin:2015eia,Martin:2017lqn} are available. For other choices of the gauge-fixing, the potential is merely known up to two loops~\cite{Martin:2018emo}. As $\beta_\xi \propto \xi$, the Landau gauge remains intact under RG evolution and does not produce hidden contributions to the resummation of $\lambda_\text{eff}(h)$.
  Although the effective potential depends on the gauge-fixing, its extrema are actually independent of it, following the Nielsen identity~\cite{Nielsen:1975fs}
\begin{equation}
\left(\xi\frac{\partial}{\partial \xi} + C(h,\xi) \frac{\partial}{\partial  h}\right)V_\text{eff}(h,\xi) = 0\,,
\end{equation}
  where $C(h,\xi)$ is a function of the field value and gauge-fixing parameter that can be determined order by order in perturbation theory.
  As a result, the stability of the SM potential is gauge-independent, as it is determined by the sign of $V_\text{eff}(h)$ at its false vacuum.

 Finally, we briefly compare our ansatz for the effective potential with procedures used previously, e.g.~\cite{Casas:1994qy,Buttazzo:2013uya,Andreassen:2014gha,Bednyakov:2015sca}, where slight differences arise due to the definition of \eq{Gamma} and the resummation  \eq{betabar}.
 Specifically,  the ansatz used in \cite{Casas:1994qy,Buttazzo:2013uya,Andreassen:2014gha,Bednyakov:2015sca} consists of evaluating the effective potential at an RG scale $\mu = h$. Starting from an initial scale $\mu_0$, all couplings $\alpha_i(\mu_0)$ are evolved towards $\alpha_i(h)$ using their conventional RG flow equations $\beta_i$.
  Field values $h \equiv h(\mu_0)$ are also renormalized and their RG evolution reads
  \begin{equation}
    h(h) = h(\mu_0) \,e^{\Gamma(h)} \quad \text{ with } \quad \Gamma(h) = \int_{\mu_0}^{h} \frac{\mathrm{d} \mu}{\mu} \gamma\,.
  \end{equation}
  Thus, the corresponding effective potential is given by the ansatz 
  \begin{equation}\label{eq:literature-ansatz}
    V_\text{eff} = \frac14 \tilde{\lambda}_\text{eff}(\alpha(h), \Gamma(h)) \,h^4 e^{4 \Gamma(h)} \,.
  \end{equation}
  Here $\tilde{\lambda}_\text{eff}$ is obtained by evaluating a fixed-order calculation of the effective potential at the RG scale that corresponding to the field value $\mu=h$.
  To this end, both couplings and field values are run from the reference scale $\mu=\mu_0$ to $\mu=h$. This amounts to a replacement 
  \begin{equation}
    \alpha_i(\mu_0) \mapsto \alpha_i(h),\; \text{ and } \;  \ln\frac{h(\mu_0)}{\mu_0} \mapsto  \ln\frac{h(h)}{h(\mu_0)}= \Gamma 
  \end{equation}
  in the fixed-order expression of the effective potential. As such, the approach is equivalent to our ansatz \eq{Veff-ansatz} up to the loop order of the potential used as input. Notice however that \eq{literature-ansatz} does not  resum logarithms of the field amplitude  contained in $\Gamma$
    \begin{equation}
      \Gamma(h) = \gamma( \alpha_i(\mu_0) ) \ln\frac{h}{\mu_0} + \mathcal{O}\left(\left[\ln\frac{h}{\mu_0}\right]^2\right)\,.
    \end{equation} 
    As unknown higher loop orders explicitly contain $\Gamma$, these contributions will be  logarithmically enhanced for larger field values $h$.
    Our ansatz, instead, avoids  large logarithms 
    and resums them completely in the modified RG evolution of \eq{betabar}.
    
    We conclude that our ansatz for the effective potential slightly improves  upon previous formulations as, in addition, large logarithms in the Higgs field amplitude are consistently resummed, taking full advantage of the effective potential's RG invariance. We have confirmed that, at the highest available loop orders used in this work, that quantitative  differences are minute, and smaller than the errors  set by input data.

\section{Some Terminology and Conventions}\label{sec:PS}

In this work, we are primarily  interested in whether BSM models stabilize the SM effective potential, or continue to exhibit instabilities, or even worse Landau poles, prior to the Planck scale $\MPl$.  Then, possible realizations include outright stability  at the Planck scale ($0\le \alpha_\lambda$), SM-like meta-stability (mildly negative $\alpha^{\rm SM}_\lambda\le\alpha_\lambda<0$), and vacuum instability $(\alpha_\lambda<\alpha^{\rm SM}_\lambda<0)$. To differentiate between  scenarios, and for want of some terminology, we refer to the different possibilities as follows:
\begin{itemize}
\item \textit{Strict Planck Safety}: We characterize BSM models as (strictly) Planck-safe, provided   the tree-level potential including both SM and BSM fields is stable for all scales up to the Planck scale, $\mu_0 \leq \mu \leq \MPl$. We also demand that the RG running does not lead to  Landau poles in any of the other couplings.

\item \textit{Soft Planck Safety}: We characterize BSM models as (softly) Planck-safe, provided  their  tree-level potential including both SM and BSM fields is stable at the Planck scale, but we allow  $\alpha_\lambda$ to be  negative at intermediate scales, but not more negative than in the SM,
\begin{equation} \label{eq:softlambda}
-10^{-4} \lesssim \min \alpha_\lambda(\mu), \quad \alpha_\lambda(\MPl) >0 \, .
\end{equation}
As the stability conditions of the  portal coupling $\delta$ involve $\sqrt{\lambda}$ we require
\begin{equation} \label{eq:softdelta}
\alpha_\delta(\mu) > 0 \quad \text{if} \quad -10^{-4} \lesssim \alpha_\lambda(\mu) \lesssim 0   
\end{equation}
in accordance with the stability conditions \eq{Vstab1}, \eq{vstab} in the limit $\lambda \to 0$.
We also demand that the running does not lead to   Landau poles elsewhere.

\item \textit{SM-like Theories}: We characterize BSM models as SM-like provided the Higgs quartic remains mildly negative at the Planck scale, yet less negative as in the SM,
\begin{equation}
-10^{-4} \lesssim \alpha_\lambda(\MPl) \lesssim 0 \,,
\end{equation}
and without  spurious Landau poles prior to $\MPl$.

\end{itemize}

\begin{figure}
\renewcommand*{\arraystretch}{0}
\includegraphics[trim={1.5cm 0 0 0}, clip, width=1.1\columnwidth]{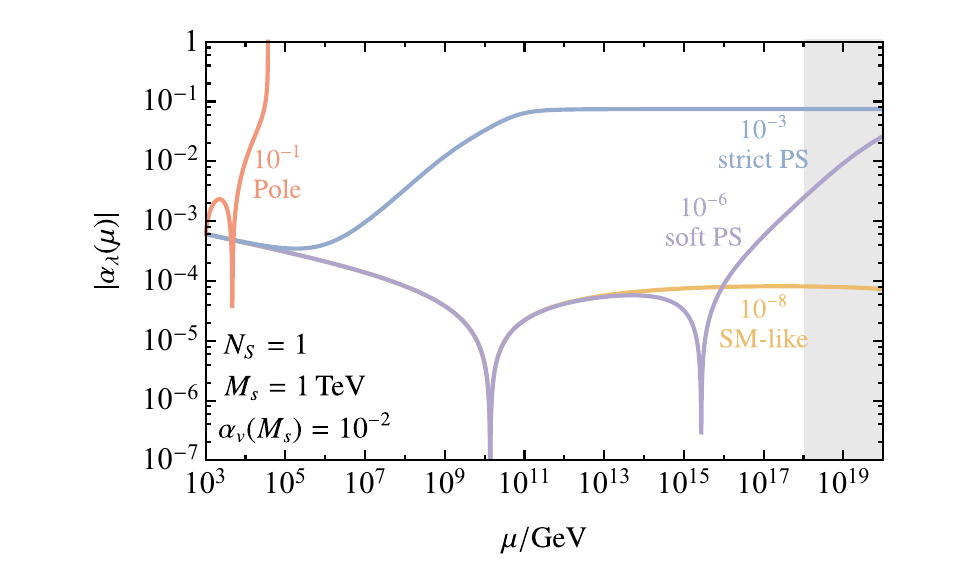}
\caption{A selection of trajectories for the Higgs quartic coupling $|\alpha_\lambda(\mu)|$, illustrating the main  scenarios depending on the Higgs portal coupling $\alpha_\delta$ at the matching scale $M_s$, exemplarily  for a 
model with a   real BSM scalar ($M_s=1 \,\text{TeV}$, $\alpha_v(M_s)=10^{-2}$, see also \eq{VHS1}). 
For feeble $\alpha_\delta=10^{-8}$ (yellow, metastability), the theory remains SM-like.
Mildly increasing $\alpha_\delta=10^{-6}$ (purple), the theory becomes (softly) Planck-safe, with $\alpha_\lambda$ changing sign twice prior to the Planck scale. 
Increasing $\alpha_\delta$ even further, $\alpha_\delta=10^{-3}$ (dark blue), the model becomes strictly Planck safe, with $\alpha_{\lambda}(\mu) >0$  up to $\MPl$. 
For strongly coupled settings, $\alpha_\delta=10^{-1}$ (red), a Landau pole arises right above the matching scale, and the running cannot be continued to the Planck scale.}
  \label{fig:PSTypes}
\end{figure}

With this classification in mind, we have adopted the following color-coding in key figures of this paper.
If an RG flow features a subplanckian Landau pole it is always labeled with \textit{Pole} (red). 
A pole-free RG flow with $ \min \alpha_\lambda(\mu) < -10^{-4}$ is marked as \textit{Higgs unstable} (brown). 
A RG evolution without poles and with $\min \alpha_\lambda(\mu) > -10^{-4}$ but violating BSM stability conditions \eq{Vstab1}, \eq{vstab}, or \eq{softdelta} is titled \textit{Vacuum unstable} (gray). 
If BSM stability conditions in \eq{Vstab1}, \eq{vstab} or \eq{softdelta} are fulfilled and $-10^{-4} \lesssim \min \alpha_\lambda(\mu) \lesssim 0$ as stated in \eq{softlambda} the corresponding RG flow is  \textit{SM-like} if $-10^{-4} < \alpha_\lambda(\MPl) \leq 0$ (yellow) or (softly) Planck safe, if $\alpha_\lambda(\MPl) > 0$ (light blue). 
If all tree-level stability conditions \eq{Vstab1} or \eq{vstab} are fulfilled for all scales $\mu_0 \leq \mu \leq \MPl$, models are (strictly) Planck safe (dark blue).
In \fig{PSTypes}, we illustrate our terminology exemplary for a SM extension with a real scalar field and various portal couplings. 

We add that these possibilities do not exhaust the space of viable models, and further possibilities exist:

1. $ \alpha_\lambda$ could be even lower, i.e.~more unstable than in the  SM. This might still be viable if the tunneling rate into the false vacuum is sufficiently large compared to the age of the universe. A verification of this is beyond the scope of this work.

2. Concerning sub-planckian Landau poles, it is conceivable that an additional new physics sector tames it, an example would be  a unified theory, or that even a walking effect in the non-perturbative regime comes to rescue
when the coupling gets large.
While the former is outside of the model framework considered the  latter is outside of the region where we have control.

\section{Scalar Mixing}
\label{sec:Mix}

The  kinetic and mass terms for a single, free scalar field $S$ read
\begin{equation} \label{eq:LS}
\mathcal{L}_S = \frac{\N}{2} (\partial_\mu S)^\dagger \partial^\mu S - \frac{\N}{2} m_S^2 S^\dagger S \, , 
\end{equation}
where $\N=2$ ($\N=1$) for a complex (real) scalar field which is  canonically normalized.
When spontaneous symmetry breaking in the BSM sector occurs, a VEV $v_s$ and a real scalar  $s$ emerge via \eq{brokenfields}
along with other components which depend on details of the scalar sector.
In particular, a $O(N_S)$ symmetric BSM  model additionally yields a scalar field $\phi$ in the fundamental representation of the remaining $O(N_S-1)$ symmetry: 
\begin{equation}
    O(N_S): \quad S =  \begin{pmatrix}
        v_s + s \\
        \phi^1 \\
        \vdots \\
        \phi^{N_S - 1}
    \end{pmatrix} \,.
\end{equation}
As the $SU(N_F)_L \times SU(N_F)_R$ symmetric scalar is complex, there is also a pseudo-real singlet mode $\tilde{s}$ in the spectrum. 
The vacuum  $V^-$ implies  a spontaneous  breaking  of  $SU(N_F)_L \times SU(N_F)_R \to SU(N_F-1)_L \times SU(N_F-1)_R$, and the additional fields $\phi_{L,R}$ appear in the fundamental of the $SU(N_F-1)_{L,R}$ subgroup and as singlet in the other, as well as $\Phi$ which is bifundamental
\begin{equation}
    S = \begin{pmatrix}
        \frac{1}{\sqrt{2}} (v_s + s + i \tilde{s}) \! & \! \phi_R^1 \! & \! \dots \! & \! \phi_R^{N_F - 1} \!\\
        \phi_L^1 \! & \! \Phi^{1,1} \! & \! \dots \! & \! \Phi^{1,N_F-1} \!\\
        \vdots \! & \! \vdots \! & \! \! & \! \vdots \! \\
        \phi_L^{N_F-1} \! & \! \Phi^{N_F-1, 1} \! & \! \dots \! & \! \Phi^{N_F-1,N_F-1} \!
    \end{pmatrix}\,.
\end{equation}
In the $V^+$ configuration, the chiral symmetry collapses to the vectorial part  $SU(N_F)_L \times SU(N_F)_R \to SU(N_F)_V$, giving rise to the real and pseudoreal adjoints $R$ and $I$, respectively. Thus the unbroken scalar decomposes to
\begin{equation}
    S_{ij} = \frac{\delta_{ij}}{\sqrt{2 N_F}} \left( v_s + s + i \tilde{s} \right) + \left(R^a + i I^a\right) t^a_{ij} \,,
\end{equation}
where $t^a_{ij}$ are the (traceless) generators of $SU(N_F)_V$.
Expanding the unbroken, model-dependent scalar potentials 
\eq{VHS1} and \eq{VHSi} in terms of these scalar components yields lengthy expressions. Fortunately, to investigate the mixing with the SM Higgs only the modes $h$ and $s$ matter. This part of the potential can be expressed in a model-independent fashion via \eq{Vbroken},
where the specific expressions in the  different models are obtained from \eq{ModelMatch}.

The potential \eq{Vbroken} is minimized for $\text{d}V(h,s)/\text{d}h=\text{d}V(h,s)/\text{d}s=0$, which yields
\begin{equation}
\mu_H^2=\lambda v_h^2 +\frac{\delta}{\N} v_s^2, \qquad \mu_S^2 = \frac{4\Delta}{\N^2} v_s^2 +\frac{\delta}{\N} v_h^2.
\end{equation}
The quadratic terms in $V_{h,s}$ can be identified with mass terms for the real scalar fields $h$ and $s$ as
\begin{equation}
V(h,s) \subset + \frac{1}{2} m_h^2 h^2 + \frac{1}{2} m_{hs}^2 hs +\frac{1}{2} m_s^2 s^2  \, , 
\end{equation}
\begin{equation} \label{eq:massterms}
\begin{aligned}
m_h^2 =& \frac{\text{d}^2 V(h,s)}{\text{d}h^2}|_{h=s=0} = 2\lambda v_h^2 \, , \\
m_{hs}^2 =& 2 \frac{\text{d}^2 V(h,s)}{\text{d}h \text{d}s}|_{h=s=0} = \frac{4}{\N} \delta v_h v_s \, ,  \\
m_s^2 =& \frac{\text{d}^2 V(h,s)}{\text{d}s^2}|_{h=s=0} = \frac{8}{\N^2} \Delta v_s^2 \, ,
\end{aligned}
\end{equation}
and
\begin{equation}
v_h^2 = \frac{m_h^2}{2 \lambda},\quad v_s^2 = \frac{\N^2 m_s^2}{8 \Delta}, \quad m_{hs}^2 = \frac{\delta}{\sqrt{\lambda \Delta}}m_h m_s \, . 
\end{equation}
The mass terms can be compactly written in matrix form
\begin{equation}
M^2 = \begin{pmatrix} m_s^2 & m_{hs}^2/2\\ m_{hs}^2/2 & m_h^2 \end{pmatrix} = 
\begin{pmatrix} \frac{8}{\N^2} \Delta v_s^2 & \frac{2}{\N} \delta v_h v_s \\ \frac{2}{\N} \delta v_h v_s & 2 \lambda v_h^2 \end{pmatrix} \, .
\end{equation}
The scalars $h$ and $s$ mix into mass eigenstates $h'$ and $s'$
\begin{equation} \label{eq:rot1}
\begin{pmatrix} s' \\ h' \end{pmatrix} = O \begin{pmatrix} s \\ h \end{pmatrix}
\end{equation}
with the orthogonal mixing matrix $O$
\begin{equation} \label{eq:rot2}
O = \begin{pmatrix} \cos\beta & \sin \beta \\ -\sin\beta & \cos \beta \end{pmatrix} \, . 
\end{equation}

The mixing angle $\beta$ is  readily obtained by exploiting that the  (1,2) element of the diagonalized mass matrix
\begin{equation} \label{eq:Mp2}
M^{\prime 2} = O M^2 O^T = \begin{pmatrix} m_{s^\prime}^2 & 0\\  0 & m_{h^\prime}^2 \end{pmatrix} 
\end{equation}
vanishes. This yields
\begin{equation}
\begin{aligned}
\tan 2\beta =& \frac{m^2_{hs}}{m_s^2 - m_h^2}\\
=& \frac{\frac{4}{\N}\delta v_h v_s}{\left(\frac{8}{\N^2} \Delta v_s^2 - 2 \lambda v_h^2 \right)}\\
=& \frac{\delta}{\sqrt{\lambda \Delta}} \frac{m_h m_s}{m_s^2-m_h^2}\, ,
\end{aligned}
\end{equation} 
cf. \eq{beta}. 
Evaluating  $M^2=O^T M^{\prime 2} O$,  \eq{Mp2}, in components yields
\begin{align}
m_s^2& =\cos^2 \beta  \, m_{s^\prime}^2 +\sin^2 \beta \, m_{h^\prime}^2  \, ,  \\
m_h^2& =\cos^2 \beta  \, m_{h^\prime}^2 +\sin^2 \beta \, m_{s^\prime}^2  \, ,  \label{eq:mh2} \\
\sin 2 \beta &=\frac{m_{hs}^2}{m_{s^\prime}^2 -m_{h^\prime}^2 } \, . 
\label{eq:sin2b}
\end{align}
The latter equation demonstrates that for  BSM scalars heavier than the Higgs in Planck-safe models which require $\delta$ positive, hence $m^2_{hs}>0$,
the mixing angle must be positive, too.
The first two equations show that  $m_{h^\prime}^2  \leq m_s^2, m_h^2 \leq m_{s^\prime}^2 $ (for the mass ordering  $m_{h^\prime} < m_{s^\prime}$  which we assume), and that  in the small angle approximation  $0< \beta \ll 1$ the diagonal mass terms coincide,
$m_{s}^\prime\simeq m_s$, $m_{h}^\prime\simeq m_h$. This approximation works well for $m_{s}^\prime$ but breaks down for $m_{h}^\prime$  once 
$\beta m_{s}^\prime \gtrsim m_{h}^\prime$.
Note that \eq{sin2b}   implies 
\begin{equation}
|\delta| < 4 \sqrt{\lambda \Delta}\, \frac{m_{s^\prime}^2- m_{h^\prime}^2}{m_{s} m_{h}}.
\end{equation}

The Higgs quartic $\lambda$  can  be obtained  from  \eq{mh2} as
\begin{equation} \label{eq:lambda}
\begin{aligned}
\lambda = &  \frac{m_h^2}{2 v_h^2} = \lambda_\text{SM} + \frac{\sin^2 \beta}{2 v_h^2} (m_{s^\prime}^2 -m_{h^\prime}^2 )\\
\simeq & \; \lambda_\text{SM} + \frac{\delta^2}{4 \Delta} \left(1 + 
\frac{\mhPrsq}{\msPrsq}
+\mathcal{O}\left(\frac{m_{h^\prime}^4}{m_{s^\prime}^4} \right) 
\right) \, , 
\end{aligned}
\end{equation}
where $\lambda_\text{SM}=\mhPrsq/(2 v_h^2)$, and in the last step we  expanded in small $\beta$ \eq{beta-approx}.
In the main text  physical  masses are denoted by
\begin{equation}
m_{h^\prime} = M_h =125 \,\text{GeV}, \qquad m_{s^\prime} = M_s  \, . 
\end{equation}

We  briefly comment on a contribution  $ - \mu_{det} (det(S)+det(S^\dagger))$ to the scalar potential \eq{VHSi},
which is allowed by the global symmetries. For $N_F=3$  $\mu_{det}$ becomes a dimensionful parameter, which is negligible for the
RG-analysis. On the other hand, it induces in $V^+$ an additional term in  $V(h,s)$,  $-\mu_{det} (s+v_s)^3/(\sqrt{2}\sqrt{N_F}^3)$,
which contributes to 
$\kappa_3$ as
$\mu_{det} \sin^3 \beta /(\sqrt{2}\sqrt{N_F}^3)$.  As it is of  third order in the mixing angle, which is small,  (\ref{eq:betabound}),
and further suppressed by $N_F^{3/2}$, it is negligible for  the phenomenological analysis.
However,  $\mu_{det}$ modifies the minimization condition
$ \mu_S^2 = \Delta  v_s^2 +\frac{\delta}{2} v_h^2   -\frac{v_s \mu_{det}}{\sqrt{2 N_F}} $, and
 breaks the one-to-one correspondence \eq{massterms}  between mass and VEV for the BSM scalar,
$m_s^2=2  \Delta v_s^2 -\frac{v_s \mu_{det}}{\sqrt{2 N_F}} $. The value of $v_s$ depends now on $\mu_{det}$, 
for a given BSM mass. While this does not affect the sign of $m^2_{hs}$ and therefore $\beta$,
it makes the analysis of parameters more involved by adding another degree of freedom. We therefore neglect $\mu_{det}$ for the purpose of this work.
In $V^-$ no such contribution to  \eq{Vbroken} arises.

\section{Unitarity Constraints \label{sec:uni}}

Tree-level perturbative unitarity in $2 \to 2$ scattering of the physical SM and BSM Higgs modes 
$h$ and $s$ with a potential
\begin{equation}
V^{(4)}(h,s) \supset \frac{\lambda}{4} h^4 + \frac{\Delta}{\N^2} s^4 + \frac{\delta}{2\N} h^2 s^2\,
\end{equation}
requires \cite{Dawson:2021jcl}
\begin{equation} \label{eq:unitarity-uni}
\alpha_\lambda \lesssim \frac{1}{6\pi}, \quad \alpha_\Delta \lesssim \frac{\N^2}{24 \pi}, \quad  \alpha_\delta  \lesssim \frac{\N}{4\pi}, \quad
\end{equation}
assuming negligible scalar mixing, see \App{Mix}.

\bibliographystyle{jhep}
\bibliography{ref}

\providecommand{\href}[2]{#2}\begingroup\raggedright\begin{thebibliography}{100}

\bibitem{Buttazzo:2013uya}
D.~Buttazzo, G.~Degrassi, P.~P. Giardino, G.~F. Giudice, F.~Sala, A.~Salvio et~al., \emph{{Investigating the near-criticality of the Higgs boson}}, \href{https://doi.org/10.1007/JHEP12(2013)089}{\emph{JHEP} {\bfseries 12} (2013) 089} [\href{https://arxiv.org/abs/1307.3536}{{\ttfamily 1307.3536}}].

\bibitem{Hiller:2022rla}
G.~Hiller, T.~H\"ohne, D.~F. Litim and T.~Steudtner, \emph{{Portals into Higgs vacuum stability}}, \href{https://doi.org/10.1103/PhysRevD.106.115004}{\emph{Phys. Rev. D} {\bfseries 106} (2022) 115004} [\href{https://arxiv.org/abs/2207.07737}{{\ttfamily 2207.07737}}].

\bibitem{Hiller:2023bdb}
G.~Hiller, T.~H\"ohne, D.~F. Litim and T.~Steudtner, \emph{{Vacuum Stability as a Guide for Model Building}},  in \emph{{57th Rencontres de Moriond on Electroweak Interactions and Unified Theories}}, 5, 2023, \href{https://arxiv.org/abs/2305.18520}{{\ttfamily 2305.18520}}.

\bibitem{ATLAS:2019mfr}
{\scshape ATLAS, CMS} collaboration, \emph{{Addendum to the report on the physics at the HL-LHC, and perspectives for the HE-LHC: Collection of notes from ATLAS and CMS}}, \href{https://doi.org/10.23731/CYRM-2019-007.Addendum}{\emph{CERN Yellow Rep. Monogr.} {\bfseries 7} (2019) Addendum} [\href{https://arxiv.org/abs/1902.10229}{{\ttfamily 1902.10229}}].

\bibitem{FCC:2018byv}
{\scshape FCC} collaboration, \emph{{FCC Physics Opportunities}: {Future Circular Collider Conceptual Design Report Volume 1}}, \href{https://doi.org/10.1140/epjc/s10052-019-6904-3}{\emph{Eur. Phys. J. C} {\bfseries 79} (2019) 474}.

\bibitem{CEPCStudyGroup:2018ghi}
{\scshape CEPC Study Group} collaboration, \emph{{CEPC Conceptual Design Report: Volume 2 - Physics \& Detector}},  \href{https://arxiv.org/abs/1811.10545}{{\ttfamily 1811.10545}}.

\bibitem{ILC:2019gyn}
{\scshape ILC} collaboration, \emph{{The International Linear Collider. A Global Project}},  \href{https://arxiv.org/abs/1901.09829}{{\ttfamily 1901.09829}}.

\bibitem{Casarsa:2023vqx}
M.~Casarsa, D.~Lucchesi and L.~Sestini, \emph{{Experimentation at a muon collider}},  \href{https://arxiv.org/abs/2311.03280}{{\ttfamily 2311.03280}}.

\bibitem{Workman:2022ynf}
{\scshape Particle Data Group} collaboration, \emph{{Review of Particle Physics}}, \href{https://doi.org/10.1093/ptep/ptac097}{\emph{PTEP} {\bfseries 2022} (2022) 083C01}.

\bibitem{CMS:2019esx}
{\scshape CMS} collaboration, \emph{{Measurement of $\mathrm{t\bar t}$ normalised multi-differential cross sections in pp collisions at $\sqrt s=13$ TeV, and simultaneous determination of the strong coupling strength, top quark pole mass, and parton distribution functions}}, \href{https://doi.org/10.1140/epjc/s10052-020-7917-7}{\emph{Eur. Phys. J. C} {\bfseries 80} (2020) 658} [\href{https://arxiv.org/abs/1904.05237}{{\ttfamily 1904.05237}}].

\bibitem{Simon:2019axh}
F.~Simon, \emph{{Scanning Strategies at the Top Threshold at ILC}},  in \emph{{International Workshop on Future Linear Colliders}}, 2, 2019, \href{https://arxiv.org/abs/1902.07246}{{\ttfamily 1902.07246}}.

\bibitem{Litim:2020jvl}
D.~F. Litim and T.~Steudtner, \emph{{ARGES \textendash{} Advanced Renormalisation Group Equation Simplifier}}, \href{https://doi.org/10.1016/j.cpc.2021.108021}{\emph{Comput. Phys. Commun.} {\bfseries 265} (2021) 108021} [\href{https://arxiv.org/abs/2012.12955}{{\ttfamily 2012.12955}}].

\bibitem{Hiller:2019mou}
G.~Hiller, C.~Hormigos-Feliu, D.~F. Litim and T.~Steudtner, \emph{{Anomalous magnetic moments from asymptotic safety}}, \href{https://doi.org/10.1103/PhysRevD.102.071901}{\emph{Phys. Rev. D} {\bfseries 102} (2020) 071901} [\href{https://arxiv.org/abs/1910.14062}{{\ttfamily 1910.14062}}].

\bibitem{Hiller:2020fbu}
G.~Hiller, C.~Hormigos-Feliu, D.~F. Litim and T.~Steudtner, \emph{{Model Building from Asymptotic Safety with Higgs and Flavor Portals}}, \href{https://doi.org/10.1103/PhysRevD.102.095023}{\emph{Phys. Rev. D} {\bfseries 102} (2020) 095023} [\href{https://arxiv.org/abs/2008.08606}{{\ttfamily 2008.08606}}].

\bibitem{Bissmann:2020lge}
S.~Bi\ss{}mann, G.~Hiller, C.~Hormigos-Feliu and D.~F. Litim, \emph{{Multi-lepton signatures of vector-like leptons with flavor}}, \href{https://doi.org/10.1140/epjc/s10052-021-08886-3}{\emph{Eur. Phys. J. C} {\bfseries 81} (2021) 101} [\href{https://arxiv.org/abs/2011.12964}{{\ttfamily 2011.12964}}].

\bibitem{Bause:2021prv}
R.~Bause, G.~Hiller, T.~H\"ohne, D.~F. Litim and T.~Steudtner, \emph{{B-anomalies from flavorful U(1)$'$ extensions, safely}}, \href{https://doi.org/10.1140/epjc/s10052-021-09957-1}{\emph{Eur. Phys. J. C} {\bfseries 82} (2022) 42} [\href{https://arxiv.org/abs/2109.06201}{{\ttfamily 2109.06201}}].

\bibitem{PlanckSafeQuark}
Hiller, Höhne, Litim and Steudtner, \emph{{Planck safety from Vector-like Quarks and Flavorful Scalars}}, {\emph{(in preparation)} (2024) }.

\bibitem{Litim:2014uca}
D.~F. Litim and F.~Sannino, \emph{{Asymptotic safety guaranteed}}, \href{https://doi.org/10.1007/JHEP12(2014)178}{\emph{JHEP} {\bfseries 12} (2014) 178} [\href{https://arxiv.org/abs/1406.2337}{{\ttfamily 1406.2337}}].

\bibitem{Litim:2015iea}
D.~F. Litim, M.~Mojaza and F.~Sannino, \emph{{Vacuum stability of asymptotically safe gauge-Yukawa theories}}, \href{https://doi.org/10.1007/JHEP01(2016)081}{\emph{JHEP} {\bfseries 01} (2016) 081} [\href{https://arxiv.org/abs/1501.03061}{{\ttfamily 1501.03061}}].

\bibitem{Bond:2016dvk}
A.~D. Bond and D.~F. Litim, \emph{{Theorems for Asymptotic Safety of Gauge Theories}}, \href{https://doi.org/10.1140/epjc/s10052-017-4976-5}{\emph{Eur. Phys. J. C} {\bfseries 77} (2017) 429} [\href{https://arxiv.org/abs/1608.00519}{{\ttfamily 1608.00519}}].

\bibitem{Buyukbese:2017ehm}
T.~Buyukbese and D.~F. Litim, \emph{{Asymptotic Safety of Gauge Theories Beyond Marginal Interactions}}, \href{https://doi.org/10.22323/1.256.0233}{\emph{PoS} {\bfseries LATTICE2016} (2017) 233}.

\bibitem{Bond:2017lnq}
A.~D. Bond and D.~F. Litim, \emph{{More asymptotic safety guaranteed}}, \href{https://doi.org/10.1103/PhysRevD.97.085008}{\emph{Phys. Rev. D} {\bfseries 97} (2018) 085008} [\href{https://arxiv.org/abs/1707.04217}{{\ttfamily 1707.04217}}].

\bibitem{Bond:2017suy}
A.~D. Bond and D.~F. Litim, \emph{{Asymptotic safety guaranteed in supersymmetry}}, \href{https://doi.org/10.1103/PhysRevLett.119.211601}{\emph{Phys. Rev. Lett.} {\bfseries 119} (2017) 211601} [\href{https://arxiv.org/abs/1709.06953}{{\ttfamily 1709.06953}}].

\bibitem{Bond:2017tbw}
A.~D. Bond, D.~F. Litim, G.~Medina~Vazquez and T.~Steudtner, \emph{{UV conformal window for asymptotic safety}}, \href{https://doi.org/10.1103/PhysRevD.97.036019}{\emph{Phys. Rev. D} {\bfseries 97} (2018) 036019} [\href{https://arxiv.org/abs/1710.07615}{{\ttfamily 1710.07615}}].

\bibitem{Bond:2017wut}
A.~D. Bond, G.~Hiller, K.~Kowalska and D.~F. Litim, \emph{{Directions for model building from asymptotic safety}}, \href{https://doi.org/10.1007/JHEP08(2017)004}{\emph{JHEP} {\bfseries 08} (2017) 004} [\href{https://arxiv.org/abs/1702.01727}{{\ttfamily 1702.01727}}].

\bibitem{Kowalska:2017fzw}
K.~Kowalska, A.~Bond, G.~Hiller and D.~Litim, \emph{{Towards an asymptotically safe completion of the Standard Model}}, \href{https://doi.org/10.22323/1.314.0542}{\emph{PoS} {\bfseries EPS-HEP2017} (2017) 542}.

\bibitem{Bond:2018oco}
A.~D. Bond and D.~F. Litim, \emph{{Price of Asymptotic Safety}}, \href{https://doi.org/10.1103/PhysRevLett.122.211601}{\emph{Phys. Rev. Lett.} {\bfseries 122} (2019) 211601} [\href{https://arxiv.org/abs/1801.08527}{{\ttfamily 1801.08527}}].

\bibitem{Bond:2019npq}
A.~D. Bond, D.~F. Litim and T.~Steudtner, \emph{{Asymptotic safety with Majorana fermions and new large $N$ equivalences}}, \href{https://doi.org/10.1103/PhysRevD.101.045006}{\emph{Phys. Rev. D} {\bfseries 101} (2020) 045006} [\href{https://arxiv.org/abs/1911.11168}{{\ttfamily 1911.11168}}].

\bibitem{Fabbrichesi:2020svm}
M.~Fabbrichesi, C.~M. Nieto, A.~Tonero and A.~Ugolotti, \emph{{Asymptotically safe SU(5) GUT}}, \href{https://doi.org/10.1103/PhysRevD.103.095026}{\emph{Phys. Rev. D} {\bfseries 103} (2021) 095026} [\href{https://arxiv.org/abs/2012.03987}{{\ttfamily 2012.03987}}].

\bibitem{Bond:2021tgu}
A.~D. Bond, D.~F. Litim and G.~M. Vazquez, \emph{{Conformal windows beyond asymptotic freedom}}, \href{https://doi.org/10.1103/PhysRevD.104.105002}{\emph{Phys. Rev. D} {\bfseries 104} (2021) 105002} [\href{https://arxiv.org/abs/2107.13020}{{\ttfamily 2107.13020}}].

\bibitem{Falkowski:2015iwa}
A.~Falkowski, C.~Gross and O.~Lebedev, \emph{{A second Higgs from the Higgs portal}}, \href{https://doi.org/10.1007/JHEP05(2015)057}{\emph{JHEP} {\bfseries 05} (2015) 057} [\href{https://arxiv.org/abs/1502.01361}{{\ttfamily 1502.01361}}].

\bibitem{Khan:2014kba}
N.~Khan and S.~Rakshit, \emph{{Study of electroweak vacuum metastability with a singlet scalar dark matter}}, \href{https://doi.org/10.1103/PhysRevD.90.113008}{\emph{Phys. Rev. D} {\bfseries 90} (2014) 113008} [\href{https://arxiv.org/abs/1407.6015}{{\ttfamily 1407.6015}}].

\bibitem{Han:2015hda}
H.~Han and S.~Zheng, \emph{{New Constraints on Higgs-portal Scalar Dark Matter}}, \href{https://doi.org/10.1007/JHEP12(2015)044}{\emph{JHEP} {\bfseries 12} (2015) 044} [\href{https://arxiv.org/abs/1509.01765}{{\ttfamily 1509.01765}}].

\bibitem{Garg:2017iva}
I.~Garg, S.~Goswami, K.~N. Vishnudath and N.~Khan, \emph{{Electroweak vacuum stability in presence of singlet scalar dark matter in TeV scale seesaw models}}, \href{https://doi.org/10.1103/PhysRevD.96.055020}{\emph{Phys. Rev. D} {\bfseries 96} (2017) 055020} [\href{https://arxiv.org/abs/1706.08851}{{\ttfamily 1706.08851}}].

\bibitem{Gabrielli:2013hma}
E.~Gabrielli, M.~Heikinheimo, K.~Kannike, A.~Racioppi, M.~Raidal and C.~Spethmann, \emph{{Towards Completing the Standard Model: Vacuum Stability, EWSB and Dark Matter}}, \href{https://doi.org/10.1103/PhysRevD.89.015017}{\emph{Phys. Rev. D} {\bfseries 89} (2014) 015017} [\href{https://arxiv.org/abs/1309.6632}{{\ttfamily 1309.6632}}].

\bibitem{Elias-Miro:2012eoi}
J.~Elias-Miro, J.~R. Espinosa, G.~F. Giudice, H.~M. Lee and A.~Strumia, \emph{{Stabilization of the Electroweak Vacuum by a Scalar Threshold Effect}}, \href{https://doi.org/10.1007/JHEP06(2012)031}{\emph{JHEP} {\bfseries 06} (2012) 031} [\href{https://arxiv.org/abs/1203.0237}{{\ttfamily 1203.0237}}].

\bibitem{Gonderinger:2012rd}
M.~Gonderinger, H.~Lim and M.~J. Ramsey-Musolf, \emph{{Complex Scalar Singlet Dark Matter: Vacuum Stability and Phenomenology}}, \href{https://doi.org/10.1103/PhysRevD.86.043511}{\emph{Phys. Rev. D} {\bfseries 86} (2012) 043511} [\href{https://arxiv.org/abs/1202.1316}{{\ttfamily 1202.1316}}].

\bibitem{Costa:2014qga}
R.~Costa, A.~P. Morais, M.~O.~P. Sampaio and R.~Santos, \emph{{Two-loop stability of a complex singlet extended Standard Model}}, \href{https://doi.org/10.1103/PhysRevD.92.025024}{\emph{Phys. Rev. D} {\bfseries 92} (2015) 025024} [\href{https://arxiv.org/abs/1411.4048}{{\ttfamily 1411.4048}}].

\bibitem{Khoze:2014xha}
V.~V. Khoze, C.~McCabe and G.~Ro, \emph{{Higgs vacuum stability from the dark matter portal}}, \href{https://doi.org/10.1007/JHEP08(2014)026}{\emph{JHEP} {\bfseries 08} (2014) 026} [\href{https://arxiv.org/abs/1403.4953}{{\ttfamily 1403.4953}}].

\bibitem{Anchordoqui:2012fq}
L.~A. Anchordoqui, I.~Antoniadis, H.~Goldberg, X.~Huang, D.~Lust, T.~R. Taylor et~al., \emph{{Vacuum Stability of Standard Model$^{++}$}}, \href{https://doi.org/10.1007/JHEP02(2013)074}{\emph{JHEP} {\bfseries 02} (2013) 074} [\href{https://arxiv.org/abs/1208.2821}{{\ttfamily 1208.2821}}].

\bibitem{Bandyopadhyay:2016oif}
P.~Bandyopadhyay and R.~Mandal, \emph{{Vacuum stability in an extended standard model with a leptoquark}}, \href{https://doi.org/10.1103/PhysRevD.95.035007}{\emph{Phys. Rev. D} {\bfseries 95} (2017) 035007} [\href{https://arxiv.org/abs/1609.03561}{{\ttfamily 1609.03561}}].

\bibitem{Bandyopadhyay:2021kue}
P.~Bandyopadhyay, S.~Jangid and A.~Karan, \emph{{Constraining scalar doublet and triplet leptoquarks with vacuum stability and perturbativity}}, \href{https://doi.org/10.1140/epjc/s10052-022-10418-6}{\emph{Eur. Phys. J. C} {\bfseries 82} (2022) 516} [\href{https://arxiv.org/abs/2111.03872}{{\ttfamily 2111.03872}}].

\bibitem{Chakrabarty:2020jro}
N.~Chakrabarty, \emph{{Doubly charged scalars and vector-like leptons confronting the muon g-2 anomaly and Higgs vacuum stability}}, \href{https://doi.org/10.1140/epjp/s13360-021-02168-3}{\emph{Eur. Phys. J. Plus} {\bfseries 136} (2021) 1183} [\href{https://arxiv.org/abs/2010.05215}{{\ttfamily 2010.05215}}].

\bibitem{Hamada:2015bra}
Y.~Hamada, K.~Kawana and K.~Tsumura, \emph{{Landau pole in the Standard Model with weakly interacting scalar fields}}, \href{https://doi.org/10.1016/j.physletb.2015.05.072}{\emph{Phys. Lett. B} {\bfseries 747} (2015) 238} [\href{https://arxiv.org/abs/1505.01721}{{\ttfamily 1505.01721}}].

\bibitem{Latosinski:2015pba}
A.~Latosinski, A.~Lewandowski, K.~A. Meissner and H.~Nicolai, \emph{{Conformal Standard Model with an extended scalar sector}}, \href{https://doi.org/10.1007/JHEP10(2015)170}{\emph{JHEP} {\bfseries 10} (2015) 170} [\href{https://arxiv.org/abs/1507.01755}{{\ttfamily 1507.01755}}].

\bibitem{Xiao:2014kba}
M.-L. Xiao and J.-H. Yu, \emph{{Stabilizing electroweak vacuum in a vectorlike fermion model}}, \href{https://doi.org/10.1103/PhysRevD.90.014007}{\emph{Phys. Rev. D} {\bfseries 90} (2014) 014007} [\href{https://arxiv.org/abs/1404.0681}{{\ttfamily 1404.0681}}].

\bibitem{Dhuria:2015ufo}
M.~Dhuria and G.~Goswami, \emph{{Perturbativity, vacuum stability, and inflation in the light of 750 GeV diphoton excess}}, \href{https://doi.org/10.1103/PhysRevD.94.055009}{\emph{Phys. Rev. D} {\bfseries 94} (2016) 055009} [\href{https://arxiv.org/abs/1512.06782}{{\ttfamily 1512.06782}}].

\bibitem{Salvio:2015jgu}
A.~Salvio and A.~Mazumdar, \emph{{Higgs Stability and the 750 GeV Diphoton Excess}}, \href{https://doi.org/10.1016/j.physletb.2016.02.057}{\emph{Phys. Lett. B} {\bfseries 755} (2016) 469} [\href{https://arxiv.org/abs/1512.08184}{{\ttfamily 1512.08184}}].

\bibitem{Son:2015vfl}
M.~Son and A.~Urbano, \emph{{A new scalar resonance at 750 GeV: Towards a proof of concept in favor of strongly interacting theories}}, \href{https://doi.org/10.1007/JHEP05(2016)181}{\emph{JHEP} {\bfseries 05} (2016) 181} [\href{https://arxiv.org/abs/1512.08307}{{\ttfamily 1512.08307}}].

\bibitem{DuttaBanik:2018emv}
A.~Dutta~Banik, A.~K. Saha and A.~Sil, \emph{{Scalar assisted singlet doublet fermion dark matter model and electroweak vacuum stability}}, \href{https://doi.org/10.1103/PhysRevD.98.075013}{\emph{Phys. Rev. D} {\bfseries 98} (2018) 075013} [\href{https://arxiv.org/abs/1806.08080}{{\ttfamily 1806.08080}}].

\bibitem{Borah:2020nsz}
D.~Borah, R.~Roshan and A.~Sil, \emph{{Sub-TeV singlet scalar dark matter and electroweak vacuum stability with vectorlike fermions}}, \href{https://doi.org/10.1103/PhysRevD.102.075034}{\emph{Phys. Rev. D} {\bfseries 102} (2020) 075034} [\href{https://arxiv.org/abs/2007.14904}{{\ttfamily 2007.14904}}].

\bibitem{Chowdhury:2015yja}
D.~Chowdhury and O.~Eberhardt, \emph{{Global fits of the two-loop renormalized Two-Higgs-Doublet model with soft Z$_{2}$ breaking}}, \href{https://doi.org/10.1007/JHEP11(2015)052}{\emph{JHEP} {\bfseries 11} (2015) 052} [\href{https://arxiv.org/abs/1503.08216}{{\ttfamily 1503.08216}}].

\bibitem{Khan:2015ipa}
N.~Khan and S.~Rakshit, \emph{{Constraints on inert dark matter from the metastability of the electroweak vacuum}}, \href{https://doi.org/10.1103/PhysRevD.92.055006}{\emph{Phys. Rev. D} {\bfseries 92} (2015) 055006} [\href{https://arxiv.org/abs/1503.03085}{{\ttfamily 1503.03085}}].

\bibitem{Ferreira:2015rha}
P.~Ferreira, H.~E. Haber and E.~Santos, \emph{{Preserving the validity of the Two-Higgs Doublet Model up to the Planck scale}}, \href{https://doi.org/10.1103/PhysRevD.92.033003}{\emph{Phys. Rev. D} {\bfseries 92} (2015) 033003} [\href{https://arxiv.org/abs/1505.04001}{{\ttfamily 1505.04001}}].

\bibitem{Ferreira:2015pfi}
P.~M. Ferreira and B.~Swiezewska, \emph{{One-loop contributions to neutral minima in the inert doublet model}}, \href{https://doi.org/10.1007/JHEP04(2016)099}{\emph{JHEP} {\bfseries 04} (2016) 099} [\href{https://arxiv.org/abs/1511.02879}{{\ttfamily 1511.02879}}].

\bibitem{Bhattacharya:2019fgs}
S.~Bhattacharya, P.~Ghosh, A.~K. Saha and A.~Sil, \emph{{Two component dark matter with inert Higgs doublet: neutrino mass, high scale validity and collider searches}}, \href{https://doi.org/10.1007/JHEP03(2020)090}{\emph{JHEP} {\bfseries 03} (2020) 090} [\href{https://arxiv.org/abs/1905.12583}{{\ttfamily 1905.12583}}].

\bibitem{Swiezewska:2015paa}
B.~Swiezewska, \emph{{Inert scalars and vacuum metastability around the electroweak scale}}, \href{https://doi.org/10.1007/JHEP07(2015)118}{\emph{JHEP} {\bfseries 07} (2015) 118} [\href{https://arxiv.org/abs/1503.07078}{{\ttfamily 1503.07078}}].

\bibitem{Chakrabarty:2016smc}
N.~Chakrabarty and B.~Mukhopadhyaya, \emph{{High-scale validity of a two Higgs doublet scenario: metastability included}}, \href{https://doi.org/10.1140/epjc/s10052-017-4705-0}{\emph{Eur. Phys. J. C} {\bfseries 77} (2017) 153} [\href{https://arxiv.org/abs/1603.05883}{{\ttfamily 1603.05883}}].

\bibitem{Schuh:2018hig}
P.~Schuh, \emph{{Vacuum Stability of Asymptotically Safe Two Higgs Doublet Models}}, \href{https://doi.org/10.1140/epjc/s10052-019-7426-8}{\emph{Eur. Phys. J. C} {\bfseries 79} (2019) 909} [\href{https://arxiv.org/abs/1810.07664}{{\ttfamily 1810.07664}}].

\bibitem{Bagnaschi:2015pwa}
E.~Bagnaschi, F.~Br\"ummer, W.~Buchm\"uller, A.~Voigt and G.~Weiglein, \emph{{Vacuum stability and supersymmetry at high scales with two Higgs doublets}}, \href{https://doi.org/10.1007/JHEP03(2016)158}{\emph{JHEP} {\bfseries 03} (2016) 158} [\href{https://arxiv.org/abs/1512.07761}{{\ttfamily 1512.07761}}].

\bibitem{ATLAS:2012yve}
{\scshape ATLAS} collaboration, \emph{{Observation of a new particle in the search for the Standard Model Higgs boson with the ATLAS detector at the LHC}}, \href{https://doi.org/10.1016/j.physletb.2012.08.020}{\emph{Phys. Lett. B} {\bfseries 716} (2012) 1} [\href{https://arxiv.org/abs/1207.7214}{{\ttfamily 1207.7214}}].

\bibitem{CMS:2012qbp}
{\scshape CMS} collaboration, \emph{{Observation of a New Boson at a Mass of 125 GeV with the CMS Experiment at the LHC}}, \href{https://doi.org/10.1016/j.physletb.2012.08.021}{\emph{Phys. Lett. B} {\bfseries 716} (2012) 30} [\href{https://arxiv.org/abs/1207.7235}{{\ttfamily 1207.7235}}].

\bibitem{Degrassi:2012ry}
G.~Degrassi, S.~Di~Vita, J.~Elias-Miro, J.~R. Espinosa, G.~F. Giudice, G.~Isidori et~al., \emph{{Higgs mass and vacuum stability in the Standard Model at NNLO}}, \href{https://doi.org/10.1007/JHEP08(2012)098}{\emph{JHEP} {\bfseries 08} (2012) 098} [\href{https://arxiv.org/abs/1205.6497}{{\ttfamily 1205.6497}}].

\bibitem{Bezrukov:2012sa}
F.~Bezrukov, M.~Y. Kalmykov, B.~A. Kniehl and M.~Shaposhnikov, \emph{{Higgs Boson Mass and New Physics}}, \href{https://doi.org/10.1007/JHEP10(2012)140}{\emph{JHEP} {\bfseries 10} (2012) 140} [\href{https://arxiv.org/abs/1205.2893}{{\ttfamily 1205.2893}}].

\bibitem{Alekhin:2012py}
S.~Alekhin, A.~Djouadi and S.~Moch, \emph{{The top quark and Higgs boson masses and the stability of the electroweak vacuum}}, \href{https://doi.org/10.1016/j.physletb.2012.08.024}{\emph{Phys. Lett. B} {\bfseries 716} (2012) 214} [\href{https://arxiv.org/abs/1207.0980}{{\ttfamily 1207.0980}}].

\bibitem{Andreassen:2014gha}
A.~Andreassen, W.~Frost and M.~D. Schwartz, \emph{{Consistent Use of the Standard Model Effective Potential}}, \href{https://doi.org/10.1103/PhysRevLett.113.241801}{\emph{Phys. Rev. Lett.} {\bfseries 113} (2014) 241801} [\href{https://arxiv.org/abs/1408.0292}{{\ttfamily 1408.0292}}].

\bibitem{Bednyakov:2015sca}
A.~V. Bednyakov, B.~A. Kniehl, A.~F. Pikelner and O.~L. Veretin, \emph{{Stability of the Electroweak Vacuum: Gauge Independence and Advanced Precision}}, \href{https://doi.org/10.1103/PhysRevLett.115.201802}{\emph{Phys. Rev. Lett.} {\bfseries 115} (2015) 201802} [\href{https://arxiv.org/abs/1507.08833}{{\ttfamily 1507.08833}}].

\bibitem{Chigusa:2018uuj}
S.~Chigusa, T.~Moroi and Y.~Shoji, \emph{{Decay Rate of Electroweak Vacuum in the Standard Model and Beyond}}, \href{https://doi.org/10.1103/PhysRevD.97.116012}{\emph{Phys. Rev. D} {\bfseries 97} (2018) 116012} [\href{https://arxiv.org/abs/1803.03902}{{\ttfamily 1803.03902}}].

\bibitem{PDG2024}
{\scshape Particle Data Group} collaboration, \emph{{(to be published)}}, \href{https://doi.org/XXX/PhysRevD.110.030001}{\emph{Phys. Rev. D} {\bfseries 110} (2024) 030001}.

\bibitem{Alam:2022cdv}
Z.~Alam and S.~P. Martin, \emph{{Standard model at 200~GeV}}, \href{https://doi.org/10.1103/PhysRevD.107.013010}{\emph{Phys. Rev. D} {\bfseries 107} (2023) 013010} [\href{https://arxiv.org/abs/2211.08576}{{\ttfamily 2211.08576}}].

\bibitem{Martin:2019lqd}
S.~P. Martin and D.~G. Robertson, \emph{{Standard model parameters in the tadpole-free pure $\overline{\rm{MS}}$ scheme}}, \href{https://doi.org/10.1103/PhysRevD.100.073004}{\emph{Phys. Rev. D} {\bfseries 100} (2019) 073004} [\href{https://arxiv.org/abs/1907.02500}{{\ttfamily 1907.02500}}].

\bibitem{Baikov:2012zm}
P.~A. Baikov, K.~G. Chetyrkin, J.~H. Kuhn and J.~Rittinger, \emph{{Vector Correlator in Massless QCD at Order $\mathcal{O}(\alpha^4_s)$ and the QED beta-function at Five Loop}}, \href{https://doi.org/10.1007/JHEP07(2012)017}{\emph{JHEP} {\bfseries 07} (2012) 017} [\href{https://arxiv.org/abs/1206.1284}{{\ttfamily 1206.1284}}].

\bibitem{Baikov:2016tgj}
P.~A. Baikov, K.~G. Chetyrkin and J.~H. K\"uhn, \emph{{Five-Loop Running of the QCD coupling constant}}, \href{https://doi.org/10.1103/PhysRevLett.118.082002}{\emph{Phys. Rev. Lett.} {\bfseries 118} (2017) 082002} [\href{https://arxiv.org/abs/1606.08659}{{\ttfamily 1606.08659}}].

\bibitem{Herzog:2017ohr}
F.~Herzog, B.~Ruijl, T.~Ueda, J.~A.~M. Vermaseren and A.~Vogt, \emph{{The five-loop beta function of Yang-Mills theory with fermions}}, \href{https://doi.org/10.1007/JHEP02(2017)090}{\emph{JHEP} {\bfseries 02} (2017) 090} [\href{https://arxiv.org/abs/1701.01404}{{\ttfamily 1701.01404}}].

\bibitem{Luthe:2017ttg}
T.~Luthe, A.~Maier, P.~Marquard and Y.~Schroder, \emph{{The five-loop Beta function for a general gauge group and anomalous dimensions beyond Feynman gauge}}, \href{https://doi.org/10.1007/JHEP10(2017)166}{\emph{JHEP} {\bfseries 10} (2017) 166} [\href{https://arxiv.org/abs/1709.07718}{{\ttfamily 1709.07718}}].

\bibitem{Chetyrkin:2017bjc}
K.~G. Chetyrkin, G.~Falcioni, F.~Herzog and J.~A.~M. Vermaseren, \emph{{Five-loop renormalisation of QCD in covariant gauges}}, \href{https://doi.org/10.1007/JHEP10(2017)179}{\emph{JHEP} {\bfseries 10} (2017) 179} [\href{https://arxiv.org/abs/1709.08541}{{\ttfamily 1709.08541}}].

\bibitem{Melnikov:2000qh}
K.~Melnikov and T.~v. Ritbergen, \emph{{The Three loop relation between the MS-bar and the pole quark masses}}, \href{https://doi.org/10.1016/S0370-2693(00)00507-4}{\emph{Phys. Lett. B} {\bfseries 482} (2000) 99} [\href{https://arxiv.org/abs/hep-ph/9912391}{{\ttfamily hep-ph/9912391}}].

\bibitem{Martin:2018yow}
S.~P. Martin, \emph{{Matching relations for decoupling in the Standard Model at two loops and beyond}}, \href{https://doi.org/10.1103/PhysRevD.99.033007}{\emph{Phys. Rev. D} {\bfseries 99} (2019) 033007} [\href{https://arxiv.org/abs/1812.04100}{{\ttfamily 1812.04100}}].

\bibitem{Martin:2014cxa}
S.~P. Martin and D.~G. Robertson, \emph{{Higgs boson mass in the Standard Model at two-loop order and beyond}}, \href{https://doi.org/10.1103/PhysRevD.90.073010}{\emph{Phys. Rev. D} {\bfseries 90} (2014) 073010} [\href{https://arxiv.org/abs/1407.4336}{{\ttfamily 1407.4336}}].

\bibitem{Martin:2015rea}
S.~P. Martin, \emph{{$Z$-Boson Pole Mass at Two-Loop Order in the Pure $\overline{MS}$ Scheme}}, \href{https://doi.org/10.1103/PhysRevD.92.014026}{\emph{Phys. Rev. D} {\bfseries 92} (2015) 014026} [\href{https://arxiv.org/abs/1505.04833}{{\ttfamily 1505.04833}}].

\bibitem{Martin:2016xsp}
S.~P. Martin, \emph{{Top-quark pole mass in the tadpole-free $\overline {MS}$ scheme}}, \href{https://doi.org/10.1103/PhysRevD.93.094017}{\emph{Phys. Rev. D} {\bfseries 93} (2016) 094017} [\href{https://arxiv.org/abs/1604.01134}{{\ttfamily 1604.01134}}].

\bibitem{Martin:2022qiv}
S.~P. Martin, \emph{{Three-loop QCD corrections to the electroweak boson masses}}, \href{https://doi.org/10.1103/PhysRevD.106.013007}{\emph{Phys. Rev. D} {\bfseries 106} (2022) 013007} [\href{https://arxiv.org/abs/2203.05042}{{\ttfamily 2203.05042}}].

\bibitem{Ford:1992pn}
C.~Ford, I.~Jack and D.~R.~T. Jones, \emph{{The Standard model effective potential at two loops}}, \href{https://doi.org/10.1016/0550-3213(92)90165-8}{\emph{Nucl. Phys. B} {\bfseries 387} (1992) 373} [\href{https://arxiv.org/abs/hep-ph/0111190}{{\ttfamily hep-ph/0111190}}].

\bibitem{Martin:2013gka}
S.~P. Martin, \emph{{Three-Loop Standard Model Effective Potential at Leading Order in Strong and Top Yukawa Couplings}}, \href{https://doi.org/10.1103/PhysRevD.89.013003}{\emph{Phys. Rev. D} {\bfseries 89} (2014) 013003} [\href{https://arxiv.org/abs/1310.7553}{{\ttfamily 1310.7553}}].

\bibitem{Martin:2014bca}
S.~P. Martin, \emph{{Taming the Goldstone contributions to the effective potential}}, \href{https://doi.org/10.1103/PhysRevD.90.016013}{\emph{Phys. Rev. D} {\bfseries 90} (2014) 016013} [\href{https://arxiv.org/abs/1406.2355}{{\ttfamily 1406.2355}}].

\bibitem{Martin:2015eia}
S.~P. Martin, \emph{{Four-Loop Standard Model Effective Potential at Leading Order in QCD}}, \href{https://doi.org/10.1103/PhysRevD.92.054029}{\emph{Phys. Rev. D} {\bfseries 92} (2015) 054029} [\href{https://arxiv.org/abs/1508.00912}{{\ttfamily 1508.00912}}].

\bibitem{Martin:2017lqn}
S.~P. Martin, \emph{{Effective potential at three loops}}, \href{https://doi.org/10.1103/PhysRevD.96.096005}{\emph{Phys. Rev. D} {\bfseries 96} (2017) 096005} [\href{https://arxiv.org/abs/1709.02397}{{\ttfamily 1709.02397}}].

\bibitem{Martin:2018emo}
S.~P. Martin and H.~H. Patel, \emph{{Two-loop effective potential for generalized gauge fixing}}, \href{https://doi.org/10.1103/PhysRevD.98.076008}{\emph{Phys. Rev. D} {\bfseries 98} (2018) 076008} [\href{https://arxiv.org/abs/1808.07615}{{\ttfamily 1808.07615}}].

\bibitem{Casas:1994qy}
J.~A. Casas, J.~R. Espinosa and M.~Quiros, \emph{{Improved Higgs mass stability bound in the standard model and implications for supersymmetry}}, \href{https://doi.org/10.1016/0370-2693(94)01404-Z}{\emph{Phys. Lett. B} {\bfseries 342} (1995) 171} [\href{https://arxiv.org/abs/hep-ph/9409458}{{\ttfamily hep-ph/9409458}}].

\bibitem{Davies:2019onf}
J.~Davies, F.~Herren, C.~Poole, M.~Steinhauser and A.~E. Thomsen, \emph{{Gauge Coupling $\beta$ Functions to Four-Loop Order in the Standard Model}}, \href{https://doi.org/10.1103/PhysRevLett.124.071803}{\emph{Phys. Rev. Lett.} {\bfseries 124} (2020) 071803} [\href{https://arxiv.org/abs/1912.07624}{{\ttfamily 1912.07624}}].

\bibitem{Davies:2021mnc}
J.~Davies, F.~Herren and A.~E. Thomsen, \emph{{General gauge-Yukawa-quartic $\beta$-functions at 4-3-2-loop order}}, \href{https://doi.org/10.1007/JHEP01(2022)051}{\emph{JHEP} {\bfseries 01} (2022) 051} [\href{https://arxiv.org/abs/2110.05496}{{\ttfamily 2110.05496}}].

\bibitem{Bednyakov:2021qxa}
A.~Bednyakov and A.~Pikelner, \emph{{Four-Loop Gauge and Three-Loop Yukawa Beta Functions in a General Renormalizable Theory}}, \href{https://doi.org/10.1103/PhysRevLett.127.041801}{\emph{Phys. Rev. Lett.} {\bfseries 127} (2021) 041801} [\href{https://arxiv.org/abs/2105.09918}{{\ttfamily 2105.09918}}].

\bibitem{Chetyrkin:2012rz}
K.~G. Chetyrkin and M.~F. Zoller, \emph{{Three-loop \textbackslash{}beta-functions for top-Yukawa and the Higgs self-interaction in the Standard Model}}, \href{https://doi.org/10.1007/JHEP06(2012)033}{\emph{JHEP} {\bfseries 06} (2012) 033} [\href{https://arxiv.org/abs/1205.2892}{{\ttfamily 1205.2892}}].

\bibitem{Bednyakov:2012en}
A.~V. Bednyakov, A.~F. Pikelner and V.~N. Velizhanin, \emph{{Yukawa coupling beta-functions in the Standard Model at three loops}}, \href{https://doi.org/10.1016/j.physletb.2013.04.038}{\emph{Phys. Lett. B} {\bfseries 722} (2013) 336} [\href{https://arxiv.org/abs/1212.6829}{{\ttfamily 1212.6829}}].

\bibitem{Chetyrkin:2013wya}
K.~G. Chetyrkin and M.~F. Zoller, \emph{{$\beta$-function for the Higgs self-interaction in the Standard Model at three-loop level}}, \href{https://doi.org/10.1007/JHEP04(2013)091}{\emph{JHEP} {\bfseries 04} (2013) 091} [\href{https://arxiv.org/abs/1303.2890}{{\ttfamily 1303.2890}}].

\bibitem{Bednyakov:2013cpa}
A.~V. Bednyakov, A.~F. Pikelner and V.~N. Velizhanin, \emph{{Three-loop Higgs self-coupling beta-function in the Standard Model with complex Yukawa matrices}}, \href{https://doi.org/10.1016/j.nuclphysb.2013.12.012}{\emph{Nucl. Phys. B} {\bfseries 879} (2014) 256} [\href{https://arxiv.org/abs/1310.3806}{{\ttfamily 1310.3806}}].

\bibitem{Chetyrkin:2016ruf}
K.~G. Chetyrkin and M.~F. Zoller, \emph{{Leading QCD-induced four-loop contributions to the \ensuremath{\beta}-function of the Higgs self-coupling in the SM and vacuum stability}}, \href{https://doi.org/10.1007/JHEP06(2016)175}{\emph{JHEP} {\bfseries 06} (2016) 175} [\href{https://arxiv.org/abs/1604.00853}{{\ttfamily 1604.00853}}].

\bibitem{Hoang:2020iah}
A.~H. Hoang, \emph{{What is the Top Quark Mass?}}, \href{https://doi.org/10.1146/annurev-nucl-101918-023530}{\emph{Ann. Rev. Nucl. Part. Sci.} {\bfseries 70} (2020) 225} [\href{https://arxiv.org/abs/2004.12915}{{\ttfamily 2004.12915}}].

\bibitem{Dehnadi:2023msm}
B.~Dehnadi, A.~H. Hoang, O.~L. Jin and V.~Mateu, \emph{{Top Quark Mass Calibration for Monte Carlo Event Generators -- An Update}},  \href{https://arxiv.org/abs/2309.00547}{{\ttfamily 2309.00547}}.

\bibitem{Andreassen:2017rzq}
A.~Andreassen, W.~Frost and M.~D. Schwartz, \emph{{Scale Invariant Instantons and the Complete Lifetime of the Standard Model}}, \href{https://doi.org/10.1103/PhysRevD.97.056006}{\emph{Phys. Rev. D} {\bfseries 97} (2018) 056006} [\href{https://arxiv.org/abs/1707.08124}{{\ttfamily 1707.08124}}].

\bibitem{Azzi:2019yne}
P.~Azzi et~al., \emph{{Report from Working Group 1}: {Standard Model Physics at the HL-LHC and HE-LHC}}, \href{https://doi.org/10.23731/CYRM-2019-007.1}{\emph{CERN Yellow Rep. Monogr.} {\bfseries 7} (2019) 1} [\href{https://arxiv.org/abs/1902.04070}{{\ttfamily 1902.04070}}].

\bibitem{Parker:2018vye}
R.~H. Parker, C.~Yu, W.~Zhong, B.~Estey and H.~M\"uller, \emph{{Measurement of the fine-structure constant as a test of the Standard Model}}, \href{https://doi.org/10.1126/science.aap7706}{\emph{Science} {\bfseries 360} (2018) 191} [\href{https://arxiv.org/abs/1812.04130}{{\ttfamily 1812.04130}}].

\bibitem{Morel:2020dww}
L.~Morel, Z.~Yao, P.~Clad\'e and S.~Guellati-Kh\'elifa, \emph{{Determination of the fine-structure constant with an accuracy of 81 parts per trillion}}, \href{https://doi.org/10.1038/s41586-020-2964-7}{\emph{Nature} {\bfseries 588} (2020) 61}.

\bibitem{Machacek:1984zw}
M.~E. Machacek and M.~T. Vaughn, \emph{{Two Loop Renormalization Group Equations in a General Quantum Field Theory. 3. Scalar Quartic Couplings}}, \href{https://doi.org/10.1016/0550-3213(85)90040-9}{\emph{Nucl. Phys. B} {\bfseries 249} (1985) 70}.

\bibitem{Ibrahim:2022cqs}
M.~Ibrahim, M.~Ashry, E.~Elkhateeb, A.~M. Awad and A.~Moursy, \emph{{Modified hybrid inflation, reheating, and stabilization of the electroweak vacuum}}, \href{https://doi.org/10.1103/PhysRevD.107.035023}{\emph{Phys. Rev. D} {\bfseries 107} (2023) 035023} [\href{https://arxiv.org/abs/2210.03247}{{\ttfamily 2210.03247}}].

\bibitem{Dawson:2021jcl}
S.~Dawson, P.~P. Giardino and S.~Homiller, \emph{{Uncovering the High Scale Higgs Singlet Model}}, \href{https://doi.org/10.1103/PhysRevD.103.075016}{\emph{Phys. Rev. D} {\bfseries 103} (2021) 075016} [\href{https://arxiv.org/abs/2102.02823}{{\ttfamily 2102.02823}}].

\bibitem{Heikinheimo:2017nth}
M.~Heikinheimo, K.~Kannike, F.~Lyonnet, M.~Raidal, K.~Tuominen and H.~Veerm\"ae, \emph{{Vacuum Stability and Perturbativity of SU(3) Scalars}}, \href{https://doi.org/10.1007/JHEP10(2017)014}{\emph{JHEP} {\bfseries 10} (2017) 014} [\href{https://arxiv.org/abs/1707.08980}{{\ttfamily 1707.08980}}].

\bibitem{Belle-II:2023esi}
{\scshape Belle-II} collaboration, \emph{{Evidence for $B^{+}\to K^{+}\nu\bar{\nu}$ Decays}},  \href{https://arxiv.org/abs/2311.14647}{{\ttfamily 2311.14647}}.

\bibitem{Bause:2023mfe}
R.~Bause, H.~Gisbert and G.~Hiller, \emph{{Implications of an enhanced $B \to K \nu \bar \nu$ branching ratio}}, \href{https://doi.org/10.1103/PhysRevD.109.015006}{\emph{Phys. Rev. D} {\bfseries 109} (2024) 015006} [\href{https://arxiv.org/abs/2309.00075}{{\ttfamily 2309.00075}}].

\bibitem{ATLAS:2019nkf}
{\scshape ATLAS} collaboration, \emph{{Combined measurements of Higgs boson production and decay using up to $80$ fb$^{-1}$ of proton-proton collision data at $\sqrt{s}=$ 13 TeV collected with the ATLAS experiment}}, \href{https://doi.org/10.1103/PhysRevD.101.012002}{\emph{Phys. Rev. D} {\bfseries 101} (2020) 012002} [\href{https://arxiv.org/abs/1909.02845}{{\ttfamily 1909.02845}}].

\bibitem{CMS:2020gsy}
{\scshape CMS} collaboration, \emph{{Combined Higgs boson production and decay measurements with up to 137 fb$^{-1}$ of proton-proton collision data at $\sqrt s$ = 13 TeV}}, {\emph{Report~No.~CMS-PAS-HIG-19-005} (2020) }.

\bibitem{ATLAS:2022vkf}
{\scshape ATLAS} collaboration, \emph{{A detailed map of Higgs boson interactions by the ATLAS experiment ten years after the discovery}}, \href{https://doi.org/10.1038/s41586-022-04893-w}{\emph{Nature} {\bfseries 607} (2022) 52} [\href{https://arxiv.org/abs/2207.00092}{{\ttfamily 2207.00092}}].

\bibitem{CMS:2022dwd}
{\scshape CMS} collaboration, \emph{{A portrait of the Higgs boson by the CMS experiment ten years after the discovery.}}, \href{https://doi.org/10.1038/s41586-022-04892-x}{\emph{Nature} {\bfseries 607} (2022) 60} [\href{https://arxiv.org/abs/2207.00043}{{\ttfamily 2207.00043}}].

\bibitem{Cepeda:2019klc}
M.~Cepeda et~al., \emph{{Report from Working Group 2}: {Higgs Physics at the HL-LHC and HE-LHC}}, \href{https://doi.org/10.23731/CYRM-2019-007.221}{\emph{CERN Yellow Rep. Monogr.} {\bfseries 7} (2019) 221} [\href{https://arxiv.org/abs/1902.00134}{{\ttfamily 1902.00134}}].

\bibitem{ATLAS:2021jki}
{\scshape ATLAS} collaboration, \emph{{Search for Higgs boson pair production in the two bottom quarks plus two photons final state in $pp$ collisions at $\sqrt{s}=13$ TeV with the ATLAS detector}}, {\emph{Report~No.~ATLAS-CONF-2021-016} (2021) }.

\bibitem{Huang:2016cjm}
P.~Huang, A.~J. Long and L.-T. Wang, \emph{{Probing the Electroweak Phase Transition with Higgs Factories and Gravitational Waves}}, \href{https://doi.org/10.1103/PhysRevD.94.075008}{\emph{Phys. Rev. D} {\bfseries 94} (2016) 075008} [\href{https://arxiv.org/abs/1608.06619}{{\ttfamily 1608.06619}}].

\bibitem{Nielsen:1975fs}
N.~K. Nielsen, \emph{{On the Gauge Dependence of Spontaneous Symmetry Breaking in Gauge Theories}}, \href{https://doi.org/10.1016/0550-3213(75)90301-6}{\emph{Nucl. Phys. B} {\bfseries 101} (1975) 173}.

\end{thebibliography}\endgroup

\end{document}